\documentclass[12pt,preprint]{aastex}%
\usepackage{amsfonts}
\usepackage{amssymb}
\usepackage{graphicx}
\slugcomment{Preprint}
\shorttitle{Gravity-powered Jet Dynamo}
\shortauthors{Bellan}

\begin{document}
%
\makeatletter\def\stackunder#1#2{\mathrel{\mathop{#2}\limits_{#1}}}
\makeatother
%

\title{Dust-driven Dynamos in Accretion Disks}%
%

\author{P. M. Bellan}%
%

\affil{Applied Physics, Caltech, Pasadena CA 91125, USA}%
%

\email{pbellan@caltech.edu}%
%

\begin{abstract}%
Magnetically driven astrophysical jets are related to accretion and involve
toroidal magnetic field pressure inflating poloidal magnetic field flux
surfaces. Examination of particle motion in combined gravitational and
magnetic fields shows that these astrophysical jet toroidal and poloidal
magnetic fields can be powered by the gravitational energy liberated by
accreting dust grains that have become positively charged \ by emitting
photo-electrons. Because a dust grain experiences magnetic forces after
becoming charged, but not before, charging can cause irreversible trapping of
the grain so dust accretion is a consequence of charging. Furthermore,
charging causes canonical angular momentum to replace mechanical angular
momentum as the relevant constant of the motion. The resulting effective
potential has three distinct classes of accreting particles distinguished by
canonical angular momentum, namely (i) \textquotedblleft
cyclotron-orbit\textquotedblright, (ii) \textquotedblleft
Speiser-orbit\textquotedblright, and (iii) \textquotedblleft zero canonical
angular momentum\textquotedblright\ particles. Electrons and ions are of class
(i) but depending on mass and initial orbit inclination, dust grains can be of
any class. Light-weight dust grains develop class (i) orbits such that the
grains are confined to nested poloidal flux surfaces, whereas grains with a
critical weight such that they experience comparable gravitational and
magnetic forces can develop class (ii) or class (iii)\ orbits, respectively
producing poloidal and toroidal field dynamos.
\end{abstract}
\ \ \ \ \ \ %

\ 

\pagebreak

\section{Introduction}

Magnetohydrodynamically driven plasma jets having topology and dynamics
analogous to astrophysical jets have been produced in laboratory experiments
by
\citet{Hsu2002,Bellan2005}
and by
\citet{Lebedev2005}%
; see discussion by
\citet{Blackman2007}%
. The feature of azimuthal symmetry, common to both the lab experiments and to
real astrophysical jets, has important implications for the structure of the
magnetic field. This is because an azimuthally symmetric magnetic field can be
expressed using a cylindrical coordinate system $\{r,\phi,z\}$ as$\ $
\begin{equation}
\mathbf{B=}\frac{1}{2\pi}\left(  \nabla\psi\times\nabla\phi+\mu_{0}I\nabla
\phi\right)  \ \label{B general2}%
\end{equation}
where the poloidal flux $\psi(r,z,t)$ is defined by
\begin{equation}
\psi(r,z,t)=\int_{0}^{r}2\pi r^{\prime}dr^{\prime}B_{z}(r^{\prime},z,t)
\label{psi def2}%
\end{equation}
and the poloidal electric current $I(r,z)$ is defined by
\begin{equation}
I(r,z,t)=\int_{0}^{r}2\pi r^{\prime}dr^{\prime}J_{z}(r^{\prime},z,t).
\label{Idef2}%
\end{equation}
The definition of $I(r,z)$ is consistent with Ampere's law for the toroidal
field, since using $\nabla\phi=\hat{\phi}/r\ $in Eq.\ref{B general2} gives the
toroidal magnetic field to be
\begin{equation}
B_{\phi}=\frac{\mu_{0}I}{2\pi r}. \label{Bphi}%
\end{equation}

Equations \ref{B general2}-\ref{Idef2} describe the magnetic field and
electric current distribution of \textit{any} axisymmetric magnetic field.
Because astrophysical jets are azimuthally symmetric, their magnetic field
must be of the form prescribed by Eqs.\ref{B general2}- \ref{Idef2} and,
indeed, it is generally believed that astrophysical jets involve large-scale
poloidal magnetic fields threading an accretion disk (e.g., see
\citet{Livio2002,Ferreira2004}%
) and in addition, toroidal magnetic fields. Application of Ampere's law to
Eq.\ref{B general2} shows that the poloidal and toroidal currents are
respectively given by
\begin{equation}
\mathbf{J}_{pol}=\frac{1}{2\pi}\nabla I\times\nabla\phi\label{Jpol2}%
\end{equation}
and
\begin{equation}
\mathbf{J}_{tor}=\ -\frac{r^{2}\nabla\cdot\left(  r^{-2}\nabla\psi\right)
}{2\pi\mu_{0}}\ \nabla\phi\label{Jtor}%
\end{equation}
showing that poloidal magnetic fields are produced by a toroidal electric
current and toroidal magnetic fields are produced by a poloidal current;
toroidal vectors are those vectors in the $\phi$ direction and poloidal
vectors are those in any combination of the $r$ and $z$ directions.
\ Knowledge of the two stream-function quantities $I(r,z,t)$ and $\psi(r,z,t)$
is thus necessary and sufficient to determine the complete vector magnetic
field and the complete vector current density.

The term `magnetic axis' has traditionally been assigned different meanings in
the respective contexts of astrophysics and laboratory toroidal magnetic
confinement devices (e.g., tokamaks, reversed field pinches, or spheromaks).
Specifically, a local maximum in $r$-$z$ space of $\psi(r,z)$ is called a
magnetic axis in the context of toroidal confinement devices whereas the $z$
symmetry axis of the magnetic field is called the magnetic axis in \ the
context of astrophysics. To avoid confusion, we will call the location of a
maximum of $\psi(r,z)$ the poloidal flux magnetic axis. Poloidal magnetic
field lines follow level contours of $\psi$ and so one can envision the
projection of the magnetic field in the $r$-$z$ plane as being like a set of
roads, each at a different altitude, encircling a mountain peak at a specific
$r$-$z$ location which is the poloidal flux magnetic axis (also called an
O-point). \ Since a toroidal current at infinity is not physical, and since
the net magnetic flux enclosed by a circle with infinite radius must vanish as
field lines cannot go to infinity, $\psi$ must vanish at infinity.
Furthermore, mathematical regularity of physical quantities requires $\psi$ to
vanish on the $z$ axis
\citep{Lewis1990}%
. Thus, a non-trivial $\psi$ can only be finite in the region $0<r<\infty$,
$-\infty<z<\infty.$ The simplest situation of physical interest is therefore
where $\psi$ has a single maximum in the $r$-$z$ plane. We will consider this
situation, namely a single poloidal field magnetic axis with $\psi(r,z)$
symmetric with respect to $z.$ This situation has been previously considered
by
\citet{Lovelace2002}
and implies via Eq.\ref{Jtor} that a toroidal current circulates in an
accretion disk to produce the poloidal magnetic field
\begin{equation}
\mathbf{B}_{pol}=\mathbf{\ }\frac{1}{2\pi}\nabla\psi\times\nabla\phi.
\label{Bpol def}%
\end{equation}
An inescapable feature of this topology is that because $\nabla\psi=0$ at the
maximum of $\psi$, i.e., at the poloidal field magnetic axis, the poloidal
magnetic field has a null on the poloidal field magnetic axis. In the $z=0$
plane, the poloidal flux $\psi$ thus starts from zero at $r=0,$ increases to a
maximum at the poloidal field magnetic axis, and then decays to zero as
$r\rightarrow\infty.$ \ 

We define $a\ $to be the radius of the poloidal field magnetic axis. In
addition, we define $\left\langle B_{z}\right\rangle $ to be the
\textit{spatially-averaged} axial magnetic field linked by the poloidal field
magnetic axis and $\psi_{0}$ to be the value of the poloidal magnetic flux at
the poloidal field magnetic axis, so
\begin{equation}
\left\langle B_{z}\right\rangle =\frac{\int_{0}^{a}dr2\pi rB_{z}(r,0)}%
{\int_{0}^{a}dr2\pi r}=\frac{\psi_{0}}{\pi a^{2}}. \label{Bbar}%
\end{equation}
\qquad\qquad

The axial field $B_{z}=(2\pi r)^{-1}\partial\psi/\partial r$ reverses sign at
$r=a$ and the radial field $B_{r}=-(2\pi r)^{-1}\partial\psi/\partial z$
reverses sign at $z=0.$ $\ $An analytic representation for a physically
realizable generic flux function satisfying all these properties is derived in
Appendix \ref{Generic Flux Function}. This generic flux function is
\begin{equation}
\psi(r,z)=\ \ \frac{27\left(  r/a\right)  ^{2}\psi_{0}}{8\left(  \left(
\frac{r}{a}+\frac{1}{2}\right)  ^{2}+\left(  \frac{z}{a}\right)  ^{2}\right)
^{3/2}} \label{generic}%
\end{equation}
and has the properties that (i)\ $\psi(r,z)$ has a maximum value of $\psi_{0}$
at $r=a,$ $z=0,$ (ii) $\psi\sim r^{2}$ for $r\ll a$ and $z=0,$ (iii) $\psi\sim
r^{-1}$ for $r\gg a,z$ and (iv) for $r\ll a/2$ or $r\gg a\ $ and for $z\gg a$
the contours of $\psi$ are identical to the contours of the poloidal flux
produced by a current loop located at $r=a/2,$ $z=0.$ This flux function thus
encompasses simpler models which assume a uniform axial magnetic field $B_{z}%
$; these simpler models would correspond to the $r,z\ll a$ region here since
in this region $\psi\sim r^{2}$ which corresponds to having a uniform axial
magnetic field $B_{z}.$ This flux function could also\ be used to describe the
far-field of a dipole by assuming that $r,z\gg a.$ Since any real axial
magnetic field must always be generated by a toroidal current located at some
finite radius, any real situation will have a poloidal flux function
qualitatively similar to Eq.\ref{generic}. The flux function prescribed in
Eq.\ref{generic} is similar in essence to the flux function used in Fig.1 of
\citet{Lovelace2002}%
.

\begin{figure}[ptb]
\caption{Plot of the normalized generic flux function $\psi(r,z)/\psi_{0}\ $in
coordinates normalized to the radius of the poloidal field magnetic axis, that
is to the radial position of the maximum of $\psi(r,z)\,.$ Contours of
iso-surfaces shown on top; these correspond to projection of poloidal magnetic
field onto $r$-$z$ plane.}%
\plotone{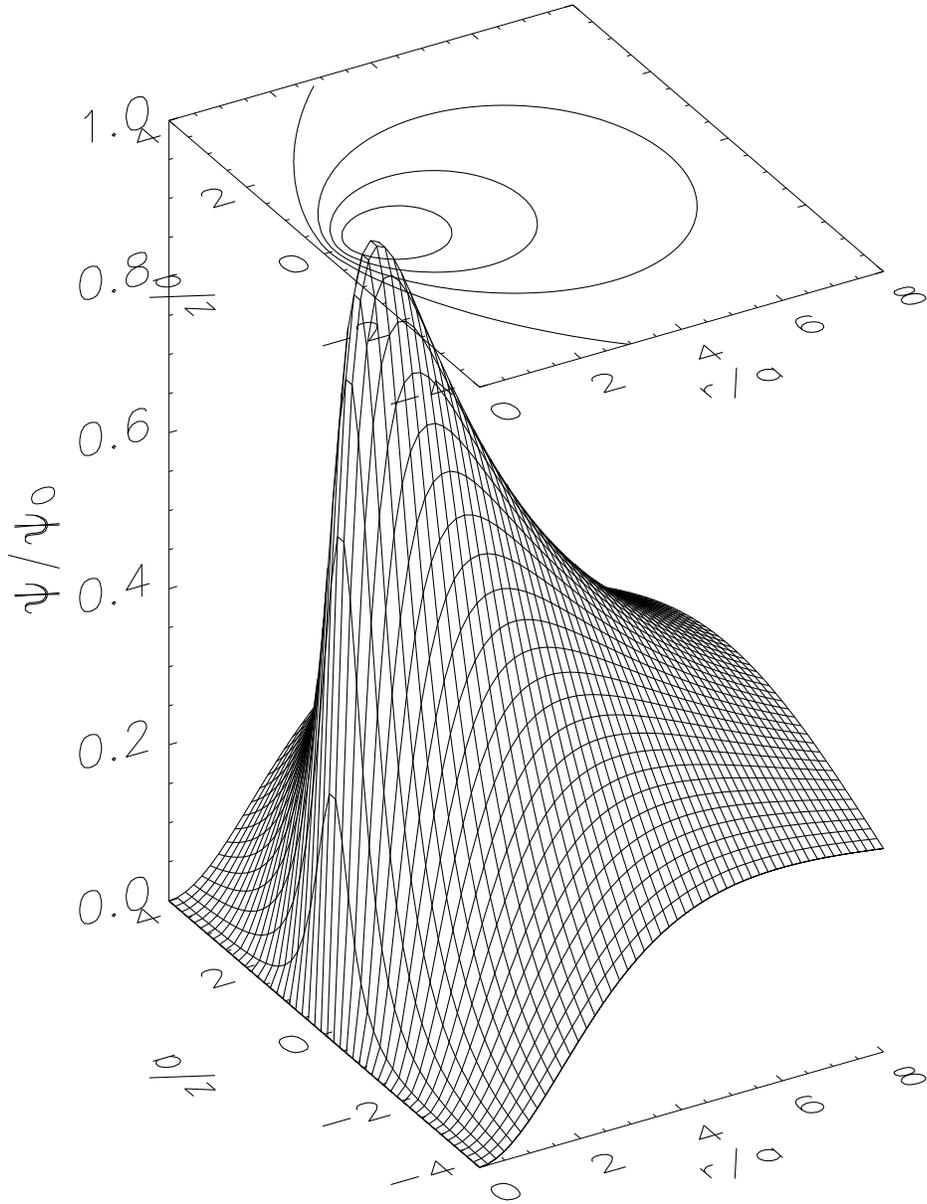}\label{GenericFluxFunction-new}%
\end{figure}

Figure \ref{GenericFluxFunction-new} plots $\psi(r,z)$ as prescribed by
Eq.\ref{generic} and shows that $\psi(r,z)$ has its maximum at the poloidal
field magnetic axis $r\ =a,$ $z=0$. This flux function corresponds to a
smoothly varying toroidal current density prescribed by Eq.\ref{Jtor}
concentrated in the vicinity of $r=a,$ $z=0.$ Since for $z=0$ and small $r$,
this function has the asymptotic dependence $\psi\simeq27\psi_{0}\left(
r/a\right)  ^{2}$, it corresponds to an approximately uniform axial magnetic
field $B_{z}\simeq27\psi_{0}/\pi a^{2}$ for $r,z\ll a.$ The $r\ll a$
inner-region $B_{z}$ is thus 27 times stronger than the average $B_{z}$ field
between $0$ and $a.$ The total toroidal current $\mathcal{I}_{\phi}$
associated with the generic flux function given by Eq.\ref{generic} is
calculated in Appendix \ref{current associated with flux function} using the
integral form of Ampere's law and found to be
\begin{equation}
\mathcal{I}_{\phi}=\frac{27\psi_{0}}{\ \pi a\mu_{0}}\ \ . \label{Iphi}%
\end{equation}

The laboratory jets involve the mutual interaction between poloidal and
toroidal magnetic fields powered by laboratory capacitor banks. The jet
acceleration mechanism results from the pressure of the toroidal magnetic
field inflating flux surfaces associated with the poloidal magnetic field. The
question arises as to what powers the toroidal and poloidal magnetic fields in
an actual astrophysical situation. Existing models of astrophysical jets are
based on the magnetohydrodynamic (MHD) approximation of plasma behavior and
typically assume (i)\ the poloidal field is pre-existing and (ii)\ the
toroidal field results from a rotating accretion disk twisting up this assumed
primordial poloidal field. The purpose of this paper is to present an
alternate model wherein it is postulated that the toroidal and poloidal field
result instead from a non-MHD dusty plasma dynamo mechanism that converts the
gravitational energy of infalling dust grains into an electrical power source
that drives poloidal and toroidal electric currents creating the respective
toroidal and poloidal fields. A brief outline of how infalling charged dust
can drive poloidal currents has been presented in
\citet{Bellan2007}%
.

This model obviously requires existence of sufficient infalling dust to
provide the jet power. Since the dust-to-gas mass ratio in the Interstellar
Medium (ISM) is 1\%, one might be tempted to argue that any jet driven by the
proposed dust infall mechanism would be limited to having\ less than 1\% of
the power available from infalling gas, a constraint that would contradict
observations. However, in
\citet{Bellan2008}
(to be referred to as Paper I), we showed that the dust-to-gas mass ratio in a
molecular cloud can be substantially enriched compared to the ISM\ value
(e.g., the dust to gas mass ratio in a molecular cloud could be enriched
20-fold compared to the 1\% ISM value). This enrichment occurs because
accreting dust slows down much more in proportion to its initial velocity than
does accreting gas so that the density amplification resulting from dust
slowing down is much greater than the corresponding density amplification of gas.

The condition for the toroidal magnetic field to inflate the poloidal magnetic
field and create a jet can be expressed as
\begin{equation}
\frac{\mu_{0}I}{\psi}>\lambda\label{lambda threshold}%
\end{equation}
where $\lambda$ is a parameter of the order of the inverse characteristic
linear dimension in the radial direction. The ratio $I/\psi$ can be thought of
as the ratio of the electric current flowing along a flux tube to the magnetic
flux content of the flux tube and is proportional to the twist of the magnetic
field. Equation \ref{lambda threshold}, well-established in spheromak
formation physics
\citep{Barnes1990,Jarboe1994,Geddes1998,Bellan2000,Hsu2005}%
, is essentially a statement that jet expansion (i.e., poloidal field
inflation)\ occurs when the toroidal magnetic field pressure force $\sim
B_{\phi}^{2}A_{1}$ acting on area $A_{1}$ exceeds the restraining force
$B_{z}^{2}A_{2}$ of the poloidal magnetic field `tension' acting on area
$A_{2}$. Here $A_{1}$ and $A_{2}$ are not exactly the same because the
toroidal and poloidal fields do not act over the same areas. \ The equivalence
between Eq. \ref{lambda threshold} and the condition $B_{\phi}^{2}>B_{z}%
^{2}A_{2}/A_{1}$ is seen by substituting $\mu_{0}I=2\pi aB_{\phi}$ from
Ampere's law and $\psi\sim B_{z}\pi a^{2}$ in Eq.\ref{lambda threshold}.

Paper I divided the regions of interest into successively smaller concentric
regions and considered dust and gas behavior in the outermost regions.
Simultaneous gas and dust accretion were considered and it was shown that the
dust could be considered as a perturbation on the gas, so that the gas
accretion problem could be solved first without considering dust and then the
solution of this gas accretion problem could be used as an input for the dust
accretion problem. Below is a listing showing which regions were considered in
Paper I, which are considered in this paper, and which will be considered in a
future paper; the nominal radii scales and star mass are from Table 3 in Paper I:

\begin{description}
\item \textit{ISM scale (considered in Paper I):} The outermost scale is that
of the Interstellar Medium (ISM). The ISM\ has a gas density $\sim$10$^{7}$
m$^{-3}$, a dust-to-gas mass ratio of 1 percent, a gas temperature
$T_{g}^{ISM}\sim100$ K$,$ and is optically thin. The ISM is assumed to be
spatially uniform and to bound a molecular cloud having radius $r_{edge}$
$\sim10^{5}$ a.u.

\item \textit{Molecular cloud scale (considered in Paper I):}\ The molecular
cloud scale has much higher density than the ISM and is characterized by force
balance between gas self-gravity\ and gas pressure. The molecular cloud scale
is sub-divided into a large, radially non-uniform low-density outer region and
a small, approximately uniform, high-density inner core region. Clouds have a
characteristic scale given by the Jeans length $r_{J}\sim1.4\times10^{4}$ a.u.
The radial dependence of gas density is provided by the Bonnor-Ebert sphere
solution which acts as the outer boundary of the Bondi accretion scale.

\item \textit{Bondi accretion scale (considered in Paper I):} The Bondi
accretion scale is $\sim r_{B}\sim4.3\times10^{3}$ a.u. which is sufficiently
small that gas self-gravity no longer matters so equilibrium is instead
obtained by force balance between gas pressure and the gravity of a central
object assumed to be a star having mass $M\sim0.4M_{\odot}$ The Bondi scale is
sub-divided into three concentric radial regions: an outermost region where
the gas flow is subsonic, a critical transition radius at exactly $r_{B}$
where the flow is sonic, and an innermost region where the gas flow is
free-falling and supersonic.

\item \textit{Collisionless dusty plasma scale (considered in this paper): }
Free-falling dust grains collide with each other in one of the above scales
and coagulate to form large-radius grains which are collisionless and
optically thin. The optically thin dust absorbs UV photons from the star,
photo-emits electrons and becomes electrically charged. The charged dust
grains are subject to electromagnetic forces in addition to gravity. Motions
of charged dust grains relative to electrons result in electric currents with
associated poloidal and toroidal magnetic fields [see preliminary discussion
in
\citet{Bellan2007}%
]. \ This region is assumed to have a scale of $10-10^{3}$ a.u.\ \ and
corresponds to the scale of $a,$ the radius of the poloidal magnetic field
axis. It is assumed that a distributed toroidal current peaked at a nominal
radius $a$ is responsible for producing a poloidal field having the generic
profile given in Fig.\ref{GenericFluxFunction-new}.

\item \textit{Jet scale (to be considered in a future publication): }The
electric currents interact with the magnetic fields to produce
magnetohydrodynamic forces that drive astrophysical jets in a manner
consistent with Eq.\ref{lambda threshold} and analogous to that reported in \
\citet{Hsu2002,Hsu2005}
and
\citet{Bellan2005}%
. This region is assumed to have a scale $\ll10^{3}$ a.u., possibly as small
as a few a.u. and will involve a deformation of the generic poloidal field
profile given in Fig.\ref{GenericFluxFunction-new} because of the pressure of
toroidal magnetic field inflating the poloidal flux surfaces.
\end{description}

\section{Outline of model}

We will show how infalling collisionless dust grains can develop special three
dimensional orbits suitable for sustaining both toroidal and poloidal dynamos.
This result is obtained by considering Hamiltonian particle dynamics in the
combination of the gravitational field of a star with mass $M$ and a
three-dimensional axisymmetric magnetic field topology consistent with
previous models of magnetically driven astrophysical jets [e.g.,
\citet{Lovelace1976}%
,
\citet{Li2001}%
,
\citet{Lovelace2002}%
, and
\citet{Lynden-Bell2003}%
]. The reason why dust grains develop these special orbits will be shown to be
due to charging of dust grains via photo-emission of electrons. The analysis
involves using Hamiltonian mechanics to generalize the centrifugal potential
so as to include magnetic force, i.e., the St\"{o}rmer effective potential is
used. St\"{o}rmer potentials have been previously used for investigating
auroral particles
\citep{Stormer1955}%
, electron and ion motion in the magnetosphere
\citep{Shebalin2004,Lemaire2003}
and most recently, charged dust grain motion in the magneto-gravitational
fields of Saturn and Jupiter
\citep{Dullin2002,Mitchell2003}%
. St\"{o}rmer potentials are also commonly used to characterize particle
orbits in tokamaks
\citep{Rome1979}
and St\"{o}rmer potentials were found to be important in the MHD-driven jet
experiment reported by
\citet{Tripathi2007}%
. We will restrict the analysis to showing how toroidal and poloidal field
dynamos can be sustained in steady state by these special Hamiltonian particle
orbits; the much more complicated problem of how a dynamo grows from a seed
magnetic field will not be addressed here. These special orbits are quite
different from conventional cyclotron orbits. As reviewed in Appendix
\ref{Orbit review} a dynamo cannot be sustained by particles executing
cyclotron orbits because cyclotron orbits and associated drifts are
diamagnetic, i.e., create magnetic fields that oppose the field in which the
particle is orbiting.

The importance of a Hamiltonian analysis can be appreciated by considering the
gedanken experiment where the charge to mass ratio of a particle in \ a
combined gravitational-magnetic field is assumed to be increased from zero
(neutral particle) to that of an electron or ion. The particle will thus make
a transition from Kepler to cyclotron orbital motion. The details of how this
transition occurs have been examined by
\citet{Bellan2007}
in the context of uniform-magnetic-field orbits restricted to a plane. The
present paper will address this issue in the more general context of three
dimensional particle orbits in a spatially non-uniform three dimensional
magnetic field having dipole-like topology appropriate for an accretion disk;
similar dipole topology has been previously invoked for accretion disks by
\citet{Lovelace2002}%
. Our analysis identifies five distinct classes of orbits and shows that the
class to which a given charged particle belongs depends both on its\ charge to
mass ratio and on the circumstances under which the charged particle was
created from an initially neutral particle. The interaction between the
distinct symmetries of the magnetic and gravitational fields removes the
isotropy of the incident neutral particles existent prior to charging so that
the newly formed charged particles separate into groups having qualitatively
different types of orbits. Some orbits correspond to a simple accretion, some
involve accretion and production of a dynamo driving toroidal current, and
some involve accretion and a dynamo driving poloidal current. The type of
orbit a charged particle develops depends on both the angular momentum and the
angle of incidence of the parent neutral particle.

The paper is organized with the goal of being concise while also realizing
that some readers may not be familiar with the concepts of adiabatic versus
non-adiabatic orbits, Speiser orbits, St\"{o}rmer effective potentials, and
how conservation of canonical angular momentum results in confinement of an
adiabatic particle to the vicinity of a poloidal flux surface. Rather than
reviewing these concepts in an introductory section , they are instead
discussed in appendices.

\section{Reduction of collisionality due to dust agglomeration}

Paper I showed that dust grains are collisionally decoupled from gas in the
ISM and then become collisionally coupled to gas in the Bonner-Ebert and Bondi
regions of a molecular cloud. Because of the spherical focusing of the dust
and gas inflows, the dust density increases to a level such that dust-dust
collisions become important. When dust grains collide with each other they may
agglomerate to form larger dust grains.
\citet{Przygodda2003}
and
\citet{vanBoekel2003}
have reported direct observational evidence of grain growth in circumstellar
disks while, in addition,
\citet{Jura1980}%
,
\citet{Miyake1993}%
,
\citet{Pollack1994}%
,
\citet{D'Alessio2001}%
, and
\citet{Dullemond2005}
provided detailed calculations showing a strong tendency for dust grain growth
when dust grains collide with each other. This agglomeration will increase the
dust grain radius $r_{d}$ while keeping the dust mass density $\rho_{d}$
constant. We will consider first how this agglomeration affects dust-gas
collisions and then how it affects dust-dust collisions.

Since the mean free path is much larger than the grain radius, the drag force
on a dust grain due to collisions with gas molecules is of the Epstein-type
and given by
\citep{Lamers1999}
\begin{equation}
F_{drag}=-(u_{d}-u_{g})\rho_{g}\sigma_{d}\sqrt{c_{g}^{2}+(u_{d}-u_{g})^{2}}
\label{collision freq}%
\end{equation}
where $c_{g}$ is the gas thermal velocity, $u_{d}$ is the dust grain velocity,
$\sigma_{d}$ is the dust grain cross-sectional area, and $u_{g}$ is the mean
velocity of the gas (i.e., the fluid velocity). In the innermost Bondi region
where flow is supersonic, we may approximate $c_{g}\simeq0$ and work in a
frame moving with\ $u_{g}$ by defining $\Delta u_{d}=u_{d}-u_{g}.$ The dust
equation of motion in this frame is thus%

\begin{equation}
m_{d}\frac{d\Delta u_{d}}{dt}=-\left(  \Delta u_{d}\right)  ^{2}\rho_{g}%
\sigma_{d}.\ \label{collision}%
\end{equation}
Defining $\xi$ to be distance in the direction of dust motion so $\Delta
u_{d}=d\xi/dt,$ Eq.\ref{collision} can be recast as%
\begin{equation}
\frac{d\Delta u_{d}}{d\xi}\Delta u_{d}=-\left(  \Delta u_{d}\right)
^{2}\ \sigma_{d}\frac{\rho_{g}}{m_{d}}\ .
\label{collision with integrating factor}%
\end{equation}
Integration gives%
\begin{equation}
\Delta u_{d}(\xi)=\Delta u_{d}(0)\exp\left(  -\xi/l_{dg}\right)
\label{slowing down}%
\end{equation}
where the dust-gas collision mean free path is
\begin{equation}
l_{dg}=\frac{m_{d}}{\rho_{g}\sigma_{d}}. \label{mean free path}%
\end{equation}

Since the dust cross-section and mass are given respectively by
\begin{equation}
\sigma_{d}=\pi r_{d}^{2} \label{sigma_d}%
\end{equation}
and%
\begin{equation}
m_{d}=\frac{4\pi r_{d}^{3}\rho_{d}^{int}}{3}\ \label{md}%
\end{equation}
where $\ $ $\rho_{d}^{int}$ is the intrinsic density of a dust grain, the
dust-gas collision mean free path can be expressed as%
\begin{equation}
l_{dg}=\frac{4\rho_{d}^{int}}{3\rho_{g}}r_{d}\ \label{ldg rd}%
\end{equation}
which shows that dust agglomeration increases the dust-gas mean free path and
so will tend to make dust collisionless with respect to gas.

Let us now consider how agglomeration affects dust-dust collisions. We first
note that the condition for dust-dust collisions to be significant is closely
related to the condition for the dust to be optically thick: if $l$ is the
characteristic length of a configuration, the condition for collisions to be
significant is $\rho_{d}\sigma_{d}l/m_{d}>1$ whereas the condition for the
dust to be optically thick is $Q_{eff}\rho_{d}\sigma_{d}l/m_{d}>1$ where $\ $
$Q_{eff}$ is an extinction efficiency parameter that depends on the ratio of
the dust radius to the light wavelength. The dust-dust collision mean free
path is thus%
\begin{equation}
l_{dd}=\frac{m_{d}}{\rho_{d}\sigma_{d}}=\frac{4\rho_{d}^{int}}{3\rho_{d}}%
r_{d}\ \label{ldd}%
\end{equation}
so if, as argued in Paper I, \ the dust mass density $\rho_{d}$ has been
enriched to be a significant fraction of the gas mass density $\rho_{g},$ the
dust-dust collision mean free path $l_{dd}$ will be the same order of
magnitude as the dust-gas mean free path $l_{dg}.$ Agglomeration will thus
tend to increase both the dust-dust and dust-gas collision mean free paths,
and furthermore will cause the dust to become optically thin. We will assume
that dust grains agglomerate when the dust number density $\ n_{d}=\rho
_{d}/m_{d}$ becomes sufficiently large for dust-dust collisions to occur and
that this agglomeration results in an increase in $r_{d}$ until the dust
grains become collisionless and optically thin again. We will not attempt to
follow the dynamics of the agglomeration process, relying instead on the
analysis in the papers cited above. Our starting point then will be assuming
the existence of collisionless dust grains exposed to star light, having
radius $r_{d}$ larger than in the ISM, and as discussed in Paper I, having a
dust to gas mass density ratio substantially enriched compared to the 1\%
value in the ISM.

\section{Review: Neutral particle motion in a gravitational
field\label{Kepler}}

For reference and in order to define terms to be used later in a more complex
context, we first review the elementary problem of the motion of a neutral
particle of mass $m$ in the gravitational field of a star of mass $M$. The
particle we have in mind could be a a dust grain with radius $r_{d}$
sufficiently large to be collisionless over the distance from its starting
point to the star.

\begin{figure}[ptb]
\caption{Lab frame has Cartesian coordinates $x,y,z$ and the magnetic field is
axisymmetric with respect to the lab frame $z$ axis. The orbital plane of a
neutral particle is normal to the neutral particle angular momentum vector
$\mathbf{L}$ which is tilted by an angle $\theta$ with respect to the $z$
axis. The orbital plane Cartesian coordinates are $x^{\prime},y^{\prime}$
where the $x^{\prime}$ axis is coincident with the $x$ axis. \ The neutral
particle makes a circular Kepler, elliptical Kepler, or cometary orbit in its
orbital plane (cometary orbit shown).}%
\label{Orbital-Plane-coordinate-system}%
\plotone{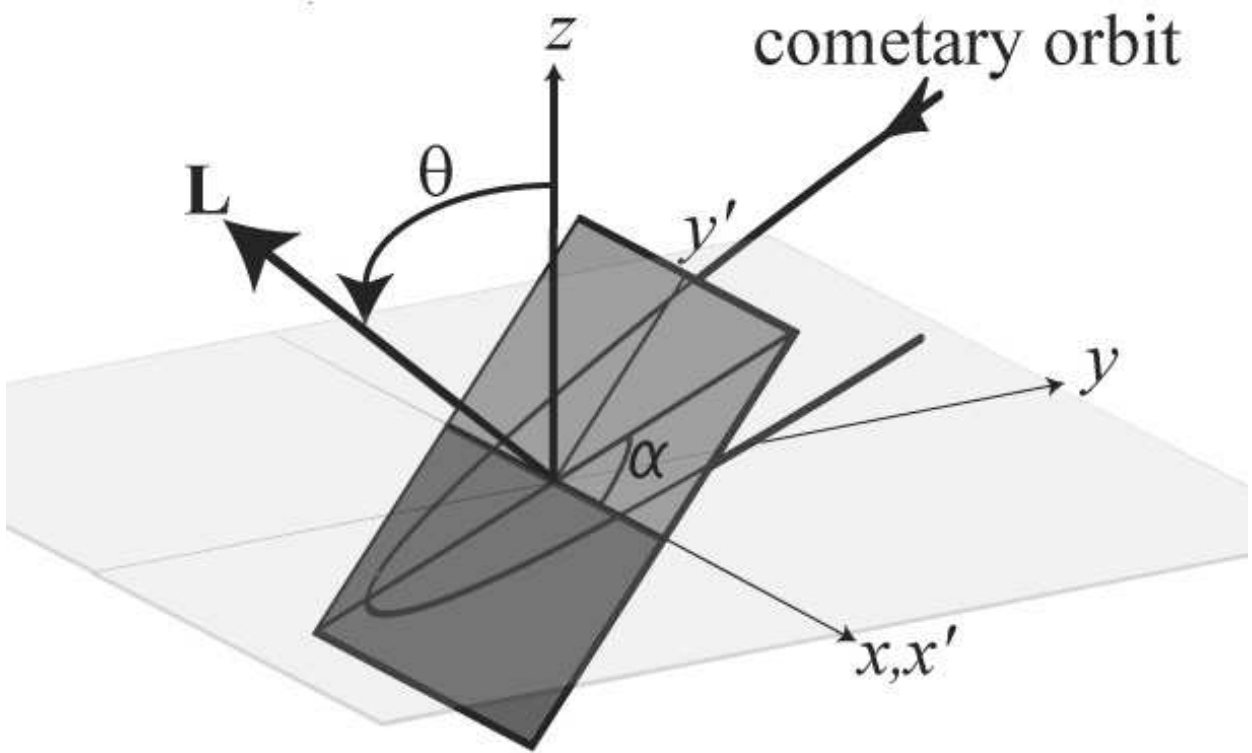}\end{figure}

The equations governing the motion of this neutral particle are spherically
symmetric whereas the motions of a charged particle in an azimuthally
symmetric electromagnetic field are cylindrically symmetric. An axisymmetric
magnetic field is assumed to exist in the lab frame and the $z$ axis is
defined by the direction of this magnetic field at the origin. Although the
neutral particle trajectory is unaffected by this magnetic field, we
nevertheless use the magnetic field coordinate system to define the lab frame.
Depending on what is being emphasized, the lab frame will be characterized by
either a cylindrical coordinate system $\{r,\phi,z\}\ $or by a Cartesian
coordinate system \thinspace$\{x,y,z\}$ so that $x=r\cos\phi,$ and
$y=r\sin\phi.$ Because the force is central, the neutral particle angular
momentum vector $\mathbf{L}=m\mathbf{r\times\dot{r}}$ is invariant and so the
neutral particle moves in an orbital plane normal to $\mathbf{L}$. The lab and
orbital planes are sketched in Fig.\ref{Orbital-Plane-coordinate-system}.\ The
$x$ axis of the lab frame is defined to be in the direction of the unit vector
$\hat{x}=$ $\hat{z}\times\mathbf{L}/L$ and the $y$ axis of the lab frame is
defined to be in the direction of the unit vector $\hat{y}=\hat{z}%
\times\left(  \hat{z}\times\mathbf{L}/L\right)  $. The orbital plane is tilted
with respect to the lab frame by an angle $\theta$ about the $x$ axis. The
$x^{\prime}$ axis of the orbital frame is defined to be coincident with the
$x$ axis of the lab frame and the $y^{\prime}$ axis of the orbital plane is an
uptilted version of the $y$ axis of the lab frame.

$\theta=0$ corresponds to prograde motion in the lab frame (i.e., the neutral
particle moves in the same sense as the toroidal current that produces the
magnetic field $B_{z}$ on the $z$ axis), $\theta=\pi$ corresponds to
retrograde motion in the lab frame, and $\theta=\pi/2$ corresponds to a polar
orbit. For purposes of following the trajectory in the orbital plane it is
convenient to use cylindrical coordinates $\rho,\eta$ defined in the
orbital\ plane such that $x^{\prime}=\rho\cos\eta$ and $y^{\prime}=\rho
\sin\eta.$ The Hamiltonian for a neutral particle moving in its orbital plane
can then be written as
\begin{equation}
H=\frac{1}{2}mv_{\rho}^{2}+\frac{L^{2}}{2m\rho^{2}}-\frac{mMG}{\rho}
\label{vt2}%
\end{equation}
where
\begin{equation}
L=m\rho v_{\eta}, \label{p}%
\end{equation}
the magnitude of the mechanical angular momentum vector, is an invariant
positive scalar. The Kepler angular frequency at a reference radius $a\ $\ is
defined as
\begin{equation}
\Omega_{0}=\sqrt{MG/a^{3}}. \label{Kepler0}%
\end{equation}
The value of $a$ is chosen to be the radius of the poloidal magnetic field
axis. Normalized quantities are defined as%
\begin{equation}%
\begin{array}
[c]{c}%
\bar{\rho}=\rho/a,\tau=\Omega_{0}t,\bar{v}_{\rho}=\frac{v_{\rho}}{\Omega_{0}%
a}\\
\bar{L}=\ \frac{L}{m\Omega_{0}a^{2}},\quad\bar{H}=\frac{H}{m\Omega_{0}%
^{2}a^{2}}.
\end{array}
\label{norm}%
\end{equation}
Equation \ref{vt2} can then be expressed in dimensionless form as
\begin{equation}
\bar{H}=\frac{\bar{v}_{\rho}^{2}}{2}+\frac{\bar{L}^{2}}{2\bar{\rho}^{2}}%
-\frac{1\ }{\bar{\rho}}. \label{nondim}%
\end{equation}

The last two terms depend on $\bar{\rho}$ and so constitute an effective
potential
\begin{equation}
\bar{\chi}(\bar{\rho})=\frac{\bar{L}^{2}}{2\bar{\rho}^{2}}-\frac{1}{\bar{\rho
}}. \label{chi netural}%
\end{equation}
This effective potential depends parametrically on $\bar{L}$ which is a
property of the particle and not the environment. Two different particles at
the same position but having different values of $\bar{L}$ will have different
effective potentials and so march to a \textquotedblleft different
drummer\textquotedblright. This \textquotedblleft different
drummer\textquotedblright\ concept will re-appear later in a more elaborate
fashion when the motion of charged particles is considered.

\begin{figure}[ptb]
\caption{(a)\ Effective potential for a neutral particle having $\bar{L}=1$;
(b)\ effective potential for a charged particle with appropriate values of
canonical angular momentum and poloidal flux function.}%
\label{Effective-Potential-Neutral-Particle}%
\plotone{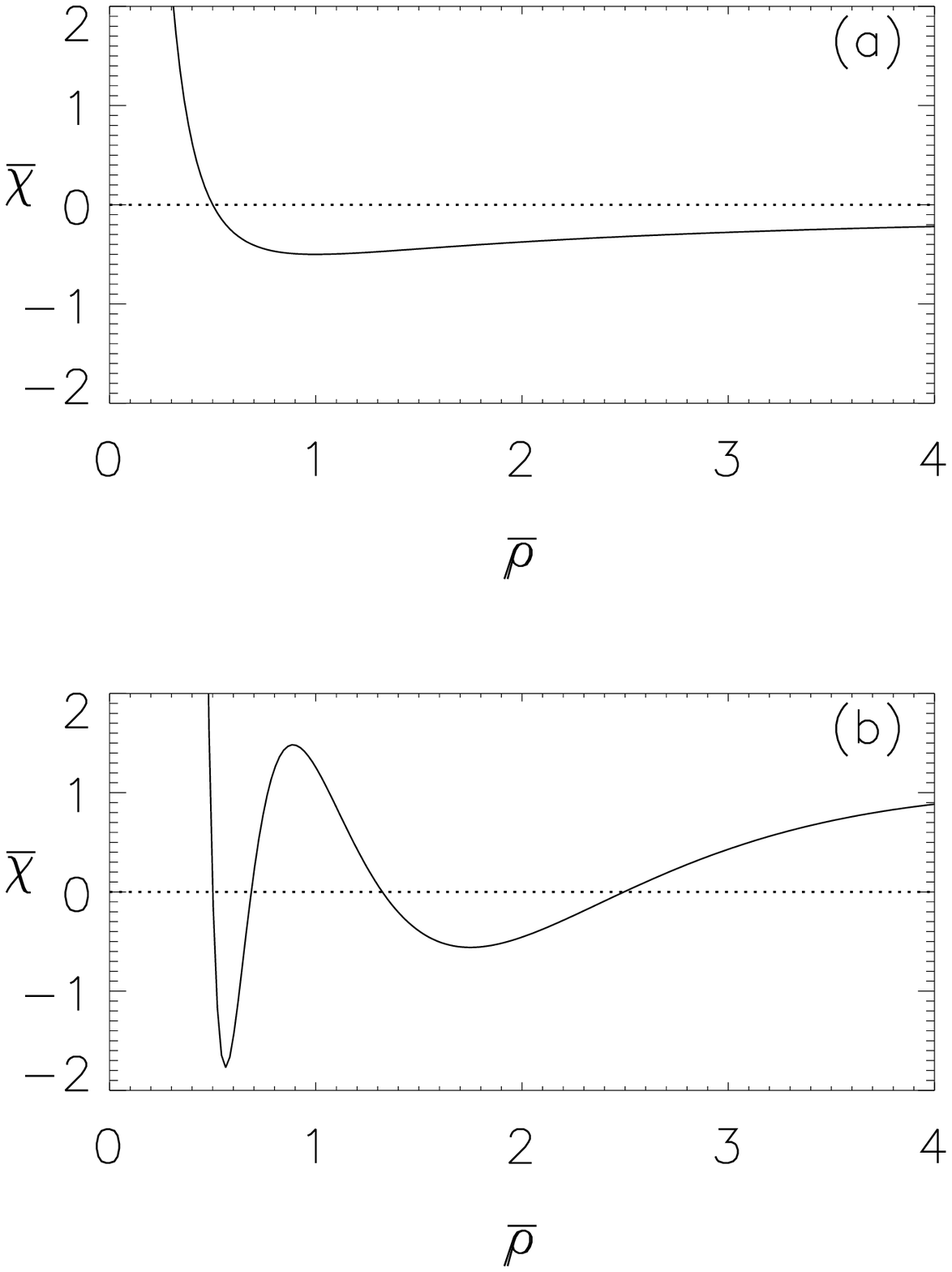}\end{figure}

$\bar{\chi}(\bar{r})$ attains its minimum value $\chi_{\min}=-1/2\bar{L}^{2}$
at the normalized radius $\bar{\rho}=\bar{L}^{2}$. \ A particle with energy
equal to this minimum has $\bar{v}_{\rho}=0$ and therefore has a circular
orbit with angular frequency $d\eta/d\tau=\ L/\Omega_{0}ma^{2}=\ \bar{L}%
/\bar{\rho}^{2}.$ Hence, if $\bar{L}=\ 1$ the minimum-energy particle traces
out a circular Kepler orbit with $d\eta/d\tau=1$ and has an energy $\bar
{H}=-1/2.$ A particle with energy $-1/2<\bar{H}<0$ cannot escape to infinity
and so has a bounded elliptical Kepler orbit. The effective potential
prescribed by Eq.\ref{chi netural} for a particle with $\bar{L}=\ 1$ is shown
in Fig.\ref{Effective-Potential-Neutral-Particle}(a).

Reflection (pericenter) of a particle \ occurs when $\bar{v}_{\rho}=0$ in
which case Eq.\ref{nondim} gives%
\begin{equation}
\bar{\rho}_{pericenter}=\frac{\bar{L}^{2}}{1+\sqrt{1+2\bar{L}^{2}\bar{H}}}.
\label{perigee}%
\end{equation}
Reflection at the pericenter \ can be considered to be the consequence of a
potential barrier preventing the particle from accessing the region $\bar
{\rho}<\bar{\rho}_{pericenter};$ the effective potential in the inaccessible
region exceeds the total available energy.$\ $Thus an unbounded particle with
$\bar{L}=1$ also has the effective potential shown Fig.
\ref{Effective-Potential-Neutral-Particle}(a), but unlike the bounded $\bar
{H}=-1/2$ Kepler particle, the unbounded particle reflects from the pericenter
potential barrier and so has a cometary orbit.

In order for an incoming unbound particle to access a given $\bar{\rho}$
without being reflected at some larger radius, the condition that $\bar
{v}_{\rho}^{2}~$cannot be negative gives the constraint on angular momentum
that
\begin{equation}
\bar{L}^{2}<2\bar{\rho}^{2}\bar{H}+2\bar{\rho}. \label{bound on p}%
\end{equation}
Since a particle with zero angular momentum will simply fall into the central
object, in order for a particle to be both unbounded and \ able to access the
radius $\bar{\rho}$ its angular momentum is constrained to lie in the range%
\begin{equation}
0<\bar{L}^{2}<2\bar{\rho}^{2}\bar{H}+2\bar{\rho}. \label{p range}%
\end{equation}

Solution of the equation of motion
\citep{Goldstein1950}
shows that the orbit can be expressed as
\begin{equation}
\frac{1}{\ \bar{\rho}}=\frac{1-\sqrt{1+2\bar{L}^{2}\ \bar{H}}\cos\left(
\eta-\alpha\right)  }{\bar{L}^{2}}\ \ \label{solution}%
\end{equation}
where $\alpha,$ which we call the clock angle in the orbital plane, is the
angle between the symmetry line of the orbit (the line passing through the
central object and the pericenter position) and the lab frame $x$ axis (which
is also the $x^{\prime}$ axis of the orbital plane).

The Cartesian orbit coordinates $\bar{x}^{\prime}=\bar{\rho}\cos\eta\ $and
$\bar{y}^{\prime}=\bar{\rho}\sin\eta$ in the orbital plane (denoted by a prime
to distinguish this plane from the lab frame)\ are%
\begin{equation}%
\begin{array}
[c]{c}%
\bar{x}^{\prime}=\frac{\bar{L}^{2}\cos\eta}{1-\sqrt{1+2\bar{L}^{2}\ \bar{H}%
}\cos\left(  \eta-\alpha\right)  }\\
\bar{y}^{\prime}=\frac{\bar{L}^{2}\sin\eta}{1-\sqrt{1+2\bar{L}^{2}\ \bar{H}%
}\cos\left(  \eta-\alpha\right)  }%
\end{array}
\label{unbounded Cartesian}%
\end{equation}
If the effective potential had a different shape, say the shape shown in
Fig.\ref{Effective-Potential-Neutral-Particle}(b) with $\bar{\chi}%
\rightarrow0$ at large $\bar{\rho},$ then a particle with $\bar{H}\geq0$ could
be trapped in one of the two minima of this effective potential. However, a
particle coming from infinity would still be unbounded and would just reflect
from some potential barrier$.$ The inability of a static Hamiltonian system to
trap a particle coming from infinity is independent of the shape of the
Hamiltonian and results from the intrinsic time reversibility of Hamiltonian dynamics.

\section{Comparison of gravitational/magnetic forces to Poynting-Robertson
force and to radiation pressure}

\qquad The analysis in this paper is based on the assumption that the
trajectory of charged dust grains results primarily from a competition between
gravitational and magnetic forces with the possibility that in certain
situations electrostatic forces and collisional drag can\ also be important.
Two other types of forces, namely those due to the Poynting-Robertson effect
and due to radiation pressure, also exist and so it is important to check to
see if these additional forces need to be taken into account. This will be
done by making a comparison with the nominal magnetic force on a charged dust
grain. The magnetic force depends on the strength of the magnetic field, a
quantity which has been estimated in self-consistent fashion in Paper III to
be in the range $10^{-8}$ to $10^{-6}$ T (i.e., 0.1 to 10 mG) for a nominal
YSO jet-disk system where the dust grains have coagulated to a nominal radius
$r_{d}=$3 $\mu$m. This estimate of the magnetic field is in rough order of
magnitude agreement with measurements reported by
\citet{Chrysostomou1994}%
, by
\citet{Roberts1997}
and by
\citet{Itoh1999}%
, and also is in agreement with the expectation that the magnetic fields in a
disk jet system should be much stronger than the nominal $10^{-10}$ T (i.e., 1
$\mu$G) magnetic fields of the ISM.

The radiation pressure acting on a dust grain at a distance $r$ from a star
with luminosity $L$ is
\begin{equation}
P_{rad}=\frac{L}{4\pi r^{2}c}Q_{rad}(r_{d}) \label{Prad}%
\end{equation}
where $Q_{rad}(r_{d})$ is the efficiency with which the photons are
absorbed/reflected by the dust grain. This pressure results in a radial
outwards force $F_{rad}=P_{rad}\sigma_{d}$. If the dust grain radius is much
larger than $\lambda_{rad}$ the wavelength of the radiation, then
$Q_{rad}\simeq1$ whereas if the dust grain radius is much smaller than the
wavelength of the radiation then $Q_{rad}\sim(\lambda_{rad}/r_{d})^{4}\ll1.$
The nominal $r_{d}\sim3$ $\mu$m dust grains assumed here are much larger than
the nominal light wavelength and so $Q_{rad}\sim1.$

Since the gravitational force $F_{g}=mMG/r^{2}$ is also in the radial
direction, the force due to radiation pressure and gravity compete; the ratio
of radiation pressure force to gravitational force on a dust grain is
\begin{equation}
\alpha=\frac{LQ_{rad}\sigma_{d}}{4\pi cm_{d}MG}=\frac{3}{16\pi}\frac{L}%
{MG\rho^{int}c}\frac{Q_{rad}(r_{d})}{r_{d}} \label{radiation pressure ratio}%
\end{equation}
where Eqs.\ref{sigma_d} and \ref{md} have been used. Assuming $r_{d}=3$ $\mu
$m, nominal luminosity $L=L_{\odot}=$ $4\times10^{26}$ watts, $M=M_{\odot},$
intrinsic dust density $\rho^{int}=2\times10^{3}$ kg m$^{-3},$ and
$Q_{rad}(r_{d})=1$ gives $\alpha=10^{-1}$ so radiation pressure can be ignored
compared to gravitational force.

The force on a dust grain due to the Poynting-Robertson effect is smaller by a
factor $v/c$ compared to the radiation pressure force, is in the toroidal
direction, opposes the Keplerian orbital motion $v_{K}=\sqrt{MG/r}$, and so
constitutes a drag force
\begin{equation}
F_{PR}=\ F_{G}\alpha\frac{v_{K}}{c}=\frac{\ Q_{rad}r_{d}^{2}}{4c^{2}\ \ }%
\sqrt{\frac{MGL^{2}}{r^{5}\ }}. \label{Poyntin-Roberston force}%
\end{equation}
The toroidal component of the magnetic force acting on a charged particle has
magnitude%
\begin{equation}
F_{mag}=Zev_{r}B \label{magnetic force}%
\end{equation}
where $Z$ is the charge. The grains typically have non-circular trajectories
with $v_{r}$ being of the order of the Kepler velocity $v_{K}$ so $F_{mag}\sim
Zev_{K}B.$ The ratio of Poynting-Robertson force to magnetic force is thus%
\begin{equation}
\frac{F_{PR}}{F_{mag}}=\ \frac{F_{G}}{F_{mag}}\alpha\frac{v_{K}}{c}.
\label{FPR to FG ratio}%
\end{equation}
Since the dust grains are assumed to be in a regime where they are acted on by
magnetic forces which are at least comparable to gravitational forces, i.e.,
$\ F_{mag}\gtrsim F_{G}$ and since $\alpha\ll1$ and $v_{K}/c\ll1$ it is seen
that the\ force due to Poynting-Robertson effect is negligible compared to
magnetic forces and so the Poynting-Robertson effect, like radiation pressure,
may be neglected.

\section{Electromagnetic particle Hamiltonian with gravity}

Hamilton-Lagrange methods are mathematically equivalent to the particle
equation of motion and so describe all physically allowed orbits (e.g.,
cyclotron, drift, Speiser, etc.). Furthermore, because the Hamilton-Lagrange
approach clarifies effects of spatial symmetries, deeper insight into orbital
dynamics is obtained than provided by direct integration of the equation of
motion. Direct integration nevertheless provides insight as well by providing
an independent verification of the predictions of Hamilton-Lagrange methods.
This two-pronged approach (Hamilton-Lagrange and direct orbit
integration)\ provides a powerful method for examining particle motion in
non-adiabatic situations.

The Lagrangian of a particle with mass $m_{\sigma}$ and charge $q_{\sigma}$ in
the combination of an axisymmetric electromagnetic field and the spherically
symmetric gravitational potential of a\ mass $M$ central object is
\begin{equation}%
\begin{array}
[c]{ccl}%
\mathcal{L} & = & \frac{m_{\sigma}}{2}\left(  v_{r}^{2}+r^{2}\dot{\phi}%
^{2}+v_{z}^{2}\right) \\
&  & +q_{\sigma}\left(  r\dot{\phi}A_{\phi}(r,z,t)+v_{z}A_{z}(r,z,t)\right) \\
&  & -q_{\sigma}V(r,z,t)+\frac{m_{\sigma}MG}{\left(  r^{2}+z^{2}\right)
^{1/2}}%
\end{array}
\label{Lagrangian}%
\end{equation}
where $V(r,z,t)$ is the electrostatic potential and a gauge with $A_{r}=0$ is
assumed. The canonical angular momentum is \
\begin{equation}
P_{\phi}\equiv\frac{\partial\mathcal{L}}{\partial\dot{\phi}}=m_{\sigma}%
r^{2}\dot{\phi}+q_{\sigma}rA_{\phi} \label{Pphi}%
\end{equation}
and, since $\ \mathbf{B}_{pol}=\nabla\times\left[  (2\pi r)^{-1}\psi\hat{\phi
}\right]  =\nabla\times\left(  A_{\phi}\hat{\phi}\right)  $ implies $\psi=2\pi
rA_{\phi}$, the canonical angular momentum can be expressed in terms of the
poloidal magnetic flux as
\begin{equation}
P_{\phi}=m_{\sigma}r^{2}\dot{\phi}+\frac{q_{\sigma}}{2\pi}\psi(r,z,t).
\label{Pphi flux}%
\end{equation}
Lagrange's equation $\dot{P}_{\phi}=\partial\mathcal{L}/\partial\phi
\ $provides the important result that $\ $
\begin{equation}
P_{\phi}=const., \label{Pphi const}%
\end{equation}
i.e., $P_{\phi}$ is a constant of the motion because the system is
axisymmetric. In the limit of a strong magnetic field, the second term in
Eq.\ref{Pphi flux} dominates the first and leads to the constraint that a
particle orbit must stay very nearly on a surface of constant $\psi;$ this is
the basis for particle confinement in axisymmetric toroidal fusion devices
(tokamaks, reversed field pinches, and spheromaks). Any deviation of a
particle from a constant $\psi$ surface is a consequence of finite $m_{\sigma
}$. When finite $m_{\sigma}$ is taken into account, it is seen that the
particle must stay within a poloidal Larmor radius of a constant $\psi$
surface, where poloidal Larmor radius means the cyclotron radius evaluated
using the local poloidal field magnitude. Equation \ref{Pphi flux} may be
solved for $\dot{\phi}$ to give
\begin{equation}
\dot{\phi}=\frac{P_{\phi}-\frac{q_{\sigma}}{2\pi}\psi(r,z,t)}{m_{\sigma}r^{2}%
}. \label{solve phidot}%
\end{equation}

The corresponding Hamiltonian is
\begin{equation}
H=\frac{m_{\sigma}}{2}\left(  v_{r}^{2}+r^{2}\dot{\phi}^{2}+v_{z}^{2}\right)
+q_{\sigma}V(r,z,t)-\frac{m_{\sigma}MG}{\left(  r^{2}+z^{2}\right)  ^{1/2}}.
\label{Hamiltonian0}%
\end{equation}
By using Eq.\ref{solve phidot} to substitute for $\dot{\phi}\,\ $in
Eq.\ref{Hamiltonian0}, the Hamiltonian can be expressed as%
\begin{equation}%
\begin{array}
[c]{ccl}%
H & = & \frac{m_{\sigma}}{2}\left(  v_{r}^{2}+v_{z}^{2}\right) \\
&  & +\frac{\left(  P_{\phi}-\frac{q_{\sigma}}{2\pi}\psi(r,z,t)\right)  ^{2}%
}{2m_{\sigma}r^{2}}\\
&  & +q_{\sigma}V(r,z,t)-\frac{m_{\sigma}MG}{\left(  r^{2}+z^{2}\right)
^{1/2}}.
\end{array}
\label{Hamiltonian}%
\end{equation}

We now consider situations where $\psi$ is time-independent and $V=0\ $so the
Hamiltonian reduces to
\begin{equation}
H=\frac{m_{\sigma}}{2}\left(  v_{r}^{2}+v_{z}^{2}\right)  +\frac{\left(
P_{\phi}-\frac{q_{\sigma}}{2\pi}\psi(r,z)\right)  ^{2}}{2m_{\sigma}r^{2}%
}-\frac{m_{\sigma}MG}{\sqrt{r^{2}+z^{2}}}. \label{H}%
\end{equation}
Since the Lagrangian does not explicitly depend on time, $H=const.$ and the
particle energy is conserved. In the $q_{\sigma}\psi=0$ limit, $P_{\phi}$
reduces to the mechanical angular momentum $p_{\phi}=mr^{2}\dot{\phi
}=m\mathbf{r\times\dot{r}\cdot}\hat{z}=L\cos\theta$ in which case the dynamics
reduces to the neutral particle orbital mechanics reviewed in Sec.\ref{Kepler}%
. Thus, if $q_{\sigma}\psi=0$ bounded orbits correspond to $H<0$ and
\ circular Kepler\ orbits correspond to $H$ having the minimum value of the
effective potential well. Unbounded $q_{\sigma}\psi=0$ orbits correspond to
$H\geq0$. Note that $p_{\phi}=L\cos\theta$ is a signed quantity, unlike $L.$

\begin{figure}[ptb]
\caption{ caption on previous page}%
\label{effective-potential-scan.eps}%
\epsscale{0.48} \plotone{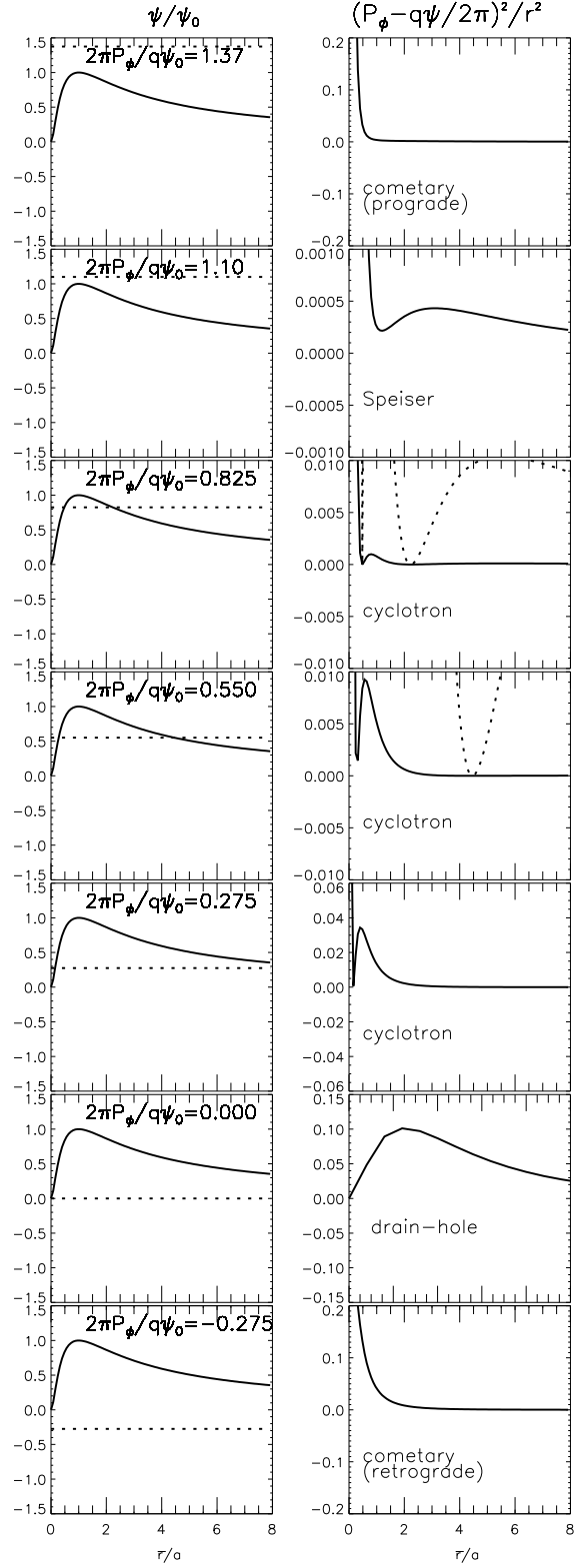}\end{figure}

If $q_{\sigma}\psi$ is finite then $\left(  P_{\phi}-q_{\sigma}\psi
(r,z)/2\pi\right)  ^{2}/2m_{\sigma}r^{2}$ is the appropriate term which
contributes to the effective potential. This term, called the St\"{o}rmer
potential, manifests \ a variety of qualitatively different spatial profiles
depending on the relationship between $P_{\phi}$ and $q_{\sigma}\psi
(r,z)/2\pi.$ These profiles are shown in
Fig.\ref{effective-potential-scan.eps} for a sequence of decreasing values of
$P_{\phi}.$ Very large positive $P_{\phi}$ gives prograde orbits similar to
unmagnetized prograde cometary orbits and very large negative $P_{\phi}$ gives
retrograde orbits similar to unmagnetized retrograde cometary orbits; in both
these cases the strong centrifugal repulsion at small $r$ causes the particle
to have an unbounded cometary orbit.

\_\_\_\_\_\_\_\_\_\_\_\_\_\_\_\_\_\_\_\_\_\_\_\_\_\_\_\_\_\_\_\_\_\_\_\_\_\_\_\_\_\_\_\_\_\_\_\_

Caption for Fig. \ref{effective-potential-scan.eps}

Left: Plot of $\psi(r,z)/\psi_{0}$ v. $r/a\,\ $for $z=0$ with sequence of
values of $2\pi P_{\phi}/q\psi_{0}$ shown as dotted line. Right: Corresponding
dependence of effective potential term $\left(  P_{\phi}-q\psi(r,z)/2\pi
\right)  ^{2}/r^{2}$ showing that potential wells develop at locations where
$2\pi P_{\phi}/q\psi_{0}$ intersects $\psi(r,z)/\psi_{0}.$ These wells
correspond to cyclotron motion if the intersection is away from the maximum of
$\psi$ and to Speiser orbits if the intersection is at or near the maximum of
$\psi.$ A potential well at $r=0$ develops if $P_{\phi}=0$ as seen in the
sixth set of plots from top; this results in drain-hole orbits. Dotted line in
right-hand second plot from top has the vertical scale multiplied by 100 to
enable visualization of the outer minimum and the dotted line on the third
plot from the top has the vertical scale multiplied by 2000 times. Note
changes of scale in right-hand plots.

\_\_\_\_\_\_\_\_\_\_\_\_\_\_\_\_\_\_\_\_\_\_\_\_\_\_\_\_\_\_\_\_\_\_\_\_\_\_\_\_\_\_\_\_\_\_\_\_

\bigskip\pagebreak

On the other hand, if $P_{\phi}$ and $q_{\sigma}\psi(r,z)/2\pi$ have
comparable magnitude, complex effective potential structures can result. For
example if at some location $P_{\phi}$ $\ $equals $q_{\sigma}\psi(r,z)/2\pi$
then $\left(  P_{\phi}-q_{\sigma}\psi(r,z)/2\pi\right)  ^{2}/2m_{\sigma}r^{2}$
\ vanishes at this location, giving a \textit{localized minimum} in the
overall effective potential.

If two separated positions exist where $P_{\phi}$ equals $q_{\sigma}%
\psi(r,0)/2\pi$ then two distinct minima exist, but if only one position
exists where $P_{\phi}$ equals $q_{\sigma}\psi(r,0)/2\pi$ then only one
minimum exists. The former situation occurs when $P_{\phi}$ lies somewhere
between $0$ and the maximum of $\psi$ and leads to cyclotron orbits with
associated grad-$B$ $\ $and curvature drifts; in this case
Eq.\ref{solve phidot} shows that the sign of $\dot{\phi}$ oscillates as the
particle oscillates back and forth across the minimum of the effective
potential, and the orbit is a cyclotron orbit. \ The situation of only one
minimum occurs when the value of $P_{\phi}$ is approximately the maximum of
$q_{\sigma}\psi/2\pi$ and gives Speiser orbits ($\dot{\phi}$ has fixed sign
and the orbit is paramagnetic). For a review of the distinction between the
diamagnetism of cyclotron orbits (and associated grad-$B$ $\ $and curvature
drifts) and the paramagnetism of Speiser orbits see Appendix
\ref{Orbit review}.

Yet another situation is where $P_{\phi}=0$. \ Because $\psi\sim r^{2}$ at
small $r$, this special case removes the singularity of $\left(  P_{\phi
}-q_{\sigma}\psi(r,0)/2\pi\right)  ^{2}/2m_{\sigma}r^{2}$ at $r=0$ and so
eliminates centrifugal force repulsion altogether. The $P_{\phi}=0$ case gives
trajectories which spiral down towards the central object\ while crossing
magnetic field lines; part of the magnetic force cancels the centrifugal force
so all that is left is gravity and a residual inward magnetic force. This
situation is completely different from either cyclotron orbits or Speiser
orbits and has only been previously discussed in the more limited context of
two dimensional situations
\citep{Bellan2007}%
. Finally, there is also the special situation discussed by
\citet{Schmidt1979}
where $P_{\phi}=-r^{2}q_{\sigma}(2\pi)^{-1}\partial\left(  \psi/r\right)
/\partial r$ in which case the charged particle executes an axis-encircling
cyclotron orbit. While possible in principle, axis-encircling cyclotron orbits
will be not be considered here because they would correspond to particles
having extreme energies (e.g., a cyclotron radius of many a.u.).

Thus, there are five qualitatively distinct types of feasible trajectories
depending on the relationship between $P_{\phi}$ and $q_{\sigma}\psi/2\pi.$ As
labeled in Fig.\ref{effective-potential-scan.eps} and in order of descending
signed value of the invariant $P_{\phi}\ $as shown by dashed horizontal line
in left column of this figure, these are:

\begin{enumerate}
\item prograde centrifugally dominated orbits [$P_{\phi}$ much larger than the
peak of $q_{\sigma}\psi(r,z)/2\pi)$]

\item Speiser orbits [$P_{\phi}$ just grazes the peak of $q_{\sigma}%
\psi(r,z)/2\pi$]

\item cyclotron orbits [$P_{\phi}$ well below the peak of $q_{\sigma}%
\psi(r,z)/2\pi$ but much greater than zero]

\item $P_{\phi}=0$ orbits [which we will call \textquotedblleft drain-hole"
orbits for reasons to be discussed later], and

\item retrograde centrifugally dominated orbits [$P_{\phi}$ negative and much
less than zero)].
\end{enumerate}

In the above list, we have removed the constraint that $z=0$ so, for example,
in case \#3 (cyclotron orbits), the locations in the $r$-$z$ plane where
$P_{\phi}$ and $q_{\sigma}\psi(r,z)/2\pi$ are equal corresponds to a specific
closed curve in the $r$-$z$ plane; i.e., a specific $\psi$ iso-surface as
shown in the projection of $\psi(r,z)$ at the top of Fig.
\ref{GenericFluxFunction-new}.

The various possible values of $P_{\phi}$ can be considered as the different
\textquotedblleft drummers\textquotedblright\ that dictate the effective
potentials governing the motion of different particles located at the same
position. A related example of this different \textquotedblleft
drummers\textquotedblright\ situation \ has been reported by
\citet{Tripathi2007}
and involves two particles at the same location having velocities with equal
magnitudes but opposite directions; the two particles have such extremely
different effective potentials that one particle is expelled from a magnetic
flux tube (hill-shaped effective potential) whereas the other remains in the
flux tube (valley-shaped effective potential).

\subsection{Mechanism for accretion of collisionless particles}

Accretion is the process of converting unbounded orbits (i.e., cometary
orbits)\ into bounded orbits. Accretion of a collisionless neutral particle is
clearly impossible if such a particle is governed by dynamics of a
time-independent Lagrangian because converting an unbounded orbit into a
bounded orbit would require changing the particle energy $H$ and such a change
is forbidden for a particle having a Lagrangian that does not explicitly
depend on time.

We now postulate an accretion mechanism as follows:\ \textit{photo-emission
acts as an effective switch which alters the form of the Hamiltonian equation
governing particle dynamics}. The particle energy $H$ and mechanical angular
momentum $mr^{2}\dot{\phi}$ \textit{do not change} during the switching, but
after photo-emission has occurred, $H$ and $mr^{2}\dot{\phi}$ become
parameters in a \textit{different Hamiltonian system} which has a different
topography of potential barriers. For example, photo-emission can transform
the neutral particle effective potential shown in
Fig.\ref{Effective-Potential-Neutral-Particle}(a) into the\ charged particle
effective potential shown in Fig.\ref{Effective-Potential-Neutral-Particle}(b).

The switching is postulated to occur when an incident neutral dust grain
absorbs sufficient energetic photons from the star. The photon absorption
causes the dust grain to photo-emit electrons and therefore become positively
charged
\citep{Lee1996,Sickafoose2000}%
. The photo-emitted electrons become free electrons equal in number to the
dust grain charge $Z$. The photo-emission process has effectively caused the
initial neutral dust grain to disintegrate into a single heavy positively
charged fragment (the charged dust grain) and $Z$ light negatively charged
fragments (the photoelectrons). The motion of each fragment is governed by the
Hamiltonian for a charged particle and this Hamiltonian is considerably
different in form from the Hamiltonian that governed the neutral particle
motion. \ 

The charge $q_{d}$ of a dust grain charged by photo-emission is given by
\begin{equation}
\frac{q_{d}}{4\pi\varepsilon_{0}r_{d}}\approx\ W_{photon}-W_{wf} \label{Z}%
\end{equation}
where $W_{photon}$ is the energy in eV of an incident photon that causes
photo-emission of a primary photo-electron and $W_{wf}$ is the work-function
in eV of the material
\citep{Shukla2002}%
.
\citet{Lee1996}
has shown that the effective photon energy is $W_{photon}\simeq8$ eV for
nominal solar parameters and the effective work function of typical dust is
$W_{wf}\simeq6$ eV so that the energy of emitted photo-electrons is $\sim2$ eV.

Combination of Eqs. \ref{md} and \ref{Z} show that the dust charge to mass
ratio will be
\begin{equation}
\frac{q_{d}}{m_{d}}=\frac{\ 3\varepsilon_{0}\left(  W_{photon}-W_{wf}\right)
}{\ \rho_{d}^{int}r_{d}^{2}}\ \label{charge to mass ratio}%
\end{equation}
which will be many of orders of magnitude smaller than the charge to mass
ratios of electrons or ions. The number $Z$ of charges on a dust grain will
be
\begin{equation}
Z=\frac{4\pi\varepsilon_{0}r_{d}\left(  W_{photon}-W_{wf}\right)  }{e}.
\label{Znumber}%
\end{equation}

Charging a dust grain to $q_{d}$ takes a finite time interval, but for
simplicity we will assume that this charging occurs at a single time defined
as $t=0.$ Charging will not change either the instantaneous position or
velocity of a particle.

Photo-emission at $t=0$ therefore decomposes an incident neutral dust grain
into positive and negative product particles each inheriting the same position
and velocity at $t=0_{+}$ that the neutral dust grain had at $t=0_{-}.$
Position and velocity can consequently be considered to be continuous
functions at $t=0$ so the canonical angular momentum with which a newly formed
charged particle is endowed is%
\begin{equation}%
\begin{array}
[c]{ccc}%
P_{\phi} & = & m_{\sigma}r_{\ast}^{2}\dot{\phi}_{\ast}+q_{\sigma}\psi(r_{\ast
},z_{\ast})/2\pi\\
& = & L_{\sigma}\cos\theta\ +q_{\sigma}\psi(r_{\ast},z_{\ast})/2\pi
\end{array}
\label{Pphiborne}%
\end{equation}
where the subscript $\ast$ denotes the value of a coordinate at the instant of
charging, i.e., at $t=0$. For simplicity we assume that photo-emission occurs
when the distance between the incident neutral and the central object is at
some critical spherical radius $R_{\ast}$ so that charging and the setting of
$t=0$ occurs when the particle crosses the surface of the fictitious $R_{\ast
}$ sphere, i.e., when $r=r_{\ast}$ and $z=z_{\ast}$ are such that
\begin{equation}
r_{\ast}^{2}+z_{\ast}^{2}=R_{\ast}^{2}. \label{ionization sphere}%
\end{equation}

Depending on the value of $L_{\sigma},$ the angle of inclination $\theta$, the
magnitude of $q_{\sigma},$ and the value of $\psi(r_{\ast},z_{\ast}),$ all
possible finite values of $P_{\phi}$ can occur, including positive, negative,
and zero. The $r$-$z$ plane topography of the effective potential can change
completely from what it was at $t=0_{-}$ because the centrifugal force
potential $p_{\phi}^{2}/2mr^{2}$ responsible for the neutral particle
potential barrier at small $r$ is replaced by the St\"{o}rmer term $(P_{\phi
}-q_{\sigma}\psi(r,z))^{2}/2\pi mr^{2}$. $\ $Figure
\ref{effective-potential-scan.eps}\ demonstrates that variation of particle
mass and variation of the incoming orbit plane inclination angle $\theta$ $\ $
results in a range of $p_{\phi}$, $q_{d}$ values and hence a range of
$P_{\phi}$ values corresponding to cyclotron, \textquotedblleft
drain-hole\textquotedblright, Speiser, or cometary orbits. If the new orbit is
cyclotron, drain-hole, or Speiser, then photo-emission has prevented the
particle from returning to infinity, i.e., the particle has accreted.
Photo-emission changes the \textquotedblleft rules of the
game\textquotedblright\ \ by effectively erecting a new potential
barrier\ which traps a previously unbound particle. The \textquotedblleft old
game\textquotedblright\ (i.e., neutral particle Keplerian motion as reviewed
in Sec.\ref{Kepler}) did not depend on particle mass or $\theta,$ but the
\textquotedblleft new game\textquotedblright\ does and leads to a mass- and
$\theta$-dependent sorting of incoming charged grains and their associated
photo-emitted electrons into qualitatively different classes of orbits.

This process whereby neutral particles enter a magnetic field from outside,
become charged, and then become subject to magnetic forces is called `neutral
beam injection' in the context of tokamak physics and `pickup' in the context
of solar physics. Neutral beam injection is used routinely for tokamak heating
and current drive
\citep{Simonen1988,Akers2002}%
. Pickup is important in the solar wind
\citep{Gloeckler2001}%
, in planetary atmospheres
\citep{Hartle2006}%
, and in producing source particles for comic rays
\citep{Ellison1998}%
. However, to the best of the author's knowledge, charging of incoming neutral
particles has not been previously proposed as a means for accreting matter
around a star and instead accretion of matter around a star has always been
argued to be the result of the \ viscosity of neutral particles, i.e.,
collisions of neutral particles with each other, as discussed for example in
\citet{Lynden-Bell1974}%
,
\citet{Shakura1976}%
, and
\citet{Pringle1981}%
. The viscosity-based neutral particle accretion models suffer from not
knowing what to do with the angular momentum of incident particles; this issue
has motivated substantial work on developing the rather complicated non-linear
turbulence-based magneto-rotational instability model as a means for
transporting excess angular momentum outwards. In contrast, the model proposed
here inherently accounts for angular momentum and\ so does not need any
\textquotedblleft add-on turbulence\textquotedblright\ to transport angular
momentum outwards.

Trapping via photo-emission has the remarkable feature that the special class
of charged particles created with zero canonical angular momentum will spiral
all the way down to the central object. \ These $P_{\phi}=0$ (drain-hole)
particles falling towards $r=0$ are in what is effectively a loss cone in
canonical angular momentum space. The drain-hole particle motion constitutes a
gravity-driven dynamo
\citep{Bellan2007}
because the accumulation of these particles near $r=0$ produces a radially
outward electric field while their flow produces a radially inward electric
current (a dynamo is characterized by having opposed internal electric field
and electric current). Since $J_{r}=-\left(  2\pi r\right)  ^{-1}\partial
I/\partial z$ (see Eq.\ref{Jpol2}), creation of this radially inward current
which is symmetric with respect to $z$ implies creation of an anti-symmetric
function $I(r,z)$ which in turn implies creation of an anti-symmetric toroidal
field $B_{\phi}\ $(see Eq.\ref{Bphi}). Equally remarkable, particles for which
$P_{\phi}$ is near the maximum of $q_{\sigma}\psi/2\pi$ develop Speiser-type
\textit{paramagnetic} orbits in the vicinity of $r=a$, $z=0$ and so can
constitute the toroidal current that produces the poloidal flux (see
discussion of Speiser orbit paramagnetism in Appendix \ref{Orbit review}). The
creation of Speiser-orbit particles is \ conceptually similar to toroidal
current drive in a tokamak via tangential neutral beam injection
\citep{Simonen1988}%
. Because the drain-hole and Speiser dynamos are both axisymmetric, both
violate the essential claim of Cowling's anti-dynamo theorem
\citep{Cowling1934}
that axisymmetric dynamos cannot exist. This violation is not a problem
because Cowling's theorem is based on MHD\ and so does not take into account
drain-hole or Speiser orbits.

The Hamiltonian for an incoming neutral dust grain of mass $m_{n}$ can be
written as
\begin{equation}
H\ =\frac{m_{n}v_{r}^{2}}{2}\ +\frac{m_{n}v_{\phi}^{2}}{2}+\frac{m_{n}%
v_{z}^{2}}{2}-\frac{m_{n}MG}{\sqrt{r^{2}+z^{2}}}. \label{neutral}%
\end{equation}
This neutral dust grain absorbs energetic photons at $t=0,$ photo-emits $Z$
free electrons, and consequently becomes positively charged with a charge of
$Z.$ The mass of the neutral is related to the mass $m_{+}$ of the positively
charged dust grain by $m_{n}=m_{+}+Zm_{e}$ where $m_{e}$ is the electron mass.
Prior to this charging process, Eq.\ref{neutral} can be written as
\begin{equation}%
\begin{array}
[c]{ccc}%
H & = & \frac{\left(  m_{+}+Zm_{e}\right)  v_{r}^{2}}{2}\ +\frac{\left(
m_{+}+Zm_{e}\right)  v_{\phi}^{2}}{2}\\
&  & +\frac{\left(  m_{+}+Zm_{e}\right)  v_{z}^{2}}{2}-\frac{\left(
m_{+}+Zm_{e}\right)  MG}{\sqrt{r^{2}+z^{2}}}%
\end{array}
\label{H before ionization}%
\end{equation}
so the positively charged dust grain with mass $m_{+}$ and the \thinspace$Z$
electrons can each be thought of as executing identical neutral-type orbits
before photo-emission occurs.

\bigskip\bigskip

At the instant before charging, the neutral particle mechanical angular
momentum is
\begin{equation}
\ p_{\phi}=m_{n}r_{\ast}^{2}\dot{\phi}_{\ast}. \label{pmech}%
\end{equation}
At the instant after charging the newly created positively charged dust grain
and its associated photo-emitted electrons all have the same values of
$r_{\ast}$ and $\dot{\phi}_{\ast}.$ The canonical angular momentum of the
positively charged dust grain will therefore be%
\begin{equation}
P_{\phi}^{+}=m_{+}r_{\ast}^{2}\dot{\phi}_{\ast}+Ze\psi(r_{\ast},z_{\ast}%
)/2\pi\label{Pphi positive}%
\end{equation}
and the canonical momentum of each associated electron will be%
\begin{equation}
P_{\phi}^{e}=m_{e}r_{\ast}^{2}\dot{\phi}_{\ast}-e\psi(r_{\ast},z_{\ast})/2\pi.
\label{Phi electron}%
\end{equation}
The initial neutral dust grain will be called the \textquotedblleft
parent\textquotedblright\ particle while the positively charged dust grain
resulting from photo-emission and its associated $Z~$photo-emitted electrons
will be called \textquotedblleft child particles\textquotedblright\ that are
\textquotedblleft siblings\textquotedblright\ of each other. The set of child
particles resulting from the charging of a specific neutral dust grain will be
called a \textquotedblleft family\textquotedblright. \ The canonical momenta
$P_{\phi}^{+}$ and $P_{\phi}^{e}$ are now the appropriate orbit invariants for
$t>0$ whereas \ the mechanical angular momenta $m_{\sigma}r^{2}\dot{\phi}$ of
the individual siblings will \textit{not} be invariant for $t>0.$ Although the
mechanical angular momentum of an individual sibling is not conserved, the
total mechanical angular momentum of the family is conserved since summing
Eq.\ref{Pphi positive} and $Z$ times Eq.\ref{Phi electron} gives $\left[
p_{\phi}\right]  _{family}=P_{\phi}^{+}+ZP_{\phi}^{e}\ $where $\left[
p_{\phi}\right]  _{family}$ $=m_{n}r_{\ast}^{2}\dot{\phi}$ is the sum of the
mechanical angular momentum of the charged dust grain and all its sibling
electrons. Because the siblings can physically separate from each other, the
mechanical angular momentum $\left[  p_{\phi}\right]  _{family}$ is not a
locally defined quantity and so is not a constant of the motion of either a
single particle or, as in ideal hydrodynamics, of a fluid element. However,
\textit{the mechanical angular momentum of the entire system is conserved}
because the angular momentum of each family is globally conserved.

The kinetic energy of each sibling at the instant before photo-emission \ is
the same as the value at the instant after photo-emission. If $H$ is
decomposed into the contributions from the various \ siblings, it is seen that
each sibling's $H_{\sigma}$ is the same before and after photo-emission.
Assuming zero electrostatic potential $\ $for now, but allowing the child
particle to be at arbitrary $z$, the Hamiltonian of each sibling is
\begin{equation}%
\begin{array}
[c]{ccl}%
H_{\sigma}\  & = & \frac{m_{\sigma}v_{r}^{2}}{2}\ +\frac{m_{\sigma}v_{z}^{2}%
}{2}\\
&  & +\frac{\left(  m_{\sigma}r_{\ast}^{2}\dot{\phi}_{\ast}+\frac{q_{\sigma}%
}{2\pi}\left[  \psi(r_{\ast},z_{\ast})-\psi(r,z)\right]  \right)  ^{2}%
}{2m_{\sigma}r^{2}}\\
&  & -\frac{m_{\sigma}MG}{\sqrt{r^{2}+z^{2}}}.
\end{array}
\label{H-positive particle}%
\end{equation}

Using Eq.\ref{Bbar} we now define
\begin{equation}
\left\langle \omega_{c\sigma}\right\rangle =\frac{q_{\sigma}\left\langle
B_{z}\right\rangle }{m_{\sigma}} \label{avg omegac}%
\end{equation}
as the spatially-averaged cyclotron frequency over the area bounded by the
poloidal field magnetic axis. We now normalize all quantities to appropriate
combinations of $a$ and $\Omega_{0},$ the Kepler angular frequency at $a$
prescribed by Eq.\ref{Kepler0}. The normalized time, cylindrical coordinates,
velocities, and magnetic flux are thus
\begin{equation}%
\begin{array}
[c]{ccl}%
\tau & = & \Omega_{0}t\\
\bar{r} & = & r/a\\
\bar{z} & = & z/a\\
\bar{v}_{r} & = & v_{r}/a\Omega_{0}\\
\bar{v}_{z} & = & v_{z}/a\Omega_{0}\\
\bar{L} & = & L_{\sigma}/m_{\sigma}a^{2}\Omega_{0}\\
\bar{p}_{\phi} & = & \bar{r}^{2}d\phi/d\tau=\bar{L}\cos\theta\\
\bar{\psi}(r,z) & = & \frac{\psi(r,z)}{\psi(a,0)}\ =\frac{\psi(r,z)}%
{\left\langle B_{z}\right\rangle \pi a^{2}}\\
\bar{H} & = & \frac{H}{m\Omega_{0}^{2}a^{2}}%
\end{array}
\label{norm quantities}%
\end{equation}
in which case Eq.\ref{H-positive particle} becomes%
\begin{equation}%
\begin{array}
[c]{ccl}%
\bar{H} & = & \frac{\ \bar{v}_{r}^{2}}{\ 2}+\frac{\ \bar{v}_{z}^{2}}{\ 2}\\
&  & +\frac{\left(  \bar{L}\cos\theta\ \ +\frac{\left\langle \omega_{c\sigma
}\right\rangle }{2\Omega_{0}}\ \left[  \bar{\psi}(\bar{r}_{\ast},\bar{z}%
_{\ast})-\bar{\psi}(\bar{r},\bar{z})\right]  \right)  ^{2}}{2\ \bar{r}^{2}}\\
&  & -\ \frac{1}{\sqrt{\bar{r}^{2}+\bar{z}^{2}}}.
\end{array}
\label{norm H}%
\end{equation}
The effective potential is now
\begin{equation}%
\begin{array}
[c]{cl}%
\chi(\bar{r},\bar{z})= & \frac{\left(  \stackrel{\textnormal{mechanical}}%
{\overbrace{\bar{L}\cos\theta}}\ \ +\stackrel{\textnormal{magnetic}}{\overbrace
{\frac{\left\langle \omega_{c\sigma}\right\rangle }{2\Omega_{0}}\ \left[
\bar{\psi}(\bar{r}_{\ast},\bar{z}_{\ast})-\bar{\psi}(\bar{r},\bar{z})\right]
}}\right)  ^{2}}{2\ \bar{r}^{2}}\\
& -\stackunder{\textnormal{gravitational}}{\underbrace{\frac{1}{\sqrt{\bar{r}%
^{2}+\bar{z}^{2}}}}}%
\end{array}
\label{effective potential magnetized}%
\end{equation}
where the mechanical, magnetic, and gravitational contributions have been
labeled. Before photo-emission, the mechanical angular momentum is invariant
so $\bar{p}_{\phi}(\bar{r}_{\ast},\bar{z}_{\ast})=\bar{L}\cos\theta$ is just
the normalized mechanical angular momentum that the parent had when it was at
infinity. Invoking Eq.\ref{p range}, it is seen that
\begin{equation}
0\leq\bar{L}<\sqrt{2\bar{R}_{\ast}^{2}\bar{H}+2\bar{R}_{\ast}}%
\ \label{pstar range}%
\end{equation}
is required since incident neutral dust grains with mechanical angular
momentum outside this range would have reflected at larger radii than $\bar
{R}_{\ast}$ and so would not have been able to access the radius $\bar
{R}_{\ast}.$

The normalized canonical angular momentum with which a typical sibling charged
particle is endowed is
\begin{equation}
\bar{P}_{\phi}=\bar{L}\cos\theta+\frac{\left\langle \omega_{c\sigma
}\right\rangle }{2\Omega_{0}}\ \bar{\psi}(\bar{r}_{\ast},\bar{z}_{\ast}).
\label{norm can momentum}%
\end{equation}
The parameters underlying Fig.\ref{Effective-Potential-Neutral-Particle}(b)
can now be understood. This figure is a plot of $\chi(\bar{r},\bar{z})$ v.
$\bar{r}$ for $\bar{z}=0\ $where $\bar{P}_{\phi}$ is calculated for the
situation where $\left\langle \omega_{c\sigma}\right\rangle /\Omega_{0}=40,$
$\bar{r}_{\ast}=0.5,$ $\bar{z}_{\ast}=0$, $\theta=0$, and $\bar{L}=1.$
Charging of an $\ $ $\bar{L}=1$ dust grain via photo-emission causes the
effective potential governing the dust grain to change from the form given in
Fig.\ref{Effective-Potential-Neutral-Particle}(a) \ to the form given in
Fig.\ref{Effective-Potential-Neutral-Particle}(b).

When the magnetic term in Eq.\ref{effective potential magnetized} becomes
comparable to the mechanical term or much larger, orbital dynamics for the
siblings become very different from the orbital dynamics of the neutral parent
that existed before photo-emission. Various orbits can occur for the siblings.
Because of the complexity of these three dimensional orbits, we will first
consider orbits confined to the $\bar{z}=0$ plane and then generalize to fully
three dimensional orbits ranging over finite $\bar{z}.$

\subsection{Distribution of Cometary, Speiser, Cyclotron, and Drain-Hole
Orbits}

As reviewed in Sec.\ref{Kepler}, neutral particles orbits are degenerate with
respect to their orbital plane inclination angle $\theta$ (see
Fig.\ref{Orbital-Plane-coordinate-system}). However, once particles become
charged, they are no longer restricted to an orbital plane, and furthermore,
as seen from Eq.\ref{effective potential magnetized}, the effective potential
of a charged particle has a strong dependence on the value of $\theta$ that
its parent particle had. This dependence was manifested in the discussion of
Fig. \ref{effective-potential-scan.eps} where it was noted that particles with
$\left\vert 2\pi P_{\phi}/q_{\sigma}\psi_{0}\right\vert \gg1$ are essentially
unmagnetized and have Keplerian cometary orbits, particles with $2\pi P_{\phi
}\theta/q_{\sigma}\psi_{0}\simeq1$ have Speiser orbits, particles with
$0\ll2\pi P_{\phi}/q_{\sigma}\psi_{0}\ll1$ have cyclotron orbits, and
particles with $2\pi P_{\phi}/q_{\sigma}\psi_{0}\simeq0$ have drain-hole
orbits. This discussion can be made more quantitative by defining
$\Lambda\equiv2\pi P_{\phi}/q_{\sigma}\psi_{0};$ note that $\Lambda$
corresponds to the horizontal dashed lines in the left hand column of
Fig.\ref{effective-potential-scan.eps}. Equation \ref{norm can momentum} can
then be recast as
\begin{equation}
\Lambda=\frac{2\Omega_{0}}{\left\langle \omega_{c\sigma}\right\rangle }\bar
{L}\cos\theta\ +\ \ \bar{\psi}(\bar{r}_{\ast},\bar{z}_{\ast}). \label{lambda}%
\end{equation}
Thus particles with $\left\vert \Lambda\right\vert \gg1$ have Keplerian
cometary orbits, particles with $\Lambda\simeq1$ have Speiser orbits,
particles with $0\ll\Lambda\ll1$ have cyclotron orbits, and particles with
$\Lambda\simeq0$ have drain-hole orbits.

Assuming $\bar{H}\ll1$ and $\bar{R}_{\ast}\sim1,$ Eq.\ref{pstar range} implies
that only particles with $0<\bar{L}<\sqrt{2}$ can access a given location.
Because there will be a distribution of all possible $\bar{L}$'s within this
allowed range, we consider a particle with the mean of these allowed values as
being representative and so assume that $\bar{L}=\sqrt{2}/2$ is the normalized
angular momentum of this representative nominal particle.

Since $\left\langle \omega_{c\sigma}\right\rangle =q_{\sigma}\left\langle
B_{z}\right\rangle /m_{\sigma}$, $\left\langle B_{z}\right\rangle =\psi
_{0}/\pi a^{2}$, and $\Omega_{0}=\sqrt{MG/a^{3}}$ this nominal particle will
have
\begin{equation}
\Lambda=K\cos\theta\ +\ \ \bar{\psi}(\bar{r}_{\ast},\bar{z}_{\ast})
\label{def lambda}%
\end{equation}
where
\begin{equation}
K=\frac{m_{\sigma}}{q_{\sigma}}\frac{\ \pi\ \sqrt{2aMG\ }}{\psi_{0}%
}\ \label{lambda rd}%
\end{equation}
parameterizes the competition between gravitational and magnetic forces. Using
Eq. \ref{charge to mass ratio} to give the charge to mass ratio it is seen
that
\begin{equation}
K=\ \frac{\ \pi\rho_{d}^{int}\sqrt{2aMG\ }}{3\varepsilon_{0}\left(
W_{photon}-W_{wf}\right)  \psi_{0}}r_{d}^{2}; \label{K rd}%
\end{equation}
thus $K$ increases when $r_{d}$ increases as a result of dust grain coagulation.

Speiser and drain hole particles occur when gravitational and magnetic forces
are comparable in magnitude, i.e., when $K$ is of order unity. For a given
star mass $M,$ poloidal flux magnetic axis radius $a,$ and magnetic flux
$\psi_{0},$ this means that Speiser and drain hole particles will occur when
coagulation has caused the dust grains to have a certain critical radius which
is of order%
\begin{equation}
r_{d}^{crit}\sim\frac{1}{\left(  2aMG\right)  ^{1/4}}\sqrt{\frac
{3\varepsilon_{0}\left(  W_{photon}-W_{wf}\right)  \psi_{0}}{\pi\rho_{d}%
^{int}\ }}. \label{rdnom}%
\end{equation}
$\ $If $r_{d}\gg r_{d}^{crit}$ then gravity will dominate and the dust grains
will behave like neutral particles whereas if $r_{d}\ll r_{d}^{crit}$ then
magnetic forces will dominate and charged dust grains will mainly have
cyclotron orbits. Since coagulation causes $r_{d}$ to increase monotonically,
there should always be some time when $r_{d}$ $\sim r_{d}^{crit}$\ and $K$ is
of order unity. This argument indicates that the dust-driven dynamo mechanism
should take place as a well-defined temporal stage in the accretion process;
before this stage $r_{d}$ is too small and after this stage $r_{d}$ is too large.

\begin{figure}[ptb]
\caption{Distribution of orbits as function of $K,\theta.$ Radius $K$ is
proportional to $r_{d}^{2}.$ Prograde orbits have $\theta=0$, retrograde
orbits have $\theta=\pi$, and polar orbits have $\left\vert \theta\right\vert
=\pi/2$.}%
\label{orbit-distributionAI-PS-DCS2.0-Ascii5-25.eps}%
\epsscale{1.0} \plotone{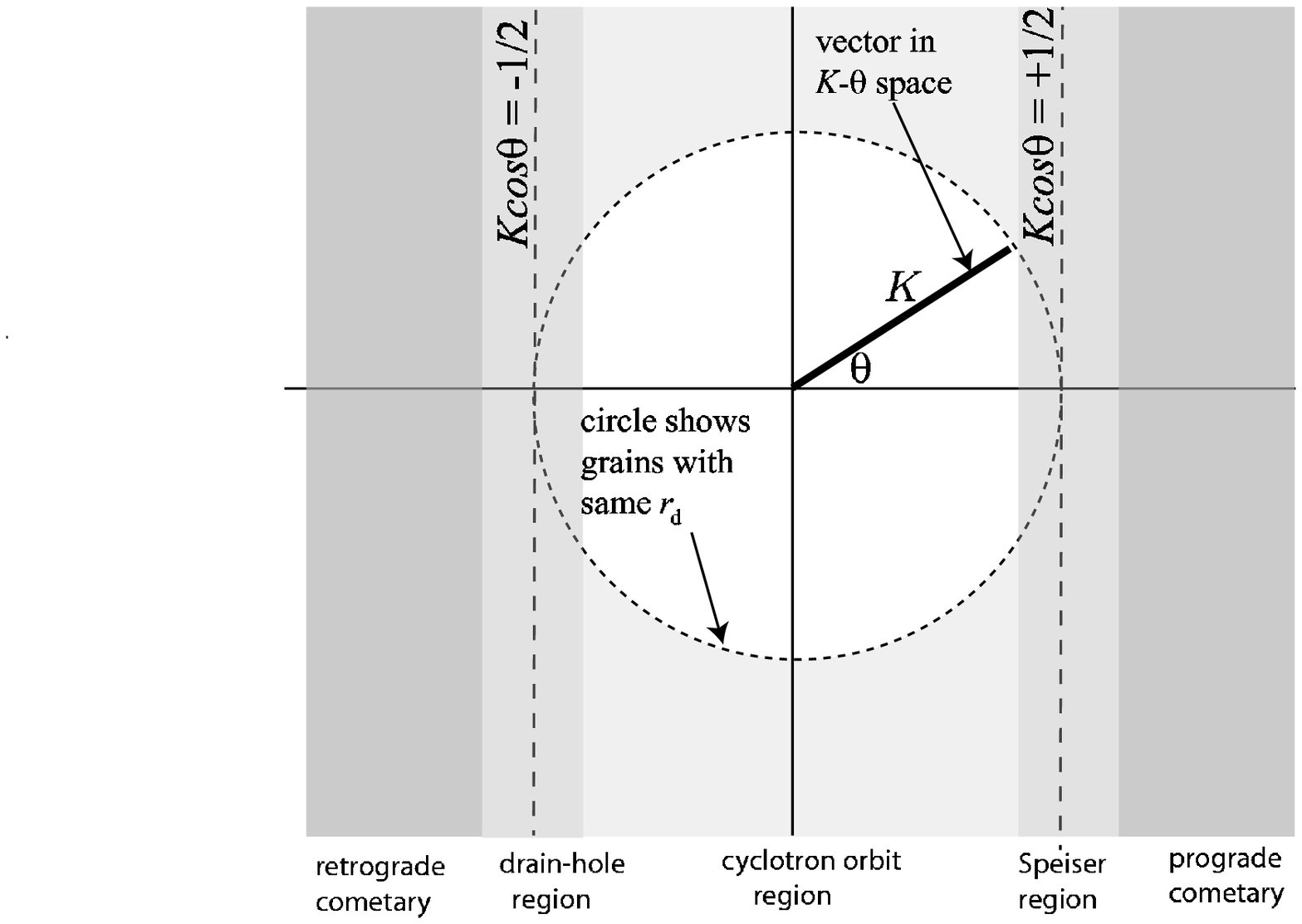}\end{figure}

Since $\bar{\psi}(\bar{r}_{\ast},\bar{z}_{\ast})$ ranges between $0$ and $1$,
let us consider the nominal situation where $\bar{\psi}(\bar{r}_{\ast},\bar
{z}_{\ast})=1/2$ in which case%
\begin{equation}
\Lambda_{nom}=K\cos\theta\ +\frac{1}{2}. \label{lambda nom}%
\end{equation}
If $K\simeq1/2,$ Speiser particles result for $\cos\theta=1$ (i.e., neutral
parent was prograde) and drain-hole particles result for $\cos\theta
=-1\ $(i.e., neutral parent was retrograde ). If $K\ll1/2,$ the orbits will be
cyclotron. Finally if $K\gg1/2,$ the orbits will be cometary if $\cos\theta$
is not close to zero. The categorization implied by Eq.\ref{lambda nom} is
shown schematically in Fig. \ref{orbit-distributionAI-PS-DCS2.0-Ascii5-25.eps}.

\subsection{Light-weight particles ($K\ll1$) \label{light-weight trapping}}

\subsubsection{Generic accretion mechanism}

\begin{figure}[ptb]
\caption{An incident neutral particle ($\bar{H}=0,$ $\bar{\rho}_{pericenter}%
=0.1,$ $\theta=30,\alpha=0^{0}$) becomes charged due to photoemission of
electrons at $R_{\ast}=2.$ The child particle mass is such that $\left\langle
\omega_{c\sigma}\right\rangle /\Omega_{0}=200$ and the child particle becomes
magnetically trapped, staying within a poloidal Larmor orbit of a constant
$\psi$ surface. In this example, the child particle is mirror trapped and so
cannot enter the strong magnetic field region at small $\bar{r}.$ (a) $\bar
{x}-\bar{y}$ plane, charging occurs where orbit abruptly changes, poloidal
field magnetic axis shown as dashed circle (b)\ $\bar{r}-\bar{z}$ plane
showing magnetic mirroring of child particle at large magnetic field (poloidal
flux contours $\bar{\psi}(\bar{r},\bar{z})$ shown as dashed lines). The orbit
the parent neutral particle would have continued to have if it had not become
charged is shown by dotted line (both projections).}%
\label{mirror..pericenter=0.1_theta=30_alpha=0_rho0=4_rstar=2_omegaratio=200.eps}%
\epsscale{0.5} \plotone{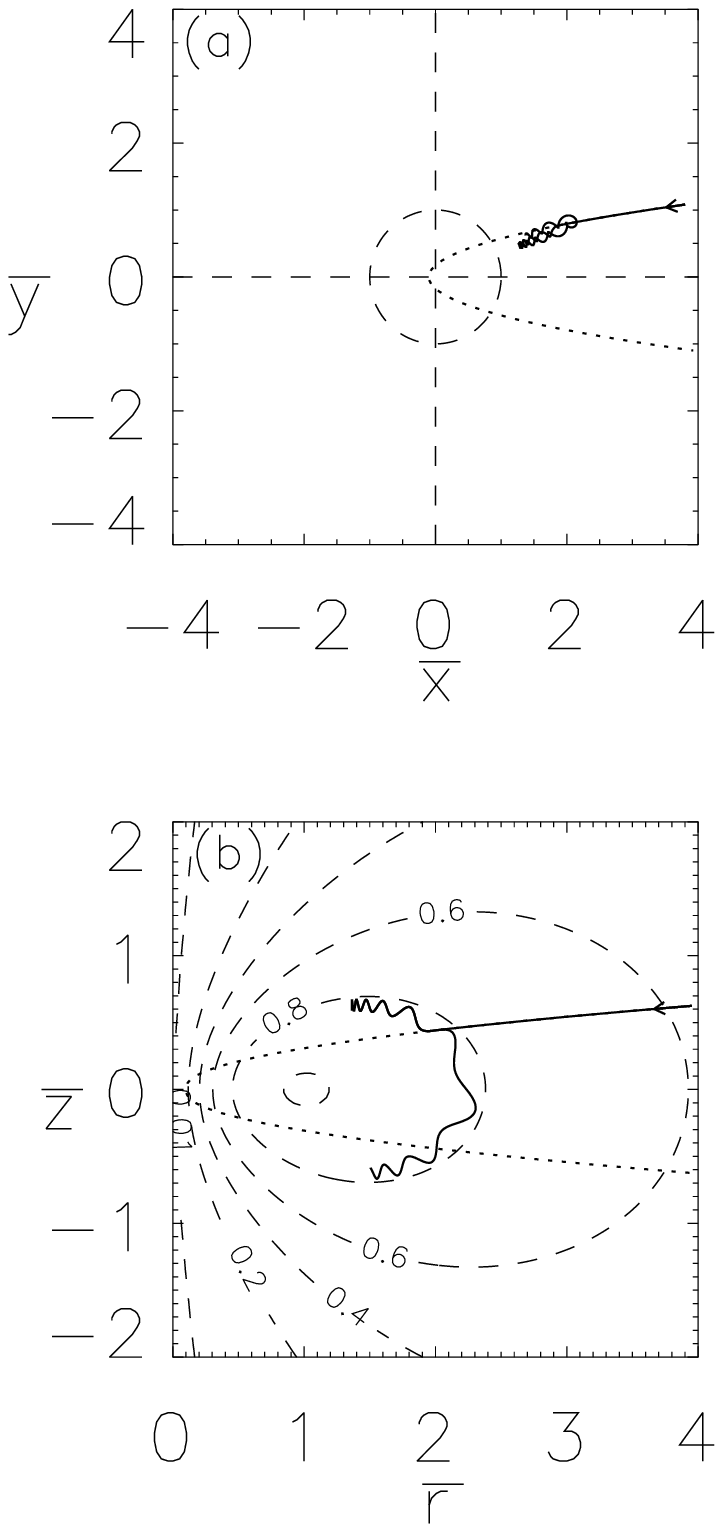}\end{figure}

\begin{figure}[ptb]
\caption{(a)\ Mechanical angular momentum $\bar{p}_{\phi}$ v. time $\tau$ and
(b)\ kinetic energy (KE), potential energy (PE) and canonical angular momentum
$\bar{P}_{\phi}$ v. time for the calculation shown in Fig.
\ref{mirror..pericenter=0.1_theta=30_alpha=0_rho0=4_rstar=2_omegaratio=200.eps}%
. Mechanical angular momentum $\bar{p}_{\phi}\ \ $ is conserved before
charging, but oscillates after charging; canonical angular momentum $\bar
{P}_{\phi}$ is much larger than mechanical angular momentum because of strong
magnetic field and is conserved after charging. }%
\label{mirror..pericenter=0.1_theta=30_alpha=0_rho0=4_rstar=2_omegaratio=200momenta.eps}%
\epsscale{0.5} \plotone{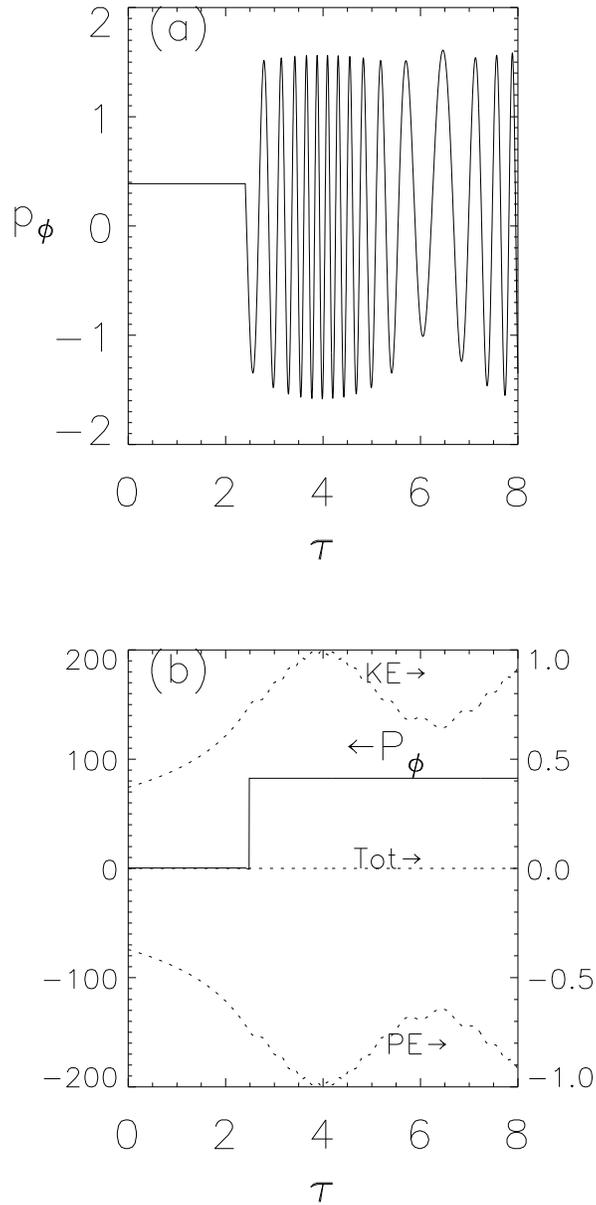}\end{figure}

Potential barriers occur at locations where $\chi(\bar{r},\bar{z})>$ $\bar
{H}.$ We first consider motion of a sibling particle constrained to stay in
the $\bar{z}=0$ plane (i.e., the particle starts with $\bar{v}_{z}=0$ and no
forces exist that push it off the $\bar{z}=0$ plane). In this case the
effective potential is a function of $\bar{r}$ only and is%
\begin{equation}
\chi(\bar{r},0)=\frac{\left(  \stackrel{\textnormal{mechanical}}{\overbrace{\bar
{L}\cos\theta}}\ \ +\stackrel{\textnormal{magnetic}}{\overbrace{\frac{\left\langle
\omega_{c\sigma}\right\rangle }{2\Omega_{0}}\ \left[  \bar{\psi}(\bar{r}%
_{\ast},0)-\bar{\psi}(\bar{r},0)\right]  }}\right)  ^{2}}{2\ \bar{r}^{2}%
}-\stackrel{\textnormal{gravitational}}{\overbrace{\frac{1}{\bar{r}}}}.
\label{chi z plane}%
\end{equation}
If $\chi(\bar{r},0)$ exceeds $\bar{H}$ at some radius $\bar{r}>$ $\bar
{r}_{\ast}$, the particle becomes trapped within a finite extent region as
indicated in Fig.\ref{Effective-Potential-Neutral-Particle}(b) or equivalently
by the third and fourth rows, right hand column of Fig.
\ref{effective-potential-scan.eps}. \ If $\bar{r}_{\ast}$ is small or large
compared to unity so $r_{\ast}$ is not near the peak of $\bar{\psi},$ then
$\bar{\psi}(\bar{r}_{\ast},0)\ll1.$ Because the particle is assumed to be
light-weight (i.e., \thinspace$r_{d}\ll r_{d}^{crit}$), its average-field
cyclotron frequency $\left\langle \omega_{c\sigma}\right\rangle $ will be much
larger than the Kepler frequency $\Omega_{0}.$ Since $\bar{L}$ is of order
unity, the light-weight particle will have $\left\vert \left\langle
\omega_{c\sigma}\right\rangle /2\Omega_{0}\right\vert \gg\bar{L}$ in which
case%
\[
\max\left\{  \frac{\ \left(  \bar{L}\cos\theta+\frac{\left\langle
\omega_{c\sigma}\right\rangle \ }{2\Omega_{0}}\left[  \bar{\psi}(\bar{r}%
_{\ast},0)-\bar{\psi}(\bar{r},0)\right]  \right)  ^{2}}{2\bar{r}^{2}%
\ }\right\}  \simeq\max\left\{  \frac{1}{2\bar{r}^{2}\ }\left(  \frac
{\left\langle \omega_{c\sigma}\right\rangle \ }{2\Omega_{0}}\bar{\psi}(\bar
{r},0)\right)  ^{2}\right\}
\]
so the peak of $\chi(\bar{r},0)$ will occur where $\bar{\psi}(\bar{r},0)$
takes on its maximum value, namely unity. The maximum of $\chi(\bar{r},0)$ for
a light-weight particle is thus
\begin{equation}
\chi(1,0)\simeq\frac{1}{2}\left(  \frac{\left\langle \omega_{c\sigma
}\right\rangle }{2\Omega_{0}}\right)  ^{2}-1 \label{generic barrier}%
\end{equation}
where \ the $-1$ term comes from the gravitational potential at $\bar
{r}=1,z=0.$ This gives the necessary condition for trapping light-weight
charged dust grains to be
\begin{equation}
\frac{\left\vert \left\langle \omega_{c\sigma}\right\rangle \right\vert
}{\ \Omega_{0}}>2\sqrt{2}. \label{trapping}%
\end{equation}
Because a typical light-weight grain has $\left\vert \left\langle
\omega_{c\sigma}\right\rangle \right\vert \gg2\sqrt{2}\Omega_{0},$ a
light-weight grain confined to the $z=0$ plane will become trapped in a
finite-sized region upon being charged, i.e., it will have accreted. The same
will be true for the associated sibling photo-electrons since they also have
$\left\vert \left\langle \omega_{ce}\right\rangle \right\vert \gg2\sqrt
{2}\Omega_{0}.$ Figure
\ref{mirror..pericenter=0.1_theta=30_alpha=0_rho0=4_rstar=2_omegaratio=200.eps}%
\ shows a direct numerical integration of the equation of motion demonstrating
this basic accretion mechanism in three dimensions: a light-weight neutral
dust grain disintegrates at a certain location into an $\left\vert
\left\langle \omega_{c\sigma}\right\rangle \right\vert \gg2\sqrt{2}\Omega_{0}$
positively charged dust grain \ (there would also be $Z$ associated
photo-electrons which for clarity are not shown in the figure but would also
have cyclotron-type orbits). An actual dust grain would start with an
infinitesimal energy $0<\bar{H}\ll1;$ the calculation here uses $\bar
{H}=0\,\ $as representative of this infinitesimal $\bar{H}$ since the
difference between an orbit with $\bar{H}=0$ and an orbit with infinitesimal
$\bar{H}$ is insignificant at any finite distance. The newly created charged
particles are trapped to the vicinity of a $\bar{\psi}(\bar{r},\bar
{z})=const.$ poloidal flux surface (poloidal flux surfaces are shown by dashed
lines in
Fig.\ref{mirror..pericenter=0.1_theta=30_alpha=0_rho0=4_rstar=2_omegaratio=200.eps}%
(b)). Because of $\mu$ conservation, the charged particles can also be
mirror-trapped, so while on the the constant $\bar{\psi}$ surface, they
reflect from regions of this surface where the magnetic field is strong.
Figure
\ref{mirror..pericenter=0.1_theta=30_alpha=0_rho0=4_rstar=2_omegaratio=200momenta.eps}%
(a) plots the time dependence of $\bar{p}_{\phi}$ for the particle shown in
Fig.\ref{mirror..pericenter=0.1_theta=30_alpha=0_rho0=4_rstar=2_omegaratio=200.eps}%
.

Figure
\ref{mirror..pericenter=0.1_theta=30_alpha=0_rho0=4_rstar=2_omegaratio=200momenta.eps}%
(b) plots the canonical angular momentum $\bar{P}_{\phi}$ (solid line) and the
kinetic/potential energies (dashed lines labeled `KE' and `PE' in figure). It
is seen that $\bar{p}_{\phi}$ is conserved before charging whereas $\bar
{P}_{\phi}$ is the conserved quantity after charging. Also, the total energy
(kinetic + potential, dashed line labeled `Tot' in figure) remains zero.
Strictly speaking, this plot should be considered as referring to the neutral
dust grain until charging, and then to the charged dust grain after charging
so the jump in $\bar{P}_{\phi}$ at the charging time seen in the figure does
not violate the requirement that $\bar{P}_{\phi}$ is a constant of the motion
for\ a specific particle.

\begin{figure}[ptbptb]
\caption{Same parameters as Fig.
\ref{mirror..pericenter=0.1_theta=30_alpha=0_rho0=4_rstar=2_omegaratio=200.eps}%
, except $R_{\ast}=0.8.$ Charged particle is now not mirror-trapped.}%
\label{notmirror..pericenter=0.1_theta=30_alpha=0_rho0=4_rstar=0.8_omegaratio=200.eps}%
\epsscale{0.5} \plotone{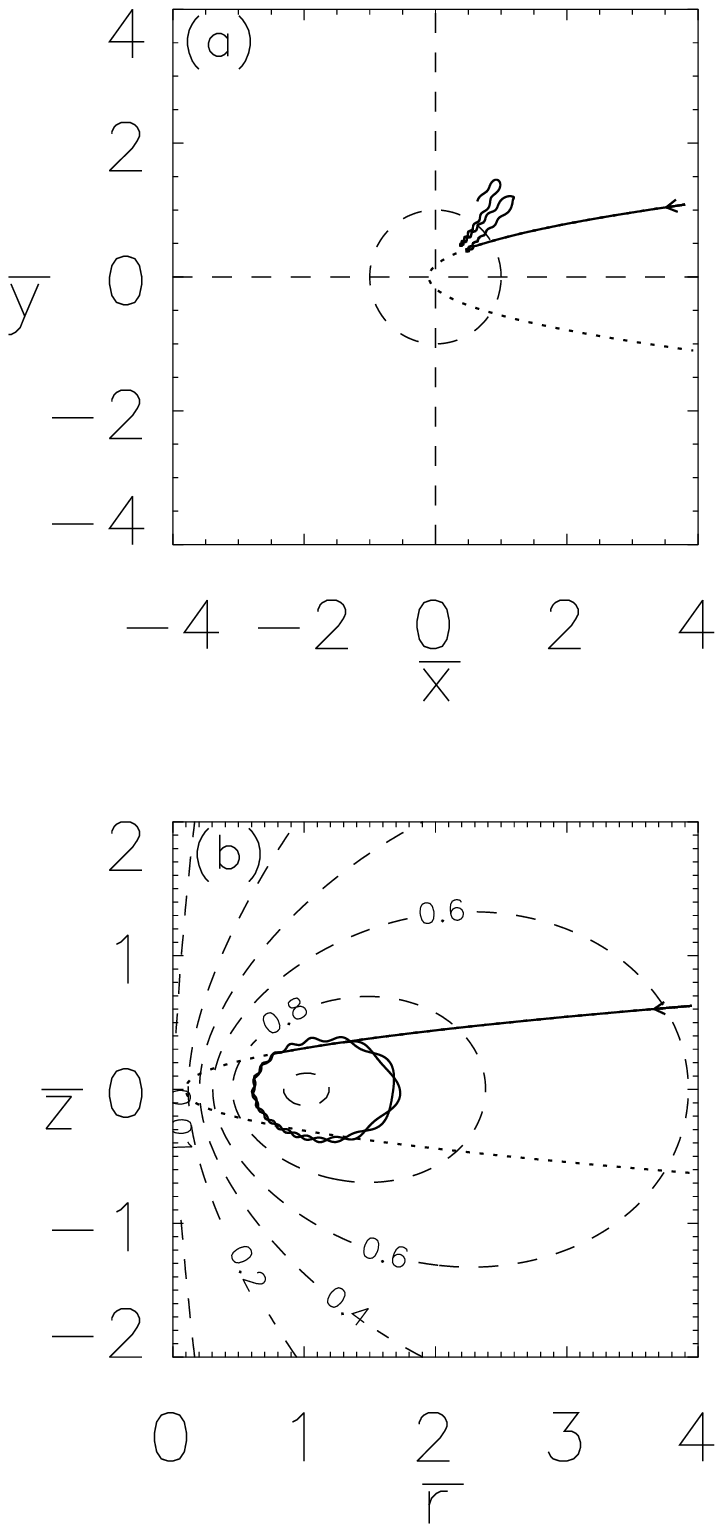}\end{figure}

Figure
\ref{notmirror..pericenter=0.1_theta=30_alpha=0_rho0=4_rstar=0.8_omegaratio=200.eps}
shows the three dimensional orbit of a light-weight charged dust grain with
slightly different parameters so that it is not mirror trapped. The derivation
of the non-dimensional equation of motion used here is given in Appendix
\ref{Equation of motion}. 

\subsubsection{Width of light-weight particle trapping well and relation to
cyclotron orbits \label{trapping well width}}

If $\left\vert \left\langle \omega_{c\sigma}\right\rangle \right\vert
/2\Omega_{0}\gg\ \left\vert \bar{p}_{\phi}(\bar{r}_{\ast},\bar{z}_{\ast
})\right\vert $, the magnetic term in Eq.\ref{effective potential magnetized}
dominates the mechanical term as soon as $\bar{r}$ deviates slightly from
$r_{\ast}.$ This implies existence of a narrow trench-like potential
well\ with minimum very close to $r_{\ast}.$ The effective potential shown
Fig.\ref{Effective-Potential-Neutral-Particle}(b) has such a trench; this
situation involves a particle confined to the $\bar{z}=0$ plane and the trench
is at $\bar{r}=\bar{\rho}=0.55.$ This situation is also evident in the third
and fourth rows, right hand column of Fig. \ref{effective-potential-scan.eps}.
If $\left\vert \left\langle \omega_{c\sigma}\right\rangle \right\vert
\gg\Omega_{0}$ the gravitational term is completely overwhelmed by the
magnetic term so the trench bottom in the $z=0$ plane is where
\begin{equation}
\ \bar{\psi}(\bar{r})=\frac{\Omega_{0}}{\left\langle \omega_{c\sigma
}\right\rangle }\bar{p}_{\phi}(\bar{r}_{\ast},\bar{z}_{\ast})\ +\bar{\psi
}(r_{\ast}).\ \label{well bottom}%
\end{equation}
Taylor expansion of $\ \bar{\psi}(\bar{r})$ near $r_{\ast}$ gives
\begin{equation}%
\begin{array}
[c]{cc}%
\ \bar{\psi}(\bar{r})= & \bar{\psi}(\bar{r}_{\ast})+\ \left(  \bar{r}-\bar
{r}_{\ast}\right)  \left(  \frac{\partial\bar{\psi}}{\partial\bar{r}}\right)
_{\bar{r}=\bar{r}_{\ast}}\\
& +\frac{1}{2}\left(  \bar{r}-\bar{r}_{\ast}\right)  ^{2}\left(
\frac{\partial^{2}\bar{\psi}}{\partial\bar{r}^{2}}\right)  _{\bar{r}=\bar
{r}_{\ast}}+...
\end{array}
\label{Taylor}%
\end{equation}
If $\bar{r}_{\ast}$ is not close to unity, then $\bar{\psi}(\bar{r}_{\ast})$
is not close to its maximum value so the leading term in the Taylor expansion
is the one involving $\partial\bar{\psi}/\partial\bar{r}.$ Using
Eq.\ref{Taylor} to substitute for $\ \bar{\psi}(\bar{r})$ in
Eq.\ref{well bottom} gives the trench bottom to be at%
\begin{equation}
\ \ \bar{r}\ \ =\bar{r}_{\ast}+\frac{2}{\left(  \partial\bar{\psi}%
/\partial\bar{r}\right)  _{\bar{r}=\bar{r}_{\ast}}}\frac{\Omega_{0}%
}{\left\langle \omega_{c\sigma}\right\rangle }\bar{p}_{\phi}(\bar{r}_{\ast
},\bar{z}_{\ast})\ \label{more well bottom}%
\end{equation}
so the trench bottom is close to $\bar{r}_{\ast}$ because $\left\vert
\left\langle \omega_{c\sigma}\right\rangle \right\vert \gg\Omega_{0}$ is being
assumed. If $\bar{r}_{\ast}<1$ then $\left(  \partial\bar{\psi}/\partial
\bar{r}\right)  _{\bar{r}=\bar{r}_{\ast}}$ is positive and vice versa since
$\bar{\psi}$ has its maximum value at $\bar{r}=1.$ The trench bottom will thus
be outside of $\bar{r}_{\ast}$ if $\bar{r}_{\ast}<1$ so $\left\langle
\omega_{c\sigma}\right\rangle \left(  \partial\bar{\psi}/\partial\bar
{r}\right)  _{\bar{r}=\bar{r}_{\ast}}$is positive and vice versa if $\bar
{r}_{\ast}>1$. The sign of $\ \bar{\psi}(\bar{r})-\bar{\psi}(\bar{r}_{\ast})$
oscillates as the particle bounces back and forth in the trench. Using
Eq.\ref{solve phidot} expressed in normalized variables, \ and noting
that\ $\bar{P}_{\phi}=\bar{p}_{\phi}(\bar{r}_{\ast},\bar{z}_{\ast})+\bar{\psi
}(\bar{r}_{\ast},\bar{z}_{\ast})\left\langle \omega_{c\sigma}\right\rangle
/2\Omega_{0}\ \ $it is seen that the azimuthal velocity
\begin{equation}%
\begin{array}
[c]{ccl}%
\frac{d\phi}{d\tau} & = & \frac{\bar{P}_{\phi}\ -\frac{\left\langle
\omega_{c\sigma}\right\rangle }{2\Omega_{0}}\bar{\psi}(\bar{r})}{\bar{r}^{2}%
}\\
& = & \frac{\bar{p}_{\phi}(\bar{r}_{\ast},\bar{z}_{\ast})\ -\frac{\left\langle
\omega_{c\sigma}\right\rangle }{2\Omega_{0}}\left[  \bar{\psi}(\bar{r}%
)-\bar{\psi}(\bar{r}_{\ast})\right]  }{\bar{r}^{2}}%
\end{array}
\label{phidot}%
\end{equation}
has an oscillating polarity. The combined oscillation of $\bar{r}$ and
\ $d\phi/d\tau$ corresponds to the particle tracing out Larmor orbits with
gyro-center at the trench bottom
\citep{Schmidt1979}%
.

To summarize: In the $\left\vert \left\langle \omega_{c\sigma}\right\rangle
\right\vert /\Omega_{0}\gg1$ situation (i.e., light-weight particles) Eq.
\ref{effective potential magnetized} provides an effective potential whereby
the magnetized charged particle is confined to the vicinity of a constant
$\psi$ surface, just like a charged particle in a tokamak
\citep{Rome1979}%
. The particle motion over the constant $\psi$ surface can be understood as a
sum of parallel to $\mathbf{B}$ motion, cyclotron motion, and particle drifts
(curvature, grad $B,$ etc.) as given by Eq.\ref{particle drifts}. Furthermore,
regions where $\mu B$ \ is large can constitute an additional potential
barrier (i.e., magnetic mirror) that prevents those subsets of particles
having inadequate velocity parallel to $\mathbf{B}$\textbf{ }from accessing
the entire constant $\psi$ surface. Because of magnetic mirroring by
$\mu\nabla B$ forces, particles in these subsets are confined to the weaker
magnetic field regions of a constant $\psi$ surface as seen in
Fig.\ref{mirror..pericenter=0.1_theta=30_alpha=0_rho0=4_rstar=2_omegaratio=200.eps}%
(b).

\subsection{Speiser orbit particles ($K\cos\theta\simeq1/2$)}

When $\bar{r}_{\ast}\simeq1$ and $\bar{z}_{\ast}\simeq0$, photoemission occurs
near the peak of $\psi(\bar{r},\bar{z})$, i.e., where $\nabla\psi$ $\simeq0$;
see Fig. \ref{effective-potential-scan.eps} second row from top where
$P_{\phi}$ is just grazing the peak of $\psi.$ The linear term in the Taylor
expansion in Eq.\ref{Taylor} is therefore negligible. Since $\partial^{2}%
\bar{\psi}/\partial\bar{r}^{2}$ is negative near the maximum of $\bar{\psi},$
Eq.\ref{Taylor} becomes
\begin{equation}
\bar{\psi}(\bar{r})\simeq\bar{\psi}(\bar{r}_{\ast})-\ \frac{1}{2}\left(
\bar{r}-\bar{r}_{\ast}\right)  ^{2}\left\vert \left(  \frac{\partial^{2}%
\bar{\psi}}{\partial\bar{r}^{2}}\right)  _{\bar{r}=\bar{r}_{\ast}}\right\vert
. \label{chi Speiser}%
\end{equation}
Equation \ref{phidot} then reduces to%
\begin{equation}
\frac{d\phi}{d\tau}\ =\frac{\bar{p}_{\phi}(\bar{r}_{\ast},0)\ +\frac
{\left\langle \omega_{c\sigma}\right\rangle }{4\Omega_{0}}\ \left(  \bar
{r}-\bar{r}_{\ast}\right)  ^{2}\left\vert \left(  \frac{\partial^{2}\bar{\psi
}}{\partial\bar{r}^{2}}\right)  _{\bar{r}=\bar{r}_{\ast}}\right\vert }{\bar
{r}^{2}} \label{preSpeiser}%
\end{equation}
and for $\bar{p}_{\phi}(\bar{r}_{\ast},0)\ $being positive (i.e., parent
particle was prograde), $d\phi/d\tau$ is always positive. This corresponds to
Speiser-type orbits because when the particles bounce back and forth across
the peak of $\psi$, they are bouncing back and forth between regions where the
poloidal magnetic field $\sim\partial\psi/\partial r$ changes sign. As
discussed in Section \ref{Speiser} of Appendix \ref{Orbit review}, this
results in paramagnetism, i.e., positively charged particles moving in the
positive $\phi$ direction and so \textit{producing} rather than opposing a
$B_{z}$ field. Creation of Speiser-orbiting particles sustains the poloidal
magnetic field against losses and will amplify an initial seed poloidal field;
creation of Speiser particles therefore constitutes a dynamo for driving
toroidal current.

\begin{figure}[ptb]
\caption{Speiser orbit resulting from parent with $\bar{H}=0$, $\bar{\rho
}_{pericenter}=0.95,$ $\theta=18^{0},$ $\alpha=0^{0}.$ Charging occurs at
$\bar{R}_{\ast}=1.2$ and the child particle is a positive particle with
$\omega_{c\sigma}/\Omega_{0}=10;\ $ (a) the $\bar{x}$-$\bar{y}$ plane orbit is
counter-clockwise corresponding to paramagnetic motion; (b) the $\bar{r}%
$-$\bar{z}$ plane orbit involves the particle continuously reflecting from the
interior of a toroidal flux tube.}%
\label{speiser_pericenter=0.95_theta=18_alpha=0_rho0=4_rstar=1.2_omegaratio=10.eps}%
\epsscale{0.5} \plotone{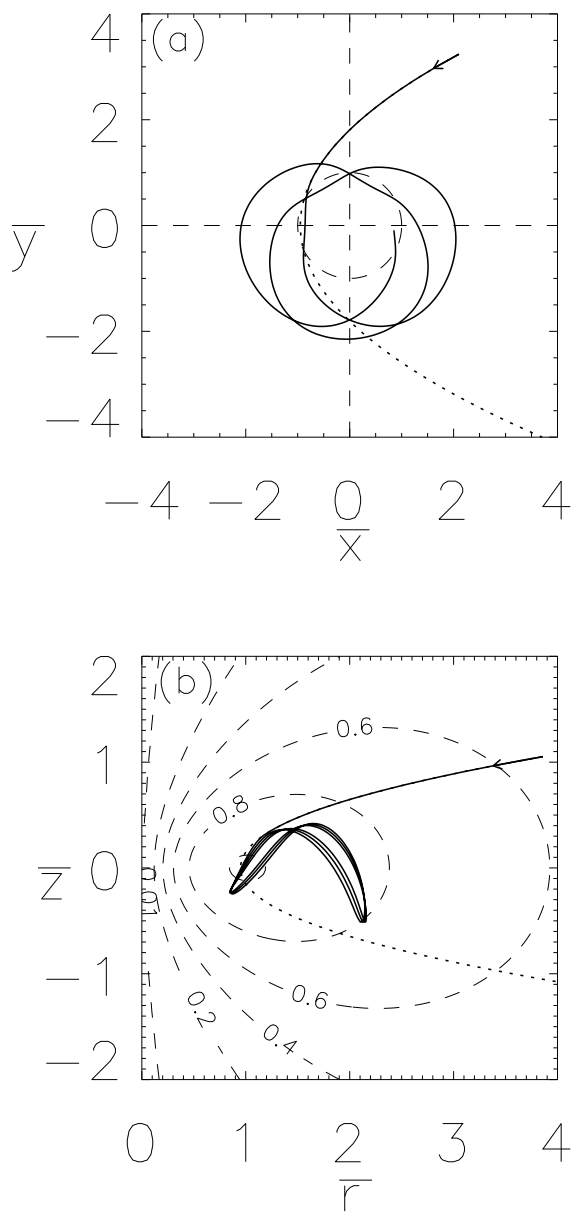}\end{figure}

Figure
\ref{speiser_pericenter=0.95_theta=18_alpha=0_rho0=4_rstar=1.2_omegaratio=10.eps}
shows the creation of a Speiser orbit by photo-emission charging of a neutral
particle near the poloidal field magnetic axis. Figure
\ref{speiser_pericenter=0.95_theta=18_alpha=0_rho0=4_rstar=1.2_omegaratio=10.eps}%
(a) shows that the orbit is paramagnetic (i.e., particle moves in positive
$\phi$ direction) while Fig.
\ref{speiser_pericenter=0.95_theta=18_alpha=0_rho0=4_rstar=1.2_omegaratio=10.eps}%
(b) shows that the orbit involves repeated reflection from the interior of a
poloidal flux surface in the manner discussed in Section \ref{Speiser} of
Appendix \ref{Orbit review}.

\subsection{Drain-hole particles ($K\cos\theta\simeq-1/2$)}

Charged particles born with $\bar{P}_{\phi}=0\ $are called `drain-hole'
particles because they behave as if they are going down a drain. The
properties of drain-hole particles restricted to the $z=0$ plane were briefly
examined in
\citet{Bellan2007}%
; here the more general 3D situation will be considered. Using
Eq.\ref{norm can momentum} and $\bar{p}_{\phi}(\bar{r}_{\ast},\bar{z}_{\ast
})=\bar{L}\cos\theta$ the $\bar{P}_{\phi}=0$ condition corresponds to
\begin{equation}
\bar{L}\cos\theta=-\frac{\left\langle \omega_{c\sigma}\right\rangle }%
{2\Omega_{0}}\ \bar{\psi}(\bar{r}_{\ast},\bar{z}_{\ast}%
)\ \label{drain-perigee}%
\end{equation}
implying that $\cos\theta$ is negative in which case the parent particle must
have been retrograde. The effective potential (see
Eq.\ref{effective potential magnetized}) for the $\bar{P}_{\phi}=0$ class of
particles reduces to
\begin{equation}
\chi(\bar{r},\bar{z})\simeq\frac{\left\langle \omega_{c\sigma}\right\rangle
^{2}}{8\Omega_{0}^{2}}\frac{\left(  \bar{\psi}(\bar{r},\bar{z})\right)  ^{2}%
}{\ \bar{r}^{2}}\ -\ \frac{1}{\sqrt{\bar{r}^{2}+\bar{z}^{2}}}
\label{chi free fall}%
\end{equation}
which has a funnel (i.e., drain-like) profile near $\bar{r}=0,$ $\bar{z}=0$
due to the second (gravitational) term and a hill on the funnel side wall with
peak near $\bar{r}=\bar{a},$ $\bar{z}=0$ due to the first (St\"{o}rmer) term.
A particle initially on the hill (i.e., near the poloidal field magnetic axis)
will fall down the hill into the drain-like funnel; see right hand column,
second row from bottom in Fig. \ref{effective-potential-scan.eps} for plot of
first term in Eq.\ref{chi free fall}. Thus, no matter where a $\bar{P}_{\phi
}=0$ particle starts in $\bar{r},\bar{z}$ space, it eventually follows a
spiral path down to\ $\bar{r}=0,\bar{z}=0;$ no centrifugal force will ever
push it back outwards because the first term in Eq.\ref{chi free fall} has no
singularity at $\bar{r}=0$ (recall that $\bar{\psi}\sim\bar{r}^{2}$ for small
$\bar{r},\bar{z}$). The sense of this downward spiraling trajectory will be in
the $-\left\langle \omega_{c\sigma}\right\rangle $ direction as shown by
Eq.\ref{phidot}. Since $\bar{\psi}\sim\bar{r}^{2}$ for small $\bar{r},\bar{z}$
it is seen from Eq.\ref{phidot} that drain-hole particles have a limiting
angular velocity
\begin{equation}
\lim_{\bar{r},\bar{z}\rightarrow0}\frac{\ d\phi}{d\tau}\ =-\ \frac
{\left\langle \omega_{c\sigma}\right\rangle }{2\Omega_{0}\ }\lim_{\bar{r}%
,\bar{z}\rightarrow0}\left(  \frac{\bar{\psi}(\bar{r},\bar{z})}{\bar{r}^{2}%
}\right)  =const. \label{drain-hole ang vel}%
\end{equation}

Combination of Eqs. \ref{p range} and \ref{drain-perigee} show that drain-hole
particles can only be created if the accessibility condition%
\begin{equation}
\frac{\left\langle \omega_{c\sigma}\right\rangle ^{2}}{4\Omega_{0}^{2}%
}\ \left(  \bar{\psi}(\bar{r}_{\ast},\bar{z}_{\ast})\right)  ^{2}<\left(
2\bar{\rho}^{2}\bar{H}+2\bar{\rho}\right)  \cos^{2}\theta
\ \label{accessibility}%
\end{equation}
is satisfied, a condition that $\left\langle \omega_{c\sigma}\right\rangle
/\Omega_{0}$ not be too large. Since $\bar{H}\simeq0$ is assumed, $\cos
\theta\simeq-1$ for drain-hole particles, and since $\bar{\rho}=\sqrt{\bar
{r}^{2}+\ \bar{z}^{2}}$ is required to be larger than the pericenter, this
condition becomes
\begin{equation}
\left\vert \bar{\psi}(\bar{r}_{\ast},\bar{z}_{\ast})\right\vert <\ \left\vert
\frac{2\Omega_{0}}{\left\langle \omega_{c\sigma}\right\rangle }\sqrt
{2\bar{\rho}_{pericenter}}\right\vert . \label{drain hole creation criterion}%
\end{equation}
Because Eq.\ref{drain hole creation criterion} requires $\left\langle
\omega_{c\sigma}\right\rangle /\Omega_{0}$ to be small, drain-hole particles,
like Speiser particles, result from dust grains with grain radii consistent in
order of magnitude with Eq.\ref{rdnom}.

Particles with $\bar{P}_{\phi}$ exactly zero (\textquotedblleft
perfect\textquotedblright\ drain-hole particles) fall down the gravitational
potential all the way to the central object at the origin $\bar{r}=0,$
$\bar{z}=0.$ Particles that are not quite perfect drain-hole particles will
have small, but finite $\bar{P}_{\phi}$ and so will reflect when close to the
central object.

\begin{figure}[ptb]
\caption{Drain-hole particle ($\bar{H}=0,$ $\bar{\rho}_{pericenter}=0.2,$
$\theta=170^{0},\alpha=0,$ $\bar{R}_{\ast}=0.8$) falls across magnetic field
all the way to the central object. This is a heavy particle (dust grain)\ and
has $\omega_{c\sigma}/\Omega_{0}=1.6;$ (a)\ shows orbit projection in $\bar
{x}$-$\bar{y}$ plane, (b) shows projection in $\bar{r}$-$\bar{z}$ plane.
Dotted line shows trajectory parent would have continued to have if it had not
become charged.}%
\label{drain-hole_pericenter=0.3_theta=170_alpha=0_rho0=4_rstar=0.8_omegaratio=1.6.eps}%
\epsscale{0.5} \plotone{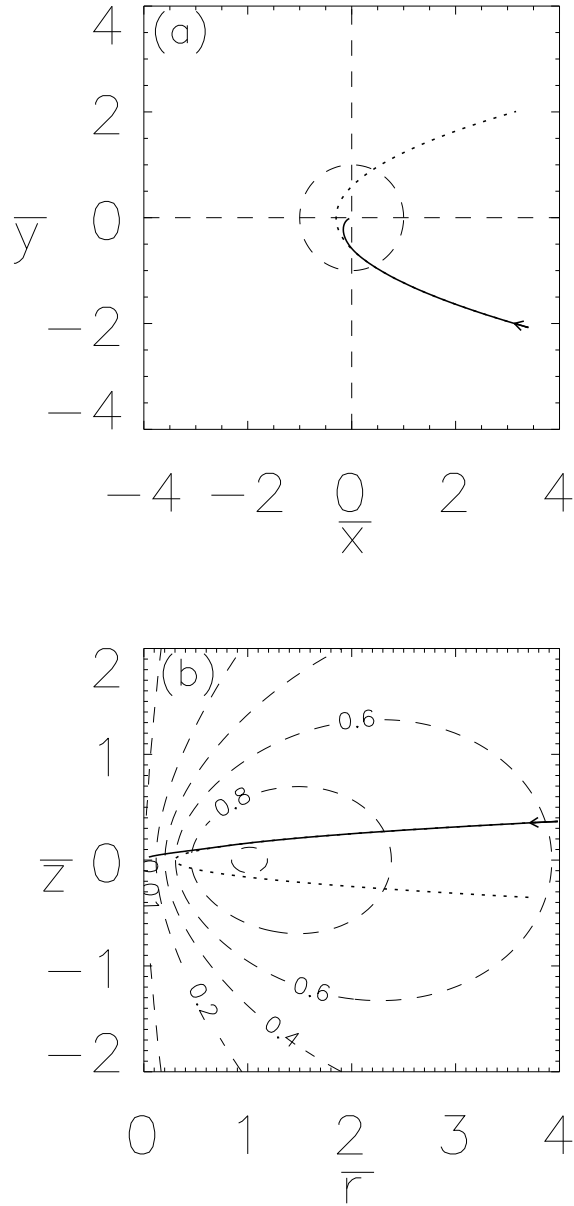}\end{figure}

\begin{figure}[ptb]
\caption{(a)\ Mechanical angular momentum $p_{\phi}$ for drain-hole particle
is not conserved when particle becomes charged, (b)\ canonical angular
momentum $P_{\phi},$ kinetic energy (KE) and potential energy (PE). The
magnitudes of the kinetic and potential energy increase without bound as the
particle falls towards the central object. The canonical angular momentum is
conserved and is near zero upon charging.}%
\label{drain-hole_pericenter=0.3_theta=170_alpha=0_rho0=4_rstar=0.8_omegaratio=1.6momenta-adobe.eps}%
\epsscale{0.5} \plotone{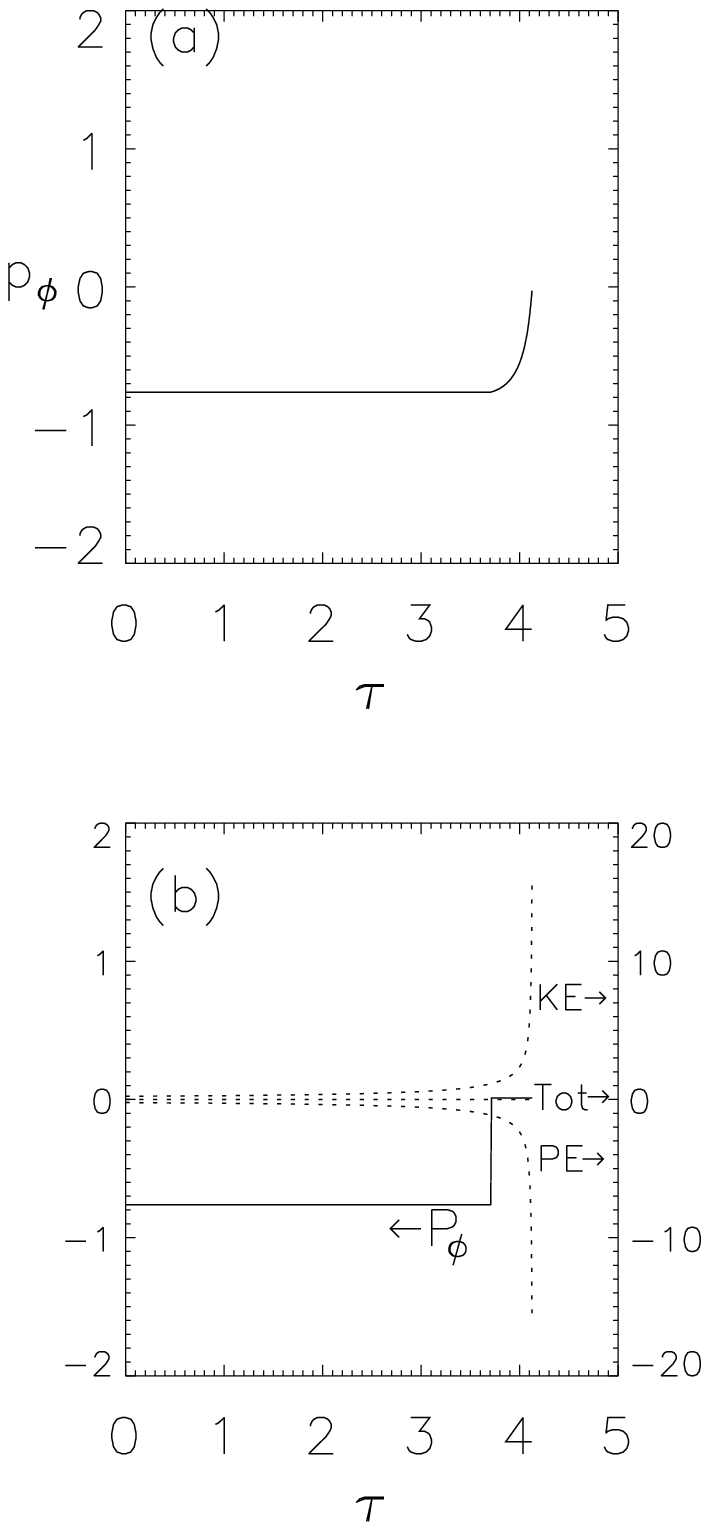}\end{figure}

Figure
\ref{drain-hole_pericenter=0.3_theta=170_alpha=0_rho0=4_rstar=0.8_omegaratio=1.6.eps}
shows a numerical calculation of a drain-hole particle orbit with $\bar{H}=0.$
The solid line in Fig.
\ref{drain-hole_pericenter=0.3_theta=170_alpha=0_rho0=4_rstar=0.8_omegaratio=1.6.eps}%
(a) shows the projection of the drain-hole orbit in the $\bar{x}$-$\bar{y}$
plane. The orbit the neutral particle would have had if it had not encountered
any photons and so remained neutral is shown as a dotted line. Figure
\ref{drain-hole_pericenter=0.3_theta=170_alpha=0_rho0=4_rstar=0.8_omegaratio=1.6.eps}%
(b)\ shows the projection in the $\bar{r}$-$\bar{z}$ plane with surfaces of
constant $\psi$ indicated (the dotted line again shows the orbit the neutral
particle would have had if it had not encountered any photons). The drain-hole
particle has a retrograde orbit (clockwise sense resulting from its angle of
inclination $\theta>90^{0}$). Figure
\ref{drain-hole_pericenter=0.3_theta=170_alpha=0_rho0=4_rstar=0.8_omegaratio=1.6momenta-adobe.eps}%
(a) shows that the mechanical angular momentum is not constant after charging
while Fig.
\ref{drain-hole_pericenter=0.3_theta=170_alpha=0_rho0=4_rstar=0.8_omegaratio=1.6momenta-adobe.eps}%
(b) shows that the canonical angular momentum remains constant at zero after
charging. Figure
\ref{drain-hole_pericenter=0.3_theta=170_alpha=0_rho0=4_rstar=0.8_omegaratio=1.6momenta-adobe.eps}%
(b) also shows how the magnitudes of the potential and kinetic energies
increase without bound as the particle descends towards the central object
while the total energy stays zero (kinetic, potential and total energies shown
as dashed lines).

When drain-hole particles approach the central object, the gravitational term
in Eq.\ref{chi free fall} dominates (recall that $\bar{\psi}\sim\bar{r}^{2}$
at small $\bar{r}$ and near $\bar{z}=0$). It therefore makes sense to use
spherical coordinates in this region in which case the Hamiltonian is
approximately
\begin{equation}
0\simeq\frac{1}{2}\left(  \frac{d\bar{R}}{d\tau}\right)  ^{2}-\frac{1}{\bar
{R}} \label{free-fall}%
\end{equation}
where $\bar{R}$ is the spherical radius and $\bar{H}\simeq0$ has been assumed.
Equation \ref{free-fall} shows that the free-fall velocity scales as
\begin{equation}
\ \left\vert \frac{d\bar{R}}{d\tau}\right\vert \ =\sqrt{\frac{2}{\bar{R}\ }}
\label{free-fall velocity}%
\end{equation}
and particle flux conservation over a spherical surface $4\pi\bar{R}^{2}$
shows that $4\pi\bar{R}^{2}n_{dh}(\bar{R})d\bar{R}/d\tau=const.$ where
$n_{dh}(\bar{R})$ is the density of drain-hole particles. Thus, if the
incoming drain-hole particles do not accumulate, spherical focusing combined
with the accelerating free-fall velocity shows that the density of drain-hole
particles scales as%
\begin{equation}
n_{dh}(\bar{R})\sim\frac{1}{\bar{R}^{2}d\bar{R}/d\tau}\sim\frac{1}{\bar
{R}^{3/2}}. \label{drain-hole density}%
\end{equation}

\begin{figure}[ptb]
\caption{Drain hole dust grains fall across poloidal field lines towards
central object leaving behind stranded electrons which are confined to
poloidal flux surface on which they are born. Drain hole particles accumulate
near central object creating large positive charge there. This repels positive
particles (drain hole particles, ions) to flow axially away from $z=0$ plane
and also attracts stranded electrons which can flow on poloidal flux surface.
The result is a clockwise poloidal current flow pattern in upper-half $r$-$z$
plane, giving a positive $B_{\phi}$ in region linked by poloidal current and a
negative $B_{\phi}$ in lower-half $r$-$z$ plane where poloidal current flow is
counter-clockwise.}%
\label{f12.eps}%
\plotone{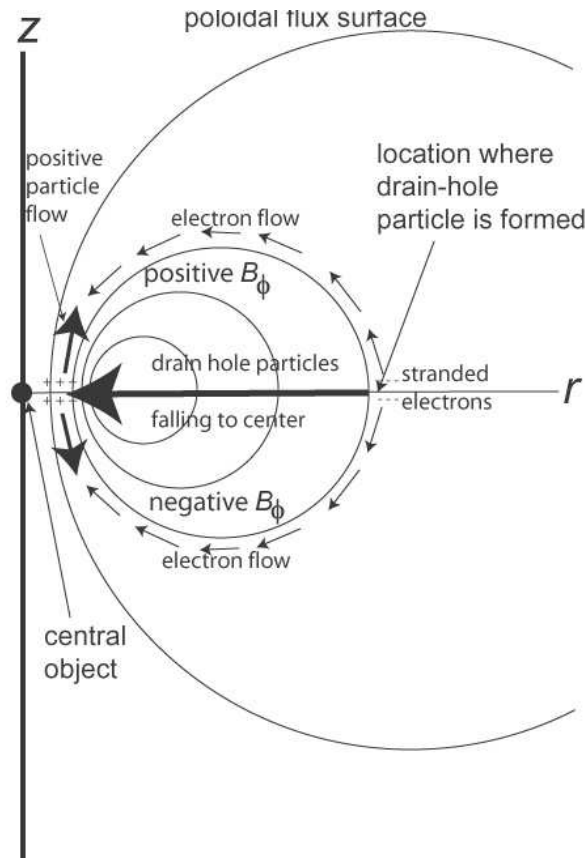}\end{figure}

Accumulation of the drain-hole particles in the vicinity of the central object
will also increase the density of drain-hole particles with time. There is
thus both a temporal increase and a geometrically-induced increase of the
drain-hole particle density as $\bar{R}$ decreases. Since the sibling
\ electrons were left stranded at large $\bar{r},$ what results is the
establishment of a large positive charge density near the central object and
an equal-magnitude negative charge density at large $\bar{r}$. The flow
pattern of the drain hole particles and the location of the stranded electrons
is sketched in Fig.\ref{f12.eps}. Eventually the positive space charge near
the central object becomes so large that it produces a repulsive electrostatic
electric field that balances the gravitational force acting on any additional
drain hole particles. The large positive potential near $\bar{r}=0$ will tend
to drive\ axial electric currents flowing away from the $\bar{z}=0$ plane
resulting in the loss or neutralization of some of the drain-hole particles.
The axial electric current could result from attraction of electrons near
$\bar{r}=0$ towards the $\bar{z}=0$ plane or from expulsion of positive
particles away from the $\bar{z}=0$ plane. Either electron attraction or
positive particle repulsion will deplete the positive space charge density
near $\bar{r}=0.$ There will then have to be a replenishing flow of additional
drain hole particles into the $\bar{r}=0$ region to compensate for this
depletion of positive space charge.\ Being very low mass, the stranded
electrons have very small Larmor orbit radius and so are constrained to stay
essentially right on the poloidal flux surface on which they were
photo-emitted (see Secs.\ref{light-weight trapping} and
\ref{trapping well width}). The electron flow is thus at a much larger
$|\bar{z}|$ than the drain-hole particle radially inward flow which is
concentrated near the $\bar{z}=0$ plane. The vertical separation between the
respective radially inward flows of positive and negative particles means that
bipolar toroidal magnetic fields will be generated in the interstitial regions
between the electron flow and the drain-hole flow (see positive and negative
$B_{\phi}$ regions in Fig.\ref{f12.eps}).

\begin{figure}[ptb]
\caption{Flow of conventional electric current for drain hole particles and
their associated stranded electrons. The electric field on the $z=0$ plane is
radially outwards while the current flow is radially inwards so
$\mathbf{J\cdot E}$ is negative, indicating that the infall of the drain-hole
particles constitutes a dynamo. The $\mathbf{J\times B}$ force (which is
essentially due to the gradient of $B_{\phi}^{2}$ and which is strongest at
small $r$) drives a bipolar axial jet.}%
\label{f13.eps}%
\plotone{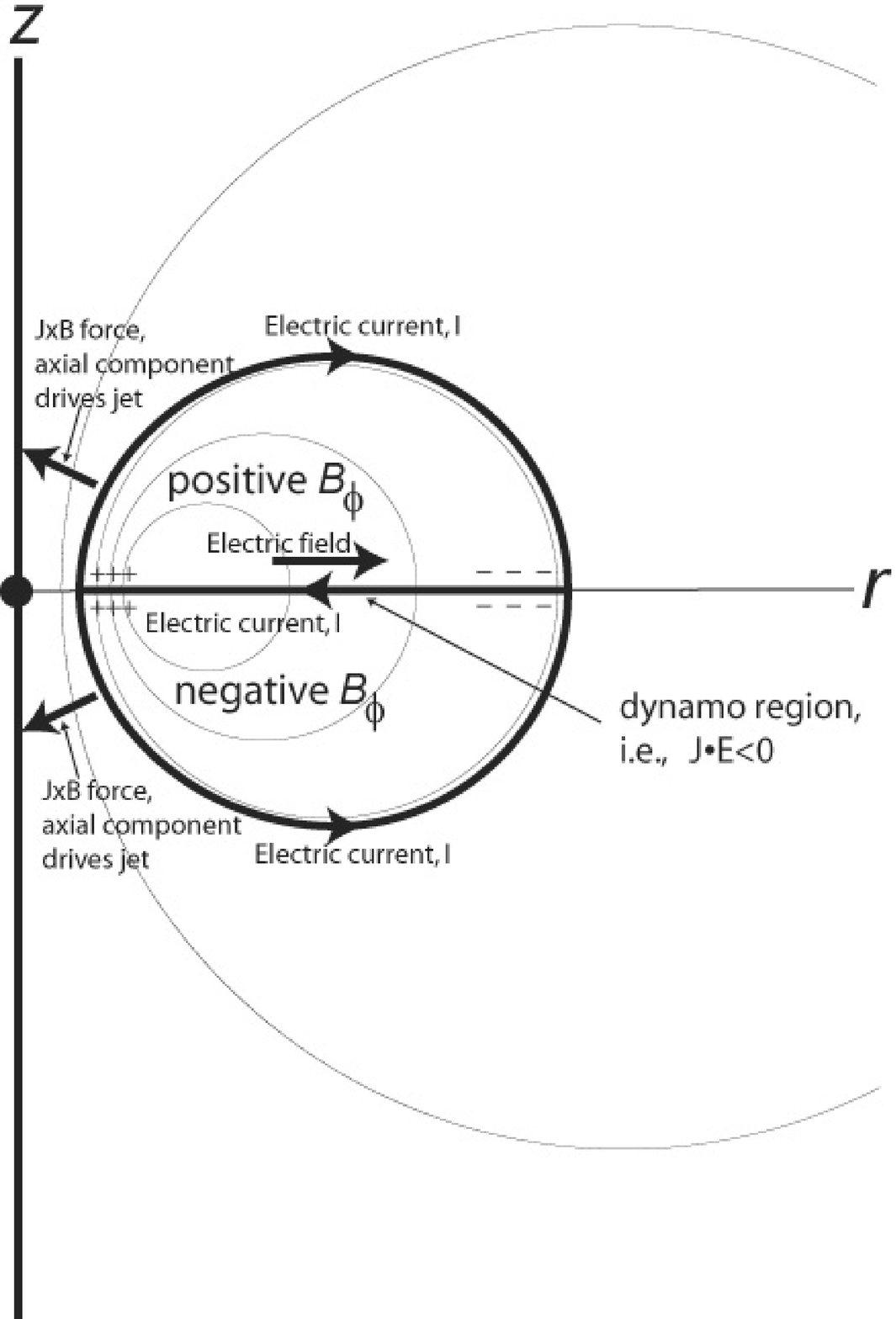}\end{figure}

If no electric current is allowed to flow, the situation is like a
free-standing battery not connected to any load, i.e., a situation where there
is a voltage differential across the battery terminals, but no current flows.
However, if bipolar axial currents are allowed to flow, then the situation is
like a battery connected to a load and the resulting radially inward
drain-hole particle current in the $\bar{z}=0$ plane is like the internal
current in a battery. The overall current flow pattern sketched in
Fig.\ref{f12.eps} results from a combination of drain hole particle and
electron motion. This pattern is sketched in Fig.\ref{f13.eps} as a
conventional electric current. The geometry of the current flow pattern and
electromotive force driving this current is identical to the geometry and flow
patterns in the laboratory configuration simulating astrophysical jets
described in
\citet{Hsu2002,Hsu2005}
and
\citet{Bellan2005}%
. The electric field due to the drain hole particles corresponds to the
electric field produced by the capacitor bank used in the laboratory
experiment. This geometry\ and symmetry is also identical to that proposed by
\citet{Lovelace1976}%
, the only difference being the means by which the radial electric field is
produced. The magnetic fields in the lab and astrophysical plasmas have the
same toroidal/poloidal topology.

The drain-hole current is\ thus powered by gravity and has $J_{r}$ radially
inward with $E_{r}$ radially outward so that $\mathbf{J\cdot E}$ is negative,
consistent with the condition for a dynamo. The drain-hole particles have
retrograde motion so their mechanical angular momentum is negative. This
negative mechanical angular momentum is removed by the braking torque
$\mathbf{r\times F=}$ $\left(  r\hat{r}\right)  \times\left(  J_{r}\hat
{r}\times B_{z}\hat{z}\right)  =-rJ_{r}B_{z}\hat{z}$ which is positive since
$J_{r}$ is negative and $B_{z}$ is positive.

The radially inward current is symmetric with respect to $z.$ This property
provides enough information to determine the symmetry properties of $I(r,z).$
The radially inward drain-hole current means that $J_{r}<0$ and $J_{z}=0$ in
the $z=0$ plane. Since Eq.\ref{Jpol2} shows that $J_{r}=-(2\pi r)^{-1}\partial
I/\partial z$ and $J_{z}=$ $(2\pi r)^{-1}\partial I/\partial r$ the condition
$J_{z}=0$ means $I(r,z)$ must vanish in the $z=0$ plane$.$ Furthermore $I$
must be an odd function of $z$ in order for $J_{r}=-(2\pi r)^{-1}\partial
I/\partial z$ to be finite in the $z=0$ plane. Finally $\partial I/\partial z$
should be positive in order to have $J_{r}<0.$ Thus, $I(r,z)$ should be
positive for $z>0$ and negative for $z<0$ so that, as sketched in
Fig.\ref{f13.eps}, there will be a bipolar axial current flowing along the $z$
axis outwards from the $z=0$ plane. The accumulation of drain-hole particles
constitutes the engine that drives the poloidal electric current that drives
the astrophysical jet. The $z$-symmetry of $\psi(r,z)$ and $z$-antisymmetry of
$I(r,z)$ has been noted previously by
\citet{Ferreira1995}%
.

Electromagnetic power flow from this dynamo can also be interpreted in terms
of the Poynting flux $\mathbf{S=E\times B}/\mu_{0}.$ Azimuthal symmetry
applied to Faraday's law shows that $E_{\phi}$ is zero for a steady-state
situation in which case the $z$-component of the Poynting flux reduces to
$S_{z}=E_{r}B_{\phi}/\mu_{0}.$ Because $E_{r}$ and $B_{\phi}$ are both
positive for $z>0$ whereas $E_{r}$ is positive while $B_{\phi}$ is negative
for $z<0$, it is seen that $S_{z}$ is positive for $z>0$ and negative for
$z<0.$ Thus, the Poynting flux associated with this dust-driven dynamo injects
energy into bipolar astrophysical jets flowing normally outward from the $z=0$ plane.

One can ask just how close to exactly zero $\bar{P}_{\phi}$ has to be in order
for a particle to behave as a drain-hole particle. Exact $\bar{P}_{\phi}=0$
would enable a particle to spiral down all the way to the center of the
central object, an obviously unrealistic situation because the particle would
vaporize as it approached the stellar surface. A more realistic question then
is how small does $\bar{P}_{\phi}$ have to be in order for a drain-hole
particle to fall to some specified normalized radius $\bar{R}_{small}\ \ $that
is much less than unity. $\bar{R}_{small}$ would presumably be of the order of
the radius at which the astrophysical jet starts and so would be of the order
of the thickness of the accretion disk or somewhat smaller. Since the
dimensionless form of Eq.\ref{H} is
\begin{equation}
\bar{H}=\ \frac{1}{2}(\bar{v}_{r}^{2}+\bar{v}_{z}^{2})\ +\chi(\bar{r},\bar{z})
\label{H TP}%
\end{equation}
where the effective potential is
\begin{equation}
\chi(\bar{r},\bar{z})=\frac{1}{2\bar{r}^{2}}\left(  \bar{P}_{\phi}%
\ \ -\ \frac{\left\langle \omega_{c\sigma}\right\rangle }{2\Omega_{0}}%
\bar{\psi}(\bar{r},\bar{z})\right)  ^{2}-\frac{1}{\sqrt{\bar{r}^{2}+\bar
{z}^{2}}} \label{chi TP}%
\end{equation}
and since $\bar{H}\simeq0$, the turning point for a drain-hole particle is
where $\chi(\bar{r},\bar{z})\simeq0.$ Because $\bar{\psi}(\bar{r},\bar
{z})\rightarrow0$ at small $\bar{r},$ the inner turning point will therefore
be where $\bar{P}_{\phi}^{2}=2\bar{r}^{2}/\sqrt{\bar{r}^{2}+\bar{z}^{2}}.$
Assuming that\ the inner turning point is at $\bar{r}\ \sim\bar{R}_{small}$
and $\tilde{z}\simeq0,$ the inner turning point is where $\bar{P}_{\phi}%
^{2}=2\bar{R}_{small}.$ A sufficient condition for assuming $\bar{P}_{\phi
}\simeq0$ is thus $\bar{P}_{\phi}^{2}<2\bar{R}_{small}$, i.e.,
\begin{equation}
-\sqrt{2\bar{R}_{small}}<\bar{L}\cos\theta+\frac{\left\langle \omega_{c\sigma
}\right\rangle }{2\Omega_{0}}\ \bar{\psi}(\bar{r}_{\ast},\bar{z}_{\ast}%
)<\sqrt{2\bar{R}_{small}} \label{Pphi crit}%
\end{equation}
and particles satisfying this condition will fall to a normalized radius
$\bar{R}<$ $\bar{R}_{small}.$ For given $\bar{L},$ $\left\langle
\omega_{c\sigma}\right\rangle /2\Omega_{0},$ and $\ \bar{\psi}(\bar{r}_{\ast
},\bar{z}_{\ast})$ this corresponds to a narrow range in $\theta$ centered
about the angle at which $\bar{P}_{\phi}$ equals zero $\ $exactly. Equation
\ref{Pphi crit} can be expressed as $\ \cos\left(  \theta+\Delta
\theta/2\right)  <\cos\theta<\cos\left(  \theta-\Delta\theta/2\right)  $ where%
\begin{equation}%
\begin{array}
[c]{ccc}%
\bar{L}\cos\left(  \theta+\Delta\theta/2\right)  & = & -\sqrt{2\bar{R}%
_{small}}-\frac{\left\langle \omega_{c\sigma}\right\rangle }{2\Omega_{0}%
}\ \bar{\psi}(\bar{r}_{\ast},\bar{z}_{\ast})\\
\bar{L}\cos\left(  \theta-\Delta\theta/2\right)  & = & \sqrt{2\bar{R}_{small}%
}-\frac{\left\langle \omega_{c\sigma}\right\rangle }{2\Omega_{0}}\ \bar{\psi
}(\bar{r}_{\ast},\bar{z}_{\ast}).
\end{array}
\label{def theta minmax}%
\end{equation}
Subtracting these two equations from each other shows that the range
$\Delta\theta$ for drain-hole particles to reach $\bar{R}_{small}$ is
\begin{equation}
\Delta\theta\simeq\frac{2\sqrt{2\bar{R}_{small}}}{\bar{L}\sin\theta}.
\label{delta theta}%
\end{equation}
The solid angle of incident particles lying between $\theta$ and
$\theta+\Delta\theta$ is $2\pi\sin\theta\Delta\theta$ and so the fraction
$f_{dh}$ of all incident particles with angular momentum $\bar{L}$ that become
drain-hole particles and fall to $\bar{R}<$ $\bar{R}_{small}$ is $\ $ $\ $
\begin{equation}
f_{dh}=\ \frac{2\pi\sin\theta\Delta\theta}{4\pi}\ =\frac{\ \sqrt{2\bar
{R}_{small}}}{\bar{L}\ }. \label{drain hole fraction}%
\end{equation}

\subsection{Drain hole dynamo power}

The strength of the equilibrium radial electric field produced by drain-hole
particles can be estimated as follows: Before any drain-hole particles
accumulate at small $\bar{r},$ there is no radial electric field, but as the
drain-hole particles accumulate, the radial outward electric field will
develop. The force due to the radial outward electric field will oppose the
gravitational and magnetic forces causing the inward motion of the drain-hole
particles. The balance between these opposing forces is quantified by the
radial equation of motion. In cylindrical un-normalized coordinates the radial
equation of motion governing drain-hole particles with $v_{z}=0$ in the $z=0$
plane (where $B_{\phi}=0$ due to $z$-antisymmetry of $I(r,z)$) is
\begin{equation}
m_{d}\left(  \ddot{r}-r\dot{\phi}^{2}\right)  =q_{d}\left(  E_{r}+r\dot{\phi
}B_{z}\right)  -\frac{m_{d}MG}{r^{2}}.\label{drain-hole motion}%
\end{equation}
$B_{z}$ is approximately uniform at small $r$ so $\psi\simeq\pi r^{2}B_{z}%
\ $at small $r$ in which case the drain-hole particle condition $P_{\phi
}=m_{d}r^{2}\dot{\phi}+q_{d}\psi/2\pi=0$ implies $\ \dot{\phi}=-q_{d}%
B_{z}/2m_{d}.\ $On eliminating $\dot{\phi}\ $in Eq.\ref{drain-hole motion},
the radial equation of motion governing drain-hole particles is%
\begin{equation}%
\begin{array}
[c]{ccl}%
\ \ddot{r} & = & \frac{q_{d}}{m_{d}}E_{r}-r\frac{\ q_{d}^{2}B_{z}^{2}}%
{4m_{d}^{2}\ }\ -\frac{MG}{r^{2}}\\
& \simeq & \frac{q_{d}}{m_{d}}E_{r}\ \ -\frac{MG}{r^{2}}%
\end{array}
\label{drain-hole motion2}%
\end{equation}
where the second line is for small $r.$ When $E_{r}=0$, the drain-hole
particles fall inwards with gravitational acceleration, but as $E_{r}$ builds
up because of accumulation at small $r$ of fallen-in positively charged
drain-hole particles, Eq.\ref{drain-hole motion2} shows that this electric
field will oppose the gravitational force and retard the infall. The
drain-hole particles will continue to fall in and accumulate, thereby
increasing $E_{r}$ until the radially outward repulsive electrostatic force
due to $E_{r}$ becomes so strong as to balance gravity and cause $\ddot{r}$ to
vanish. Thus, gravitational force is balanced by the radially outward force
from the space charge electric field of the accumulated positively charged
drain-hole particles. The saturation electric field is
\begin{equation}
\ E_{r}\ \ =\frac{m_{d}MG}{q_{d}r^{2}}.\label{drain hole Er}%
\end{equation}

Using $E_{r}=-\partial V/\partial r,$ integration of Eq.\ref{drain hole Er}
from large $r$ to the jet radius $r_{jet}$ gives the voltage at the jet to be%
\begin{equation}
V_{jet}=\frac{m_{d}MG}{q_{d}r_{jet}}. \label{Vjet}%
\end{equation}

The jet electric current corresponds to the charge per second carried inward
by the drain-hole particles. The number of drain-hole particles accreting per
second is $\dot{M}_{dh}/m_{d}$ where $\dot{M}_{dh}$ is the mass accretion rate
per second of drain-hole particles and $m_{d}$ is the mass of an individual
drain-hole particle. Thus, the poloidal electric current is
\begin{equation}
I_{jet}=q_{d}\dot{M}_{dh}/m_{d}. \label{Ijet}%
\end{equation}
The jet electric power is
\begin{equation}
P_{jet}=I_{jet}V_{jet}=\frac{\ \dot{M}_{dh}MG}{\ r_{jet}}\ \ \ \label{Pjet}%
\end{equation}
which is just the rate at which gravitational potential energy is released by
drain-hole particles falling from large $r$ to the jet radius. The drain-hole
dynamo converts the gravitational energy released from accretion into
electrical power suitable for driving bipolar jets that are moving away from
the $z=0$ plane. The jet power is proportional to both the central object mass
$M$ and the drain-hole mass accretion rate $\dot{M}_{dh}$. Paper I showed that
the dust mass accretion rate can be a substantial fraction of the total mass
accretion so $P_{jet}$ can be a substantial fraction of the power of all
accreting material. The jet power accelerates the jet material to escape
velocity and so is equal to the power available from accreting drain-hole dust
grains. Thus, assuming that the axial starting point for jet particles is of
the order of $r_{jet},$ the power required to drive the jet particles to
escape velocity is $P_{jet}=\dot{M}_{jet}MG/r_{jet}$ and so the jet mass
outflow would be approximately equivalent to the drain hole particle accretion
rate, i.e., $\dot{M}_{jet}\simeq\dot{M}_{dh}.$ The particles in the jet would
not, in general, be the drain hole particles, but instead would be plasma
magnetohydrodynamically accelerated using the drain-hole accretion as the
power source. Assuming $\bar{L}\simeq1\ $ and $\bar{R}_{small}=r_{jet}%
/a\sim0.1$ in Eq.\ref{drain hole fraction}, the fraction of retrograde
particles that are drain hole and able to reach $r_{jet}$ would be
$f_{dh}=\ \allowbreak0.4;$ the fraction of combined retrograde and prograde
dust grains would thus be 0.2. The example in paper I\ showed that because of
differences in proportional slowing down, the dust accretion rate would be
enriched to be 20\% of the gas accretion rate. This gives $\dot{M}_{dh}%
/\dot{M}_{g}\sim0.2\times0.2=0.04$ and so predicts a jet power that would be
about 1/25 of the power associated with all accreting dust and gas. This ratio
of outflow power to accretion power is consistent with the estimate given by
\citet{Bacciotti2004}
using HST observations of T Tauri jets.

\section{Torque and angular momentum}

An important question repeatedly asked about accretion disks and jets is the
role played by jets in satisfying conservation of \ mechanical angular
momentum of the accreting material. It will now be shown that mechanical
angular momentum is exactly conserved in our model.

Because of axisymmetry, the canonical angular momentum of the $j^{th}$
$\ $charged dust grain
\begin{equation}
P_{\phi d,j}^{+}=m_{+}rv_{\phi}^{j}+Ze\psi(r,z)/2\pi\ \label{canmomentum}%
\end{equation}
and the canonical angular momentum of the $k^{th}$ electron
\begin{equation}
P_{\phi,k}^{e}=m_{e}rv_{\phi}^{k}-e\psi(r,z)/2\pi\label{e can momentum}%
\end{equation}
at any position $r,z$ are both invariants, i.e., $P_{\phi d}^{+}=const.$ and
$P_{\phi}^{e}=const.$ In general, the dust grains and the electrons at any
position $r$,$z$ will have quite different values of $v_{\phi}\ $but, in order
for the plasma to be macroscopically quasi-neutral, there must be
approximately $Z$ electrons adjacent to each dust grain$.$ Because electron
and dust grain trajectories differ, these neighboring electrons will typically
not be the original sibling electrons photo-emitted when the dust grain became charged.

The radial and axial velocities of a specific dust grain or electron can be
written as $v_{r}^{\sigma,j}\ =dr^{\sigma,j}/dt$ and $\ v_{z}^{\sigma
,j}\ =dz^{\sigma,j}/dt$ where $r^{\sigma,j}(t)\ $ and $z^{\sigma,j}(t)$ are
the position of the $j^{th}$ particle of species $\sigma$. Since $P_{\phi}$ is
conserved for each individual particle, the time derivatives of the $P_{\phi}%
$'s of a dust grain at a location $r,z$ and its neighboring neutralizing $Z$
electrons respectively give $dP_{\phi d,j}^{+}/dt=0$ and $dP_{\phi,k}%
^{e}/dt=0.$ Using $d\psi/dt=v_{r}\partial\psi/\partial r+v_{z}\partial
\psi/\partial z$ for the time derivative of $\psi$ measured in the particle
frame, respective time derivatives of Eqs.\ref{canmomentum} and
\ref{e can momentum} give
\begin{equation}
\frac{d}{dt}\left(  m_{+}rv_{\phi}^{+,j}\right)  =\ -\frac{Ze}{2\pi}\left(
\frac{\partial\psi}{\partial r}v_{r}^{+,j}+\frac{\partial\psi}{\partial
z}v_{z}^{,j}\right)  \label{ang momentum dust deriv}%
\end{equation}
and
\begin{equation}
\frac{d}{dt}\left(  Zm_{e}rv_{\phi}^{e,k}\right)  \ =\frac{Ze}{2\pi}\left(
\frac{\partial\psi}{\partial r}v_{r}^{e,k}+\frac{\partial\psi}{\partial
z}v_{z}^{e,k}\right)  . \label{ang momentum electron deriv}%
\end{equation}

Using $B_{r}=-(2\pi r)^{-1}\partial\psi/\partial z$ and $B_{z}=(2\pi
r)^{-1}\partial\psi/\partial r$ from Eq.\ref{Bpol def} and summing
Eqs.\ref{ang momentum dust deriv} and \ref{ang momentum electron deriv} over
the dust grains and their associated $Z$ neutralizing electrons at location
$r,z$ gives
\begin{equation}
\frac{dL_{\phi}}{dt}=r\left(  J_{z}B_{r}-J_{r}B_{z}\right)  =\ \ r\hat{\phi
}\cdot\mathbf{J}_{pol}\mathbf{\times B}_{pol} \label{MHD ang momentum}%
\end{equation}
where $J_{r}$, $J_{z}$ are the respective radial and axial current densities
and $L_{\phi}$ is the total mechanical angular momentum density taking into
account both dust grains and electrons. Thus, from the macroscopic point of
view there is a torque about the $z$ axis, namely $\hat{z}\cdot\mathbf{r\times
F}=\hat{z}\times\left(  r\hat{r}+z\hat{z}\right)  \mathbf{\cdot}\left(
\mathbf{J}_{pol}\times\mathbf{B}_{pol}\right)  =r\hat{\phi}\cdot
\mathbf{J}_{pol}\mathbf{\times B}_{pol}$ acting to change the local mechanical
angular momentum density.

On the other hand, using Eqs.\ref{Bpol def} and \ref{Jpol2} it is seen that
when this torque is integrated over the entire volume to infinity,
\begin{equation}%
\begin{array}
[c]{ccl}%
\int d^{3}r\,r\hat{\phi}\cdot\mathbf{J}_{pol}\times\mathbf{B}_{pol}\  & = &
\int\ d^{3}r\,r^{2}\nabla\phi\cdot\left(  \frac{1}{2\pi}\nabla I\times
\nabla\phi\right)  \times\left(  \frac{1}{2\pi}\nabla\psi\times\nabla
\phi\right) \\
& = & \frac{1}{4\pi^{2}}\int\ d^{3}r\,r^{2}\nabla\phi\times\left(  \ \nabla
I\times\nabla\phi\right)  \cdot\left(  \ \nabla\psi\times\nabla\phi\right) \\
& = & \frac{1}{4\pi^{2}}\int\ d^{3}r\,\nabla I\cdot\left(  \nabla\psi
\times\nabla\phi\right) \\
& = & \frac{1}{4\pi^{2}}\int\ d^{3}r\,\nabla\cdot\left(  I\left(  \nabla
\psi\times\nabla\phi\right)  \right) \\
& = & 0
\end{array}
\label{integrate torque}%
\end{equation}
since both $I$ and $\nabla\psi$ vanish at $\infty.$ Thus, the total mechanical
angular momentum of the system is exactly \ conserved because there is no net
torque applied to the whole system.

$I$ is an odd function of $z$ and $\psi$ is an even function of $z,$ and in
the jet $\mathbf{J}_{pol}$ is nearly parallel to $\mathbf{B}_{pol}.$ This
suggests the following generic form for the current
\begin{equation}
\mu_{0}I(r,z)\simeq\lambda\psi(r,z)\tanh\left(  \frac{z}{h(r)}\right)
\label{generic current}%
\end{equation}
where $h(r)$ represents the height of the accretion disk at radius $r\ $and
$\lambda,$ the current per flux, has units of inverse length. Thus far from
the $z=0$ plane, Eq.\ref{generic current} has the form $\mu_{0}I/\psi=\lambda
$sign$(z)$ so that the jet above the $z=0$ plane has the opposite handedness
of the jet below the $z=0$ plane. The parameter $\lambda$ is closely related
to the current per flux in a force-free system (i.e., a system satisfying
$\ \nabla\times\mathbf{B=}\lambda\mathbf{B}$), but differs slightly because
here $\lambda$ refers to just the ratio of the poloidal current to the
poloidal flux. Using Eq.\ref{MHD ang momentum} it is seen that the density of
MHD\ torque about the $z$ axis is of the generic form%
\begin{equation}%
\begin{array}
[c]{ccl}%
\frac{dL_{\phi}}{dt} & = & \frac{1}{4\pi^{2}r\ }\left(  -\frac{\partial
I}{\partial r}\frac{\partial\psi}{\partial z}+\frac{\partial I}{\partial
z}\frac{\partial\psi}{\partial r}\right) \\
& = & \frac{\lambda}{4\mu_{0}\pi^{2}r}\left\{  -\frac{\partial\ }{\partial
r}\left[  \psi(r,z)\tanh\left(  \frac{z}{h(r)}\right)  \right]  \frac
{\partial\psi}{\partial z}+\frac{\partial\ }{\partial z}\left[  \psi
(r,z)\tanh\left(  \frac{z}{h(r)}\right)  \right]  \frac{\partial\psi}{\partial
r}\right\}  .
\end{array}
\label{dLphitdt}%
\end{equation}
We assume that $\partial h/\partial r\ll1$ so the radial scale length at which
$h$ changes is much larger than $h.$ Also, from symmetry $\partial
\psi/\partial z=0$ on the $z=0$ midplane. Together, these conditions imply
that near the midplane the last term in Eq.\ref{dLphitdt} dominates so near
the midplane
\begin{equation}
\frac{dL_{\phi}}{dt}\simeq\frac{\lambda}{4\mu_{0}\pi^{2}hr}\frac{\psi
(r,z)}{\cosh^{2}(z/h)}\frac{\partial\psi}{\partial r}=\frac{\lambda\psi
(r,z)}{2\pi\mu_{0}h}\frac{B_{z}(r,z)}{\cosh^{2}(z/h)}. \label{approx dLphidt}%
\end{equation}
As seen from Eq.\ref{integrate torque} the torque density is proportional to
$\ -\nabla I\times\nabla\phi\cdot\nabla\psi\sim-r\mathbf{J}_{pol}\cdot
\nabla\psi$ and so is positive for poloidal current flow away from the
poloidal field magnetic axis and negative for poloidal current flow towards
the poloidal field magnetic axis$.$The \ torque density vanishes as
$r\rightarrow0$ and as $r\rightarrow\infty$ and at the poloidal field magnetic
axis because $r\mathbf{\ }\nabla\psi\rightarrow0$ at $r=0,$ $\mathbf{J}%
_{pol}\rightarrow0$ at $r=\infty$, and $\nabla\psi=0$ at the poloidal field
magnetic axis. The direction of poloidal current flow is shown in
Fig.\ref{f13.eps}. This \ torque acts on the drain hole particles and their
associated electrons since these particles are the carriers of the poloidal
current as sketched in Fig.\ref{f12.eps} and Fig.\ref{f13.eps}. Unlike the
drain-hole particles, no \ torque $r\hat{\phi}\cdot\mathbf{J\times B}$ about
the $z$ axis acts on the Speiser particles because the current associated with
the Speiser particles is in the $\phi$ direction.

\section{Conclusions}

We have shown that charging of collisionless dust grains incident upon a star
causes the dust grain orbital dynamics to change from a relatively simple
Kepler form to more complicated motion involving competition between magnetic
and gravitational forces. This competition gives rise to five qualitatively
different types of orbits. Two of these, the retrograde and prograde cometary
orbits are just perturbations of Kepler cometary orbits. The orbit of a
particle where magnetic forces overwhelm gravitational forces is just a Larmor
(cyclotron) orbit and in this case the particle is constrained to remain
within a poloidal Larmor radius of a poloidal flux surface in a manner similar
to tokamak confinement. Particles where magnetic and gravitational forces are
comparable can have two very different types of orbit depending on whether the
incident particle is prograde or retrograde. Prograde particles of this latter
type develop Speiser orbits; these orbits are paramagnetic with respect to the
poloidal magnetic field and so can be the source of the poloidal magnetic
field. Retrograde particles having comparable magnetic and gravitational
forces can have a peculiar behavior whereby centrifugal force is eliminated
with the result that the charged particle falls in towards the star along a
spiral orbit. The accumulation of these \textquotedblleft
drain-hole\textquotedblright\ particles near the star provides a radial
electric field oriented so as to drive the poloidal currents and toroidal
magnetic fields of an astrophysical jet.

This paper showed the existence of these different types of orbits, how their
orientation is suitable for generating the poloidal and toroidal magnetic
fields associated with an accretion disk and astrophysical jet, and how
questions of angular momentum conservation are inherently resolved. A future
paper will investigate the quantitative values of dust grain parameters
required to produce toroidal and poloidal fields in the accretion disk of a
young stellar object.

Finally, we offer some remarks regarding the effect of deviations from axisymmetry. The model presented here assumed perfect magnetic field axisymmetry 
field whereas actual accretion disks are observed to have varying amounts of non-axisymmetry. This situation is  analogous
to toroidal magnetic fusion devices such as tokamaks, reversed field pinches, and spheromaks all of which are modeled to first
approximation as being axisymmetric, but in reality have  deviations from axisymmetry due to waves, turbulence, instability, and errors in machine construction.
It is known from these devices that a modest breaking of symmetry  does not invalidate the results of the axisymmetric model, but rather weakens the
conclusions, e.g., instead of cyclotron-orbiting particles being perfectly confined to the vicinity of a poloidal flux surface, when there is deviation from
axisymmetry  cyclotron-orbiting particles can slowly wander away from  the  poloidal flux surface they started on.
One would expect that deviations from axisymmetry in accretion disks  would cause a similar transport of cyclotron particles across poloidal flux surfaces.
Because symmetry breaking  causes the canonical angular momentum of particles to change, it could be 
considered as being somewhat like a collision that changes the canonical angular momentum of each of two  particles involved in a collision
while conserving the total canonical angular momenta. Hence,  deviations from axisymmetry
would  cause a   jiggling   of the canonical angular momenta of individual particles so that particles  on
the borderline between being drain-hole and cyclotron or on the borderline between being Speiser and cyclotron might spend part of the  time (i.e., between jiggles) 
being one type and part of the time  being the neighboring type. Similarly cyclotron particles that are on the borderline between being
mirror-trapped and not mirror-trapped would, as they get kicked into and out of the mirror loss-cone,  spend part of the  time being mirror-trapped and 
part of the  time not being mirror-trapped. However, at any given time there would
be a certain fraction of particles of each type, i.e., a certain fraction would be cyclotron, a certain fraction would be drain-hole,   
a certain fraction would be
Speiser, and a certain fraction would be  cometary.
  
\pagebreak

\appendix

\begin{center}
\textbf{APPENDICES}
\end{center}

\section{Derivation of generic poloidal flux
function\label{Generic Flux Function}}

If all the toroidal current $\mathcal{I}_{\phi}$ is concentrated at the
poloidal location $r=R_{0}$ and $z=0$, then the toroidal current density is
$\mathbf{J}_{tor}=\ \hat{\phi}\mathcal{I}_{\phi}\delta(z)\delta(r-R_{0}).$ On defining%

\begin{equation}
k^{2}=\frac{4R_{0}r}{\left(  R_{0}+r\right)  ^{2}+z^{2}} \label{k2 def}%
\end{equation}
analytic solution of Eq.\ref{Jtor} using $\mathbf{J}_{tor}=\ \hat{\phi
}\mathcal{I}_{\phi}\delta(z)\delta(r-R_{0})$ gives
\citep{Jackson1999}
\begin{equation}
\psi(r,z)=\ \ \frac{\mu_{0}\mathcal{I}_{\phi}}{k\ }\sqrt{R_{0}r}\left[
\left(  2-k^{2}\ \right)  K(k)-2E(k)\right]  \ \label{elliptic}%
\end{equation}
where $E$ and $K$ are complete elliptic integrals. Equation \ref{elliptic}
describes the situation where all the current density is concentrated at
$r=R_{0},$ $z=0$, i.e., the current flows in a wire of zero cross-section
located at $r=R_{0},$ $z=0.$ This equation can also be used to (i)\ describe
the field observed at locations far from the poloidal field magnetic axis of a
distributed current localized in the vicinity of the poloidal field magnetic
axis and (ii)\ as the Green's function for a distributed toroidal current.
This is because for an observer who is far from $r=R_{0},$ $z=0,$ the field of
a distributed toroidal current localized near $r=R_{0},$ $z=0$ is
indistinguishable from the field of a zero cross-section wire carrying the
same total current. Equation \ref{elliptic} has a logarithmic singularity at
the wire location because the wire has infinitesimal diameter.

Two analytic limits are of interest for Eq.\ref{elliptic}. The first is where
$r\ll R_{0}$ so
\begin{equation}
k^{2}\simeq\frac{4R_{0}r}{R_{0}^{2}+z^{2}} \label{k2 approx 1}%
\end{equation}
and the second is where $r\gg R_{0}$ so
\begin{equation}
k^{2}\simeq\frac{4R_{0}r}{r^{2}+z^{2}}. \label{k2 approx 2}%
\end{equation}
The former gives the field near the loop axis and the latter gives the field
at locations far from the current loop. In both cases $k^{2}$ is small
compared to unity and so the small argument asymptotic expansions of the
complete elliptic integrals can be used, namely,%
\begin{equation}
E(k)=\frac{\pi}{2}\left(  1-\frac{k^{2}}{4}\ -\frac{3}{64}k^{4}-...\right)
,\qquad K(k)=\frac{\pi}{2}\left(  1+\frac{k^{2}}{4}+\frac{9}{64}%
k^{4}+...\right)  . \label{elliptic asymptotic}%
\end{equation}
Thus for small $k,$ it is seen that $\left(  2-k^{2}\right)  K(k)-2E(k)\simeq
\pi k^{4}/16$ in which case$\ $%
\begin{equation}
\psi(r,z)=\ \frac{\pi\mu_{0}\mathcal{I}_{\phi}}{2\ \ }\frac{R_{0}^{2}r^{2}%
}{\left(  \left(  R_{0}+r\right)  ^{2}+z^{2}\right)  ^{3/2}}%
\ \label{psi approx}%
\end{equation}
so for $r\ll R_{0}$
\begin{equation}
\lim_{r\ll a}\psi(r,z)\simeq\frac{\pi\mu_{0}\mathcal{I}_{\phi}}{2\ \ }%
\frac{R_{0}^{2}r^{2}}{\left(  R_{0}^{2}+z^{2}\right)  ^{3/2}}
\label{small r psi}%
\end{equation}
and for $r\gg R_{0}$%
\begin{equation}
\lim_{r\gg a}\psi(r,z)\simeq\frac{\pi\mu_{0}\mathcal{I}_{\phi}}{2\ \ }%
\frac{R_{0}^{2}r^{2}}{\left(  r^{2}+z^{2}\right)  ^{3/2}}. \label{large r psi}%
\end{equation}
For purposes of discussion and also numerical computation, it is convenient to
choose Eq.\ref{psi approx} to represent the poloidal flux of a generic
toroidal current \textit{everywhere}. Making this choice for the poloidal flux
function (instead of the prescription given by Eq.\ref{elliptic}) means that
$\psi(r,z)$ has a smooth hill-top at $r=2R_{0}$ rather than a logarithmic
singularity at $r=R_{0}$ and has the same behavior far from $r=R_{0},$ $z=0$
as does Eq.\ref{elliptic}.

Thus, a useful analytic representation for a nonsingular, physically
realizable flux function is obtained by recasting Eq.\ref{psi approx} \ in the
form
\begin{equation}
\psi(r,z)=\ \ \frac{27\left(  r/a\right)  ^{2}}{8\left(  \left(  \frac{r}%
{a}+\frac{1}{2}\right)  ^{2}+\left(  \frac{z}{a}\right)  ^{2}\right)  ^{3/2}%
}\psi_{0}. \label{realistic psi}%
\end{equation}
This has a maximum of $\psi_{0}$ at $r=a,$ scales as $r^{2}$ for small $r,$
and scales as $r^{-1}$ for large $r.$ Equation \ref{Jtor} can be used to
calculate the associated toroidal current density which will be sharply peaked
near $r=a$ and $z=0.$ The $\psi(r,z)$ prescribed by Eq.\ref{realistic psi} has
the features that it provides a dipole-like field far from the $z$ axis and a
nearly uniform axial field near the $z$ axis, corresponds to a realistic
distributed toroidal current, has a well-defined poloidal field magnetic axis,
is analytically tractable, and is convenient for numerical computation of
representative particle orbits in a physically relevant magneto-gravitational field.

\section{Current associated with flux function
\label{current associated with flux function}}

Using Ampere's law to relate the toroidal current and the poloidal magnetic
field it is seen that
\begin{equation}
\mathcal{I}_{\phi}=\frac{1}{\mu_{0}}\oint_{C}\mathbf{B}_{pol}\mathbf{\cdot
}d\mathbf{l} \label{Amperes law}%
\end{equation}
where the contour $C$ links the total toroidal current $\mathcal{I}_{\phi}$.
By letting the line integral go to infinity in the radial and $z$ directions
it is seen that only the portion of the line integral along the $z$ axis makes
a finite contribution so%
\begin{equation}%
\begin{array}
[c]{ccl}%
\mathcal{I}_{\phi} & = & \ \frac{1}{\mu_{0}}\int_{-\infty}^{\infty}%
B_{z}(0,z)dz\\
& = & \frac{27\psi_{0}}{16\pi\mu_{0}}\lim_{r\rightarrow0}\frac{1}{r}%
\frac{\partial\ }{\partial r}\left(  \frac{r^{2}}{a^{2}}\int_{-\infty}%
^{\infty}\ \frac{dz\ }{\ \left(  \left(  \frac{r}{a}+\frac{1}{2}\right)
^{2}+\left(  \frac{z}{a}\right)  ^{2}\right)  ^{3/2}}\right)  \ .
\end{array}
\label{Iphi 1}%
\end{equation}

Defining $b=r/a+1/2$ and $z/a=b\sinh\vartheta$ the $z$ integral can be
expressed as%
\begin{equation}%
\begin{array}
[c]{ccl}%
\int_{-\infty}^{\infty}\ \frac{dz\ }{\ \left(  \left(  \frac{r}{a}+\frac{1}%
{2}\right)  ^{2}+\left(  \frac{z}{a}\right)  ^{2}\right)  ^{3/2}} & = &
a\int_{-\infty}^{\infty}\ \frac{b\cosh\vartheta\,d\vartheta}{\ \left(
b^{2}+b^{2}\sinh^{2}\vartheta\right)  ^{3/2}}\\
& = & \frac{a}{b^{2}}\left[  \tanh\vartheta\right]  _{-\infty}^{\infty}\\
& = & \frac{2a}{\left(  r/a+1/2\right)  ^{2}}.
\end{array}
\label{integral details}%
\end{equation}
Since
\begin{equation}
\lim_{r\rightarrow0}\frac{1}{r}\frac{\partial\ }{\partial r}\left[
\frac{r^{2}}{a^{2}}\frac{2a}{\left(  r/a+1/2\right)  ^{2}}\right]  =\frac
{16}{a} \label{lim integral}%
\end{equation}
the total toroidal current is
\begin{equation}
\mathcal{I}_{\phi}=\frac{27\psi_{0}}{\ \pi a\mu_{0}}.
\label{solve Iphi integral}%
\end{equation}

\section{Review:\ Distinction between diamagnetic (adiabatic) orbits and
paramagnetic (Speiser) orbits\label{Orbit review}}

\subsection{Diamagnetism of cyclotron (Larmor) orbits}

We first review charged particle motion in a uniform magnetic field
$\mathbf{B}=B_{z}\hat{z}$ (so $\psi=B_{z}\pi r^{2}$) and no electric field;
orbital motion in more complex fields will be discussed later. The particle
motion is prescribed by the Lorentz equation
\begin{equation}
m_{\sigma}\frac{d\mathbf{v}}{dt}=q_{\sigma}\mathbf{v}\times B_{z}\hat{z}\ .
\label{Lorentz}%
\end{equation}
If the particle is restricted to the $z=0$ plane, the respective radial and
azimuthal components of Eq.\ref{Lorentz} are%
\begin{equation}
m_{\sigma}\left(  \ddot{r}-r\dot{\phi}^{2}\right)  =q_{\sigma}r\dot{\phi}B_{z}
\label{Lorentz r}%
\end{equation}%
\begin{equation}
\frac{m_{\sigma}}{r}\frac{d}{dt}\left(  r^{2}\dot{\phi}\right)  =-q_{\sigma
}\dot{r}B_{z}. \label{Lorentz phi}%
\end{equation}
We consider circular motion (i.e., cyclotron or Larmor orbits) so $\ r=const.$
in which case Eq.\ref{Lorentz phi} gives $\dot{\phi}=const.$ and
Eq.\ref{Lorentz r} then becomes%
\begin{equation}
\dot{\phi}=-\omega_{c\sigma} \label{phidot diamag}%
\end{equation}
where $\omega_{c\sigma}=q_{\sigma}B_{z}/m_{\sigma}$ is the signed cyclotron
frequency. The minus sign in Eq.\ref{phidot diamag} indicates that cyclotron
motion is \textit{diamagnetic}. Thus if a gyrating charged particle is
considered as a $\phi$-directed current, the polarity of this current is such
as to create a magnetic field which opposes the initial field $B_{z}$, i.e.,
cyclotron orbits tend to depress the value of $\psi.$ The diamagnetism of
cyclotron orbits means that cyclotron orbits cannot be the source for the
assumed poloidal magnetic field $\psi(r,z)$ nor the means by which this field
is sustained against dissipation.

\subsection{Adiabatic orbits}

When the magnetic field is non-uniform or there are electric fields, and if
these additional features are sufficiently weak that to lowest order the
Larmor orbit (cyclotron orbit) description is approximately correct, then
additional charged particle motions occur which are superimposed on the Larmor
orbits $\mathbf{v}_{L}(t);$ these additional motions are adiabatic in the
sense of classical mechanics. Defining\ $v_{\parallel}$ as the velocity
component parallel to the magnetic field and $\mathbf{v}_{\perp}$ as the
component perpendicular to the magnetic field these motions are the standard
drifts
\citep{Longmire1967,Chen1984}%
, namely the $E\times B$ drift $\mathbf{v}_{E}=\mathbf{E\times B}/B^{2},$ the
polarization drift $\mathbf{v}_{p}=m_{\sigma}q_{\sigma}^{-1}B^{-2}%
d\mathbf{E}_{\perp}/dt$, the curvature drift $\mathbf{v}_{c}=-m_{\sigma
}v_{\parallel}^{2}\hat{B}\cdot\nabla\hat{B}\times\mathbf{B/}q_{\sigma}B^{2}$,
and the grad $B$ drift $\mathbf{v}_{\nabla B}=-\mu\nabla B\times
\mathbf{B/}q_{\sigma}B^{2}$ where $\mu=m_{\sigma}v_{\perp}^{2}/2B$ is the
magnetic moment, an adiabatic invariant. There is also a `force' drift
$\mathbf{v}_{F}$ $=\mathbf{F\times B}/q_{\sigma}B^{2}$ where $\mathbf{F}$ is a
generic non-electromagnetic force, which here is gravity, so $\mathbf{F}%
=mMG\nabla\left(  r^{2}+z^{2}\right)  ^{-1/2}.$ Taking into account all these
drifts, the velocity of an adiabatic-orbit charged particle becomes%
\begin{equation}%
\begin{array}
[c]{cc}%
\mathbf{v}= & v_{\parallel}\hat{B}+\mathbf{v}_{L\sigma}(t)\ +\frac
{\mathbf{E\times B}}{B^{2}}+\frac{m_{\sigma}}{q_{\sigma}B^{2}}\frac
{d\mathbf{E}_{\perp}}{dt}\ -\frac{m_{\sigma}v_{\parallel}^{2}\hat{B}%
\cdot\nabla\hat{B}\times\mathbf{B}}{q_{\sigma}B^{2}}\\
& -\frac{\mu\nabla B\times\mathbf{B}}{q_{\sigma}B^{2}}+\mathbf{\ }%
\ \frac{m_{\sigma}MG}{q_{\sigma}B^{2}}\nabla\left(  \frac{1}{\sqrt{r^{2}%
+z^{2}}}\right)  \times\mathbf{B.}%
\end{array}
\label{particle drifts}%
\end{equation}
The last four drifts in Eq.\ref{particle drifts} explicitly involve
$q_{\sigma}$ and thus produce macroscopic currents. When these currents are
summed and, in addition, diamagnetic current is taken into account, \ the
result is equivalent to the MHD\ equation of motion where the polarization
drift plays the role of the inertial term
\citep{Goldston1995,Bellan2006}%
. The ideal MHD\ concept of frozen-in flux is directly equivalent to $\mu$
conservation because $\mu$ conservation corresponds to conservation of the
magnetic flux linked by a cyclotron orbit. Thus, the ideal MHD\ concept of
frozen-in flux is based on the adiabatic invariance of cyclotron orbits.

\begin{figure}[ptb]
\caption{Orbit of a positive particle in the $z=0$ plane located inside the
poloidal field magnetic axis (indicated by dashed circle), coordinates are
normalized to the poloidal field magnetic axis radius. $B_{z}$ is positive
inside the circle and negative outside.}%
\label{Orbit-Inside-Poloidal-Field-Magnetic-Axis}%
\epsscale{1.0} \plotone{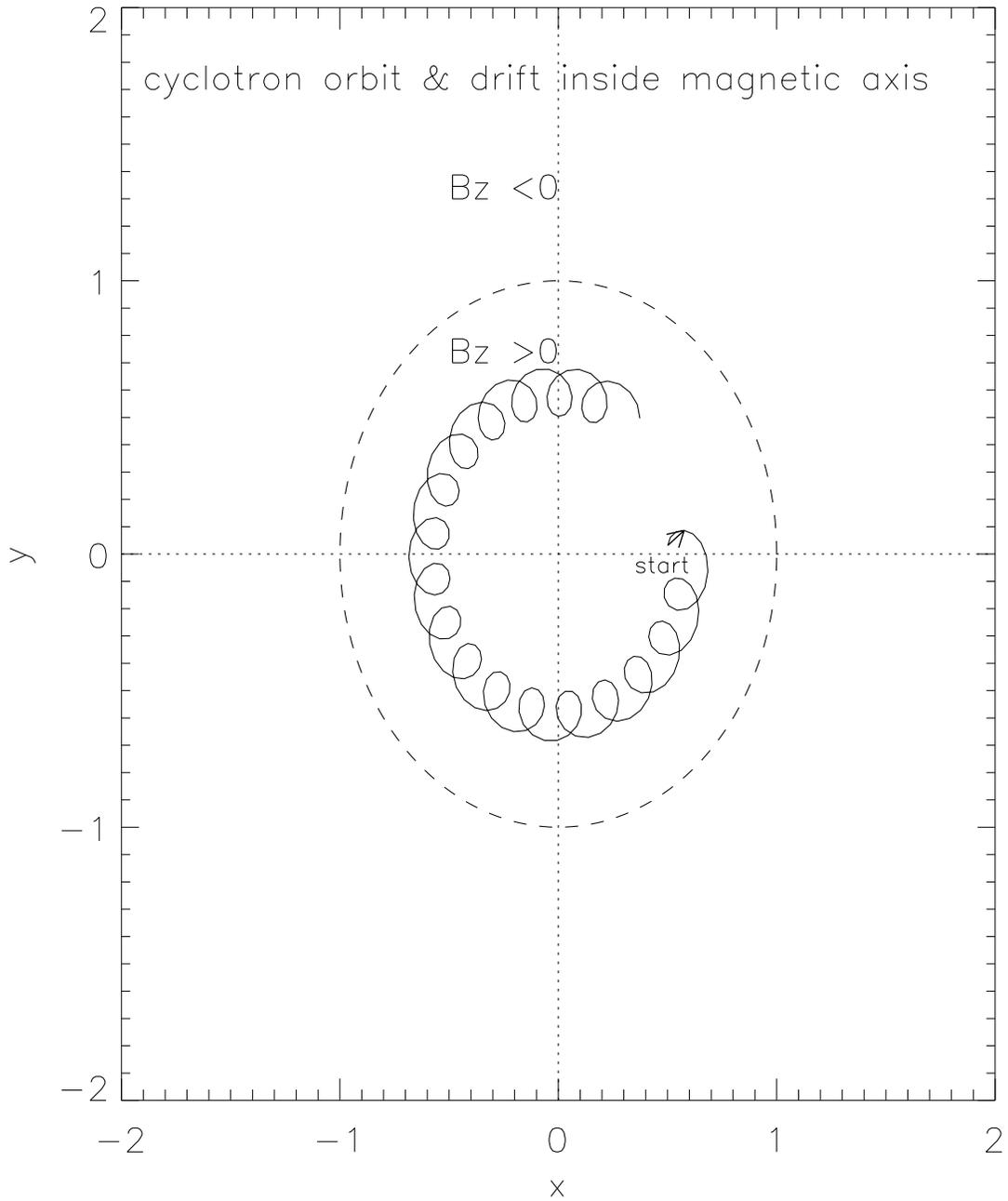}\end{figure}

\begin{figure}[ptb]
\caption{Orbit for a positively charged particle located in the $z$ plane
outside the poloidal field magnetic axis (dashed circle).}%
\label{Orbit-Outside-Poloidal-Field-Magnetic-Axis}%
\epsscale{1.0} \plotone{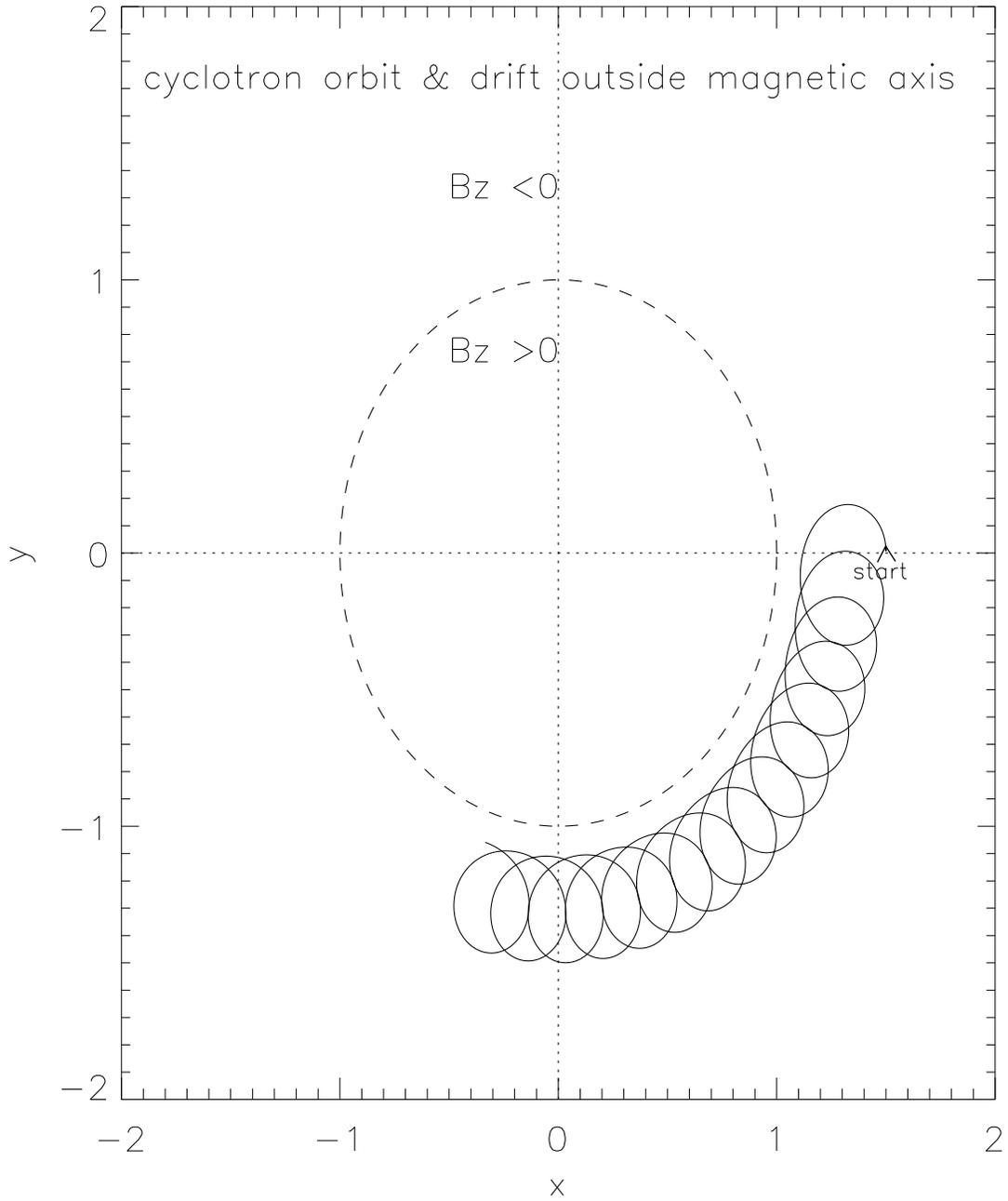}\end{figure}

The poloidal flux function specified by Eq.\ref{generic} corresponds to a
magnetic field generated by a toroidal current flowing in the positive $\phi$
direction (counterclockwise direction); the $B_{z}$ component of this field is
positive for $r<a$ and negative for $r>a$ where $a$ is the location of the
poloidal field magnetic axis. The poloidal magnetic field has both curvature
and gradients so that away from field nulls, particles should have parallel
motion and cyclotron orbits together with superimposed curvature and grad $B$
drifts. Figure \ref{Orbit-Inside-Poloidal-Field-Magnetic-Axis} shows the
numerically calculated orbit of a particle located in the $z=0$ plane in a
magnetic field prescribed by Eq.\ref{generic} and located inside the poloidal
field magnetic axis (indicated by dashed circle). It is seen that the particle
makes cyclotron orbits with a superimposed drift due to curvature and $\nabla
B$. The cyclotron orbit is clockwise consistent with the assertion that
cyclotron motion is diamagnetic. Figure
\ref{Orbit-Outside-Poloidal-Field-Magnetic-Axis} shows the situation for a
particle located at a radius outside the poloidal field magnetic axis. The
sense of the cyclotron orbit is now reversed as is the polarity of $B_{z}$ so
the cyclotron orbit is again diamagnetic. For both inside and outside
particles the drift motion is clockwise and so opposes the original toroidal
current creating the poloidal flux and so the curvature and $\nabla B$ drift
motion can also be considered diamagnetic.

The current associated with the gravitational force drift is%
\begin{equation}
\mathbf{J}_{g}=\sum_{\sigma}n_{\sigma}q_{\sigma}\mathbf{v}_{F}=\ \frac{\varrho
MG}{\ B^{2}}\nabla\left(  \frac{1}{\sqrt{r^{2}+z^{2}}}\right)  \times
\mathbf{B} \label{Jg}%
\end{equation}
where $\varrho=\sum m_{\sigma}n_{\sigma}$ is the mass density.

From a macroscopic (i.e., MHD) point of view, the force associated with the
gravitational drift current exactly balances the gravitational force component
perpendicular to the magnetic field since%

\begin{equation}%
\begin{array}
[c]{ccl}%
\mathbf{J}_{g}\times\mathbf{B} & = & \ \frac{\varrho MG}{\ B^{2}}\left(
\nabla\left(  \frac{1}{\sqrt{r^{2}+z^{2}}}\right)  \times\mathbf{B}\right)
\times\mathbf{B}\\
& = & -\varrho MG\nabla_{\perp}\left(  \frac{1}{\sqrt{r^{2}+z^{2}}}\right)
\end{array}
\label{grav force}%
\end{equation}
If $I=0$ on the $z=0$ plane (as is consistent with astrophysical jet symmetry
used by
\citet{Lovelace1976}%
), the gravitational drift is not defined on the poloidal field magnetic axis
because $\mathbf{B}$ vanishes on the poloidal field magnetic axis and the
theory of particle drifts fails. In other words, going from the first to the
second line in Eq.\ref{grav force} at the poloidal field magnetic axis would
involve dividing zero by zero (since $B=0$ on the poloidal field magnetic axis).

When summed over species, the curvature and grad $B$ drifts correspond to
currents which balance macroscopic pressure gradients (when diamagnetic
current is included) and the polarization current corresponds to the inertial
term in the MHD\ equation of motion. This analysis shows, as discussed in
\citet{Bellan2007}%
, that plasma particles undergoing cyclotron motion and drifts do not have
Keplerian orbits. It also shows that the poloidal field magnetic axis is a
special place where conventional particle drift theory fails.

\subsection{Non-adiabatic motion: the Speiser orbit\label{Speiser}}

An extreme form of magnetic non-uniformity occurs where the magnetic field
reverses direction. In this case an orbit quite distinct from the cyclotron
orbit and its associated adiabatic drifts occurs. This non-adiabatic orbit,
called a meandering or Speiser orbit
\citep{Speiser1965}%
, consists of semi-circles of counterclockwise motion interspersed with
semi-circles of clockwise motion.

\begin{figure}[ptb]
\caption{Speiser orbit. The charged particle bounces back and forth across the
field null at the poloidal field magnetic axis resulting in a counterclockwise
(i.e., paramagnetic) orbit.}%
\label{SpeiserOrbit}%
\epsscale{1.0} \plotone{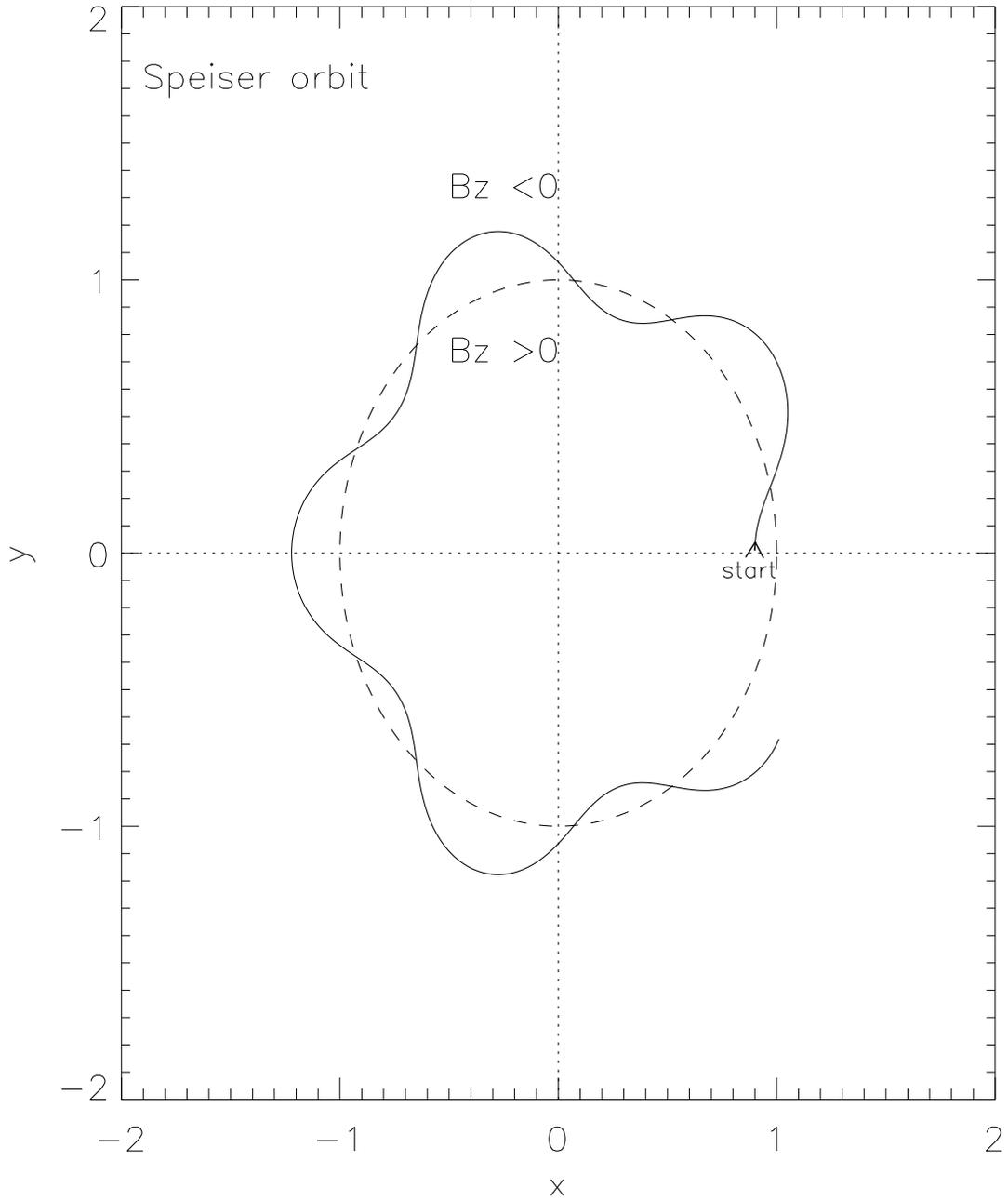}\end{figure}

A numerically calculated Speiser orbit for a positively charged particle in
the $z=0$ plane is shown in Fig.\ref{SpeiserOrbit}. The particle oscillates
across the poloidal field magnetic axis between the inside region where
$B_{z}>0$ and the outside region where $B_{z}<0.$ The result is a net
counterclockwise motion so, in contrast to cyclotron orbits, Speiser orbits
are \textit{paramagnetic.} The paramagnetism of Speiser orbits has been
considered an important aspect of current sheets in Earth's magnetotail,
[e.g., see
\citet{Zelenyi2000}%
], but to the author's knowledge this paramagnetism has not been previously
considered in the axisymmetric three-dimensional geometry discussed here which
is relevant to accretion disks and astrophysical jets. In particular, we will
show that the poloidal flux function can be considered as a consequence of
Speiser orbits such as shown in Fig.\ref{SpeiserOrbit}. \ Speiser orbits are
not consistent with the drift approximation (i.e., $E\times B$ drift, grad $B$
drift, curvature drift, etc.) \ because the drift approximation is based on
the assumption that, to lowest order, the particle is undergoing cyclotron
motion. The inconsistency between Speiser orbits and the drift approximation
is obvious when one considers that the drift approximation fails where $B$
reverses polarity whereas Speiser orbits depend on this reversal.

If motion in the $z$ direction is also allowed, then because $B_{r}$ also
reverses at the poloidal field magnetic axis, the particle can also oscillate
vertically across the poloidal field magnetic axis to make vertical Speiser
orbits. The combined $r$ and $z$ Speiser motion means that particles moving at
an arbitrary angle across the poloidal field magnetic axis will reflect from
interior surfaces of the nested poloidal flux surfaces concentric with the
poloidal field magnetic axis. These nested poloidal flux surfaces can thus be
imagined as the walls of a toroidal tunnel and the Speiser orbit particles can
be considered as reflecting from the interior walls of this toroidal tunnel
while moving in the counterclockwise direction to trace out paramagnetic
orbits and create poloidal flux.

\section{Equation of motion and its solutions\label{Equation of motion}}

The Hamiltonian orbit analysis presented here shows that photo-emission
creates new effective potential barriers. The topography of these barriers
depends on a combination of environmental factors, particle properties, and
the location of the charging. Representative orbits obtained by numerically
integrating the equation of motion$\ $have been presented and are consistent
with the predictions of the Hamiltonian theory. We outline here the derivation
of the dimensionless equation of motion; this derivation gives insights into
several fundamental issues regarding the dynamics, especially the influence of
initial conditions.

The equation of motion for a charged particle in a combined electromagnetic
and gravitational field is
\begin{equation}
m_{\sigma}\frac{d^{2}\mathbf{x}}{dt^{2}}=q_{\sigma}\left(  \mathbf{E+v\times
B}\right)  +m_{\sigma}MG\nabla\frac{1}{\sqrt{r^{2}+z^{2}}}.
\label{motion em gravity}%
\end{equation}

Using Eq.\ref{B general2} for the magnetic field, the equation of motion can
thus be written as%
\begin{equation}%
\begin{array}
[c]{cl}%
m_{\sigma}\frac{d^{2}\mathbf{x}}{dt^{2}}= & q_{\sigma}\mathbf{E}\\
& \mathbf{+}\frac{q_{\sigma}}{2\pi}\frac{d\mathbf{x}}{dt}\times\left(
\frac{\partial\psi}{\partial\mathbf{x}}\times\frac{\partial\phi}%
{\partial\mathbf{x}}+\mu_{0}I\frac{\partial\phi}{\partial\mathbf{x}}\right) \\
& +m_{\sigma}MG\nabla\left(  \frac{1}{\sqrt{r^{2}+z^{2}}}\right)  .
\end{array}
\label{eq motion}%
\end{equation}
Then, using the definitions given in Eq.\ref{norm quantities}, the equation of
motion can be expressed in dimensionless form as%
\begin{equation}
\frac{d^{2}\mathbf{\bar{x}}}{d\tau^{2}}=\ \mathbf{\bar{E}}+\frac{\left\langle
\omega_{c\sigma}\right\rangle }{2\Omega_{0}}\frac{d\mathbf{\bar{x}}}{d\tau
}\times\left(  \ \frac{\partial\bar{\psi}}{\ \partial\mathbf{\bar{x}}}%
\times\frac{\partial\phi}{\ \partial\mathbf{\bar{x}}}+\frac{\mu_{0}I}%
{a\pi\left\langle B_{z}\right\rangle }\frac{\partial\phi}{\ \partial
\mathbf{\bar{x}}}\right)  -\frac{\mathbf{\bar{x}}}{\left\vert \mathbf{\bar{x}%
}\right\vert ^{3}} \label{norm eq motion}%
\end{equation}
where
\begin{equation}
\mathbf{\bar{E}}=\frac{q_{\sigma}}{am_{\sigma}\Omega_{0}^{2}}\mathbf{E}%
=-\frac{q_{\sigma}}{a^{2}m_{\sigma}\Omega_{0}^{2}}\frac{\partial V}%
{\partial\mathbf{\bar{x}}}=-\frac{\partial\bar{V}}{\partial\mathbf{\bar{x}}}
\label{Enorm}%
\end{equation}
is the dimensionless electric field and
\begin{equation}
\bar{V}=\frac{aq_{\sigma}V\ }{\ m_{\sigma}MG}\ \label{Vnorm}%
\end{equation}
is the dimensionless electrostatic potential.

Equation \ref{norm eq motion} clearly shows that the dynamics change from
being gravitationally dominated to being magnetically dominated according to
the ratio $\left\langle \omega_{c\sigma}\right\rangle /\Omega_{0}$. The
possibility of complex interactions between gravitational and magnetic forces
when $\left\langle \omega_{c\sigma}\right\rangle /\Omega_{0}$ is of order
unity is also evident. The coefficient $\mu_{0}I/a\pi\left\langle
B_{z}\right\rangle $ is related to the pitch of a twisted field. The
Hamilton-Lagrange formalism shows that $I$ plays a subservient role for
particle orbits compared to $\psi$ because canonical angular momentum depends
on $\psi$, not $I$. However, large $I$ increases $\left\vert B\right\vert $
and so contributes to the effective potential $\mu|B|$ thereby providing
additional possibilities for localization. Thus, if $\mu\ \ $is large, a
particle is not only constrained to stay on a constant $\psi$ surface, but
\ is additionally constrained to stay out of regions on this surface where
$\mu|B|\ $is large. If poloidal currents flow, then the associated
$\mathbf{J}_{pol}\mathbf{\times B}_{tor}$ forces drive jets which inflate and
distend the $\psi$ surfaces. Thus, the orbits will depend indirectly on $I$
when the jet dynamics alter the shape of the constant $\psi$ surfaces.

Using the relations
\begin{equation}%
\begin{array}
[c]{cl}%
\frac{\partial\phi}{\ \partial\mathbf{\bar{x}}}= & \frac{\hat{\phi}}{\bar{r}%
}=\frac{-\hat{x}\bar{y}+\hat{y}\bar{x}}{\bar{x}^{2}+\bar{y}^{2}}\ \\
\frac{\partial\bar{\psi}}{\ \partial\mathbf{\bar{x}}}= & \frac{\partial
\bar{\psi}}{\ \partial\bar{r}}\hat{r}+\frac{\partial\bar{\psi}}{\ \partial
\bar{z}}\hat{z}=\frac{\partial\bar{\psi}}{\ \partial\bar{r}}\left(  \frac
{\hat{x}\bar{x}+\hat{y}\bar{y}}{\sqrt{\bar{x}^{2}+\bar{y}^{2}}}\right)
+\frac{\partial\bar{\psi}}{\ \partial\bar{z}}\hat{z}\ ,
\end{array}
\label{grad phi psi}%
\end{equation}
the normalized equation of motion can be expressed in Cartesian coordinates as%
\begin{equation}%
\begin{array}
[c]{cl}%
\frac{d^{2}\mathbf{\bar{x}}}{d\tau^{2}}= & -\frac{\partial\bar{V}}%
{\partial\mathbf{\bar{x}}}\\
& +\frac{\left\langle \omega_{c\sigma}\right\rangle }{2\bar{r}\Omega_{0}}%
\frac{d\mathbf{\bar{x}}}{d\tau}\times\left(
\begin{array}
[c]{c}%
\ \frac{\partial\bar{\psi}}{\ \partial\bar{r}}\hat{z}-\frac{\partial\bar{\psi
}}{\ \partial\bar{z}}\frac{\mathbf{\bar{r}}}{\bar{r}}\smallskip\\
+\frac{\mu_{0}I}{a\pi\left\langle B_{z}\right\rangle }\frac{\left(  -\hat
{x}\bar{y}+\hat{y}\bar{x}\right)  }{\bar{r}}%
\end{array}
\right) \\
& -\ \ \frac{\left(  \mathbf{\bar{r}}+\bar{z}\hat{z}\right)  }{\left\vert
\bar{r}^{2}+\bar{z}^{2}\right\vert ^{3/2}}%
\end{array}
\label{motion Cartesian}%
\end{equation}
where $\mathbf{\bar{r}}=\bar{x}\hat{x}+\bar{y}\hat{y}\ $ and $\bar{r}%
=\sqrt{\bar{x}^{2}+\bar{y}^{2}}$. Equation \ref{motion Cartesian} is in a form
suitable for numerical computation and has been used to provide the orbital
plots shown earlier.

At this point it is convenient to use the generic poloidal flux function given
by Eq.\ref{generic} so the unity-maximum, dipole-like, normalized flux
function will be
\begin{equation}
\bar{\psi}(\bar{r},\bar{z})=\ \ \frac{27\bar{r}^{2}}{8\left(  \left(  \bar
{r}+\frac{1}{2}\right)  ^{2}+\bar{z}^{2}\right)  ^{3/2}}\ \label{norm generic}%
\end{equation}
with%
\begin{equation}
\frac{\partial\bar{\psi}}{\ \partial\bar{r}}=\ \frac{27\bar{r}\left(
\ \bar{r}+1+4\bar{z}^{2}-\ 2\bar{r}^{2}\right)  }{16\left(  \left(  \bar
{r}+\frac{1}{2}\right)  ^{2}+\bar{z}^{2}\right)  ^{5/2}} \label{dpsibydr}%
\end{equation}
and
\begin{equation}
\frac{\partial\bar{\psi}}{\ \partial\bar{z}}=-\ \ \frac{81\bar{r}^{2}\bar{z}%
}{8\left(  \left(  \bar{r}+\frac{1}{2}\right)  ^{2}+\bar{z}^{2}\right)
^{5/2}}. \label{dpsibydz}%
\end{equation}
Thus, the normalized poloidal magnetic field components are
\begin{equation}
\bar{B}_{r}=\frac{81\bar{r}\bar{z}}{16\pi\left(  \left(  \bar{r}+\frac{1}%
{2}\right)  ^{2}+\bar{z}^{2}\right)  ^{5/2}} \label{Br norm}%
\end{equation}
and
\begin{equation}
\bar{B}_{z}=\frac{27\left(  \ \bar{r}+1+4\bar{z}^{2}-\ 2\bar{r}^{2}\right)
}{32\pi\left(  \left(  \bar{r}+\frac{1}{2}\right)  ^{2}+\bar{z}^{2}\right)
^{5/2}}. \label{Bz norm}%
\end{equation}

We now consider the problem of establishing appropriate initial conditions for
an incoming neutral particle.\ For purposes of starting a computation we
assume the particle is located at some initial radial position $\bar{\rho}%
_{0}$ in the orbital plane such that $\bar{\rho}_{0}>\bar{\rho}_{pericenter}$
where $\bar{\rho}_{pericenter}~\ $is given by Eq.\ref{perigee}. Solving
Eq.\ref{nondim} for the initial inward radial velocity gives%
\begin{equation}
\bar{v}_{\rho0}=-\sqrt{2\bar{H}-\frac{\bar{L}^{2}}{\bar{\rho}_{0}^{2}}%
+\frac{2\ }{\bar{\rho}_{0}}}. \label{initial vrhro}%
\end{equation}
\bigskip and the corresponding initial orbital frame azimuthal velocity is
\begin{equation}
\bar{v}_{\eta0}=\frac{\bar{L}}{\bar{\rho}_{0}}.
\label{initial orbital azimuthal}%
\end{equation}

Equation \ref{solution} can be solved for the initial polar angle in the
orbital frame as
\begin{equation}
\eta=\alpha+\cos^{-1}\left(  \frac{1-\bar{L}^{2}/\bar{\rho}_{0}}{\sqrt
{1+2\bar{L}^{2}\ \bar{H}}}\right)  . \label{initial eta}%
\end{equation}
Using Eq.\ref{unbounded Cartesian} the initial orbital frame Cartesian
coordinates are thus%
\begin{equation}%
\begin{array}
[c]{ccl}%
\bar{x}^{\prime} & = & \frac{\bar{L}^{2}\cos\eta_{0}}{1-\sqrt{1+2\bar{L}%
^{2}\ \bar{H}}\cos\left(  \eta_{0}-\alpha\right)  }\\
\bar{y}^{\prime} & = & \frac{\bar{L}^{2}\sin\eta_{0}}{1-\sqrt{1+2\bar{L}%
^{2}\ \bar{H}}\cos\left(  \eta_{0}-\alpha\right)  }\\
\bar{z}^{\prime} & = & 0.
\end{array}
\label{initial orbital Cartesian}%
\end{equation}
The orbital frame Cartesian velocity components are related to the orbital
frame cylindrical velocity components by%
\begin{equation}%
\begin{array}
[c]{cl}%
\bar{v}_{x^{\prime}0}= & \bar{v}_{\rho0}\cos\eta_{0}-\bar{v}_{\eta0}\sin
\eta_{0}\\
\bar{v}_{y^{\prime}0}= & \bar{v}_{\rho0}\sin\eta_{0}+\bar{v}_{\eta0}\cos
\eta_{0}\\
\bar{v}_{z^{\prime}0}= & 0.
\end{array}
\label{initial orbital vel}%
\end{equation}
We now take into account that the orbital frame Cartesian coordinate system is
rotated by the angle of inclination $\theta$ about the $x$ axis with respect
to the lab frame coordinate system. The $x$ and $x^{\prime}$ components of
both position and velocity are the same in the two frames but the $y$ and $z$
components are related by \
\begin{equation}%
\begin{array}
[c]{cl}%
\bar{y} & =-\bar{z}^{\prime}\sin\theta+\bar{y}^{\prime}\cos\theta\\
\bar{z} & =\bar{z}^{\prime}\cos\theta+\bar{y}^{\prime}\sin\theta.
\end{array}
\label{oribtal to lab transformation}%
\end{equation}
Since $\bar{z}^{\prime}$ is by definition zero in the orbital frame, the
initial lab frame Cartesian coordinates are%
\begin{equation}%
\begin{array}
[c]{cl}%
\bar{x}_{0}= & \bar{x}_{0}^{\prime}\\
\bar{y}_{0}= & \bar{y}_{0}^{\prime}\cos\theta\\
\bar{z}_{0}= & \bar{y}_{0}^{\prime}\sin\theta.
\end{array}
\label{initial lab Cartesian coord}%
\end{equation}
Since $v_{z}^{\prime}\ $is similarly zero in the orbital frame, in analogy to
Eq.\ref{initial lab Cartesian coord}, the initial lab frame Cartesian
velocities are%
\begin{equation}%
\begin{array}
[c]{cl}%
\bar{v}_{x0}= & \bar{v}_{x^{\prime}0}\\
\bar{v}_{y0}= & \bar{v}_{y^{\prime}0}\cos\theta\\
\bar{v}_{z0}= & \bar{v}_{y^{\prime}0}\sin\theta.
\end{array}
\label{initial lab Cartesian vel}%
\end{equation}

Thus, if one wishes to start the numerical computation at the radius
$\bar{\rho}_{0}\,$on the trajectory of an incoming particle with orbit
parameters $\{\bar{H}$,$\bar{L},\theta,\alpha\},$ Eqs.\ref{initial vrhro},
\ref{initial orbital azimuthal}, \ref{initial eta},
\ref{initial lab Cartesian coord} and \ref{initial lab Cartesian vel} give the
appropriate initial position and velocity lab frame Cartesian components.
Before charging, the orbits are degenerate with respect to choice of $\theta$
or $\alpha,$ but after charging there is a strong dependence on these two
angles. In particular, if $0\leq\theta<90^{0}$ the orbit is prograde and
Speiser type orbits are possible if the charging occurs near the poloidal
field magnetic axis. On the other hand if $90^{0}<\theta\leq180^{0}$ the orbit
is retrograde and drain-hole orbits are possible. Thus, a subclass of prograde
incident neutral particles transform upon charging into the
toroidal-current/poloidal-field dynamo while a subclass of retrograde neutral
particles transform upon charging into the poloidal-current/toroidal-field
dynamo that drives a bipolar astrophysical jet. Because $\theta$ and $\alpha$
also affect the angle between the velocity vector and the magnetic field at
charging, $\theta$ and $\alpha$ affect the value of $\mu$ and hence the extent
to which accreted particles with cyclotron orbits will be mirror trapped to
subregions of constant $\psi$ surfaces. For example, if $\alpha=0$ then
variation of the angle of inclination $\theta$ for a given $\ $ $\bar{\rho
}_{pericenter},$ and charging radius $\bar{R}_{\ast}$ will determine whether
the charged particles created upon disintegration of an incoming neutral
particle will be normal trapped particles, untrapped particles, drain-hole
particles, or Speiser particles.

\pagebreak

\bibliographystyle{authordate1}
\bibliography{ApJ-gravitydynamoJuly2006}
\ 
\end{document}